\documentclass[preprint,preprintnumbers,amsmath,amssymb,floatfix,nofootinbib,superscriptaddress,longbibliography]{revtex4-1}

\usepackage{amsmath}
\usepackage{amssymb}
\usepackage[dvips]{graphicx}
\usepackage{url}
\usepackage{xcolor}
\usepackage[colorlinks,citecolor=blue,urlcolor=blue,linkcolor=blue]{hyperref}

\def\tb{\bar{t}}
\def\bb{\bar{b}}
\def\lo{\mathrm{LO}}
\def\nlo{\mathrm{NLO}}
\def\nwa{\mathrm{NWA}}
\def\nlops{\mathrm{NLO+PS}}
\def\GeV{\mathrm{GeV}}
\def\TeV{\mathrm{TeV}}
\def\proc{$pp \to \ell^+\nu\ell^-\nu\ell^\pm\nu~b\bb$}
 
\def\helacnlo{\textsc{Helac-NLO}}
\def\powhegbox{\textsc{Powheg-Box}}
\def\pythia{\textsc{Pythia8}}
\def\powheg{POWHEG}
\def\mgfive{\textsc{MG5\_aMC@NLO}}
\def\mg5{\textsc{MadGraph5}}
\def\mcnlo{MC@NLO}
\def\rivet{\textsc{Rivet}}
\def\lhapdf{\textsc{LHAPDF}}
\def\fastjet{\textsc{FastJet}}
\def\madspin{\textsc{MadSpin}}
\def\feyngame{\textsc{FeynGame}}
\def\heplot{\textsc{Heplot}}
\def\nlox{\textsc{NLOX}}

\usepackage[normalem]{ulem}

\begin{document}
\title{Modeling uncertainties of $t\tb W^\pm$ multilepton signatures}
\preprint{CAVENDISH-HEP-21/12, P3H-21-063, TTK-21-37}
\author{G. Bevilacqua}
\affiliation{ELKH-DE Particle Physics Research Group, University of Debrecen, 
H-4010 Debrecen, PBox 105, Hungary}
\author{H. Y. Bi}
\affiliation{Institute for Theoretical Particle Physics and Cosmology, 
RWTH Aachen University, D-52056 Aachen, Germany}
\author{F. Febres Cordero}
\affiliation{Physics Department, Florida State University, Tallahassee, 
FL 32306-4350, U.S.A.}
\author{H. B. Hartanto}
\affiliation{Cavendish Laboratory, University of Cambridge, J.J. Thomson Avenue, 
Cambridge CB3 0HE, United Kingdom}
\author{M. Kraus}
\affiliation{Physics Department, Florida State University, Tallahassee, 
FL 32306-4350, U.S.A.}
\author{J. Nasufi}
\affiliation{Institute for Theoretical Particle Physics and Cosmology, 
RWTH Aachen University, D-52056 Aachen, Germany}
\author{L. Reina}
\affiliation{Physics Department, Florida State University, Tallahassee, 
FL 32306-4350, U.S.A.}
\author{M. Worek}
\affiliation{Institute for Theoretical Particle Physics and Cosmology, 
RWTH Aachen University, D-52056 Aachen, Germany}

\date{\today}

\begin{abstract}
In light of recent discrepancies between the modeling of $t\tb W^\pm$
signatures and measurements 
reported by the Large Hadron Collider (LHC) experimental collaborations, we 
investigate in detail theoretical uncertainties for multi-lepton signatures. We 
compare results from the state-of-the-art full off-shell calculation and its 
Narrow Width Approximation to results obtained from the on-shell $t\tb W^\pm$ 
calculation, with approximate spin-correlations in top-quark and $W$ decays, 
matched to parton showers. In the former case double-, single-, and non-resonant 
contributions together with interference effects are taken into account, while 
the latter two cases are only based on the double resonant top-quark 
contributions.
The comparison is performed for the LHC at $\sqrt{s} = 13~\TeV$ for which we 
study separately the multi-lepton signatures as predicted from the dominant NLO
contributions at the perturbative orders $\mathcal{O}(\alpha_s^3\alpha^6)$ and 
$\mathcal{O}(\alpha_s\alpha^8)$.
Furthermore, we combine both contributions and propose a simple way to approximately
incorporate the full off-shell effects in the NLO computation of on-shell 
$pp\to t\tb W^\pm$ matched to parton showers.
\end{abstract}

\maketitle
\section{Introduction}
The hadronic production of top-quark pairs in association with a $W$ boson is 
one of the most massive signatures currently accessible at the Large Hadron 
Collider (LHC) and allows to study possible deviations from the Standard Model 
(SM) dynamics of top quarks in the presence of charged electroweak (EW) gauge 
bosons. As the accompanying top quarks are rapidly decaying into a $W$ boson and 
a $b$ quark the signature of the $pp \to t\tb W^\pm$ process involves two $b$ 
jets and the subsequent decay pattern of three decaying $W$ bosons. This gives 
rise to the rare Standard Model production of same-sign lepton pairs and other 
multi-lepton signatures that are relevant to a multitude of 
searches~\cite{ATLAS:2016dlg,CMS:2016mku,ATLAS:2017tmw,CMS:2017tec} for physics 
beyond the Standard Model (BSM) as well as to the measurement of Higgs-boson 
production in association with a top-quark pair
\cite{ATLAS:2018mme,CMS:2018uxb,ATLAS:2019nvo,CMS:2020iwy} and of four 
top-quarks~\cite{ATLAS:2018kxv,CMS:2019rvj,ATLAS:2020hpj,ATLAS:2021kqb}.

Due to the importance of the $pp \to t\tb W^\pm$ process to validate the EW 
interactions of the top quark and as a dominant background to many ongoing 
measurements and searches at the LHC, it is crucial to have the process under 
excellent theoretical control. Interestingly, small tensions between the $t\tb W$ 
measurements and the SM predictions have been reported since its discovery. The
first measurements~\cite{ATLAS:2015qtq,CMS:2015uvn} during the LHC Run 1 at a
center-of-mass energy of $\sqrt{s}=8~\TeV$ already reported tensions for the
measured cross section slightly above the $1\sigma$ level. Further measurements
\cite{ATLAS:2016wgc,CMS:2017ugv,ATLAS:2019fwo} performed at $\sqrt{s}=13~\TeV$
confirmed the picture that a slight excess of $t\tb W$ events is observed.
Additional measurements of the $t\tb W$  cross section~\cite{ATLAS:2019nvo,
CMS:2020iwy,CMS:2019rvj,ATLAS:2020hpj}, from analyses aimed at the measurement of
$t\tb H$ and $t\tb t \tb$ production and utilizing data corresponding to an 
integrated luminosity of up to $\mathcal{L}=139~\textrm{fb}^{-1}$, still see a 
persisting tension with respect to the corresponding SM predictions depending on 
the considered final-state signature, where the largest deviation of up to a 
factor of $1.7$ has been found in multi-lepton signatures~\cite{ATLAS:2019nvo}.

Fueled by these deviations, tremendous progress has been made in the last decade
in the theoretical description of the $pp\to t\tb W^\pm$ processes. The first 
next-to-leading (NLO) QCD corrections have been computed in 
Refs.~\cite{Badger:2010mg,Campbell:2012dh}, while EW contributions have been 
investigated in Refs.~\cite{Dror:2015nkp,Frixione:2015zaa,Frederix:2017wme}.
Furthermore, the resummation of soft gluon effects at the next-to-next-to-leading
logarithmic (NNLL) level has been achieved recently in Refs.~\cite{Li:2014ula,
Broggio:2016zgg,Kulesza:2018tqz,Broggio:2019ewu,Kulesza:2020nfh}. Targeting a 
more realistic description of fiducial signatures, the $t\tb W^\pm$ process has 
been matched to parton showers either using the \mcnlo{} matching 
scheme~\cite{Frixione:2002ik,Frixione:2003ei} in Refs.~\cite{Maltoni:2014zpa,
Maltoni:2015ena,Frederix:2020jzp}, or the \powheg{} method~\cite{Nason:2004rx,
Frixione:2007vw} in Refs.~\cite{Garzelli:2012bn, Cordero:2021iau}. The impact of
higher-order corrections via multi-jet merging has been studied as well in 
Refs.~\cite{vonBuddenbrock:2020ter,ATLAS:2020esn,Frederix:2021agh}. In parallel 
also the inclusion of off-shell effects for top quarks and $W$ bosons as well as 
single and non-resonant contributions together with interference effects have 
been studied including NLO QCD corrections~\cite{Bevilacqua:2020pzy,Denner:2020hgg,
Bevilacqua:2020srb} as well as complete NLO SM 
corrections~\cite{Denner:2021hqi}.

The emergent picture from all the aforementioned studies is that NLO QCD 
corrections are of the order of $8\%-20\%$ depending on the chosen SM input 
parameters and cuts applied on the final states. The resummation of soft gluon 
effects in on-shell $t\tb W$ production improves only marginally the perturbative
convergence of NLO QCD predictions and increases the rate by less than $10\%$. 
The second largest contribution in the perturbative expansion of the cross 
section comprises the NLO QCD corrections to the pure EW born and amounts roughly
to a $+10\%$ correction to the total NLO cross section. Including higher-order 
corrections via multi-jet merging also increases the inclusive cross section by a
similar amount. On the other hand QCD corrections to top-quark decays are 
negative and decrease the production rate by roughly $5\%$. Finally, even 
though off-shell effects and non-resonant contributions are small at the 
integrated level, they become sizable in the tails of dimensionful observables 
where they can make a difference even up to $70\%$. Overall, none of the effects 
described above can fully explain the tension between the measurements and the SM
prediction. However, given the complexity of the signatures involved and the 
progress made to theoretically model them at both the integrated and differential
fiducial level, a study aimed at comparing existing computations and 
understanding the origin of their differences as well as the individual residual
uncertainties is of the utmost importance to eventually propose a theory 
recommendation for the comparison with data.

As existing computations are often quite different in nature and therefore 
inherently affected by distinctive theoretical uncertainties, it is not evident 
how to combine the various findings into a precise modeling of the $t\tb W$ 
process. In this paper we attempt to do a first comparison between parton-shower 
matched predictions and fixed-order full off-shell calculations, together with 
their corresponding narrow-width-approximations (NWA), in order to understand the 
different approaches in more detail. In all cases, we include NLO QCD corrections
to both the QCD and EW born level, in the following simply denoted by QCD and EW 
(see Sec.~\ref{sec:setup}). Incidentally, we also quantify for the first time the
size of full off-shell effects for the EW contribution. Our comparison focuses on
the multi-lepton signature as it is not only the cleanest signature on the 
theoretical side but also yields the strongest discrepancies when compared to 
measurements. 

The paper is organized as follows. In section~\ref{sec:setup} we summarize the 
computational setup of the calculations employed in our study, while in 
section~\ref{sec:differences} we highlight in detail the differences between the 
various theoretical approaches. In section~\ref{sec:pheno} we study the modeling
of top-quark production and decays separately for the QCD and EW contributions. 
In section~\ref{sec:improve} we show our phenomenological results and propose a
simple method to approximately capture the full off-shell effects in 
parton-shower matched calculations for on-shell $t\tb W^\pm$ production. Finally,
we give our conclusions in section~\ref{sec:conclusions}.

\section{Computational setup}
\label{sec:setup}
In our study we consider the \proc{} process at $\mathcal{O}(\alpha_s^3\alpha^6)$
and $\mathcal{O}(\alpha_s\alpha^8)$, where $\ell = e,\mu$ and all possible 
lepton-flavor combinations are considered. In the following we will refer to the 
former perturbative order as $t\tb W^\pm$ QCD and the latter as $t\tb W^\pm$ EW. 
We will provide predictions for the LHC operating at a center-of-mass energy of 
$\sqrt{s} = 13~\TeV$ and parametrize the proton content using the 
\textrm{NNPDF~3.1}~\cite{NNPDF:2017mvq} PDF set as provided by the \lhapdf{} 
interface~\cite{Buckley:2014ana}. We employ the necessary SM input parameters in 
the $G_F$ input scheme~\cite{Denner:2000bj}. For the electroweak bosons we choose
the following masses and corresponding decay widths 
\begin{equation}
\begin{alignedat}{2}
 M_W &= 80.3850~\GeV\;, \qquad \Gamma_W &&= 2.09767~\GeV\;, \\
 M_Z &= 91.1876~\GeV\;, \qquad \Gamma_Z &&= 2.50775~\GeV\;, \\
 M_H &= 125.000~\GeV\;, \qquad \Gamma_H &&= 0.00407~\GeV\;.
\end{alignedat}
\end{equation}
Together with the Fermi constant $G_F = 1.166378\cdot 10^{-5}~\GeV^{-2}$ the 
electromagnetic coupling is given by
\begin{equation}
 \alpha = \frac{\sqrt{2}}{\pi}G_F M_W^2\left( 1 -\frac{M_W^2}{M_Z^2}\right)\;.
\end{equation}
Finally, we choose for the top-quark mass
\begin{equation}
    m_t = 172.5~\GeV\;.
\end{equation}
We postpone the discussion of the top-quark width as it depends on the 
approximation used in the corresponding calculation. Let us now turn to the 
description of the various theoretical predictions employed in our study.
\begin{description}
\item[\textbf{Full off-shell}] 
Using the \helacnlo{} framework~\cite{Bevilacqua:2011xh,Cafarella:2007pc,
vanHameren:2009dr,Czakon:2009ss,Ossola:2007ax,Bevilacqua:2013iha,Czakon:2015cla} 
we generate results at fixed-order NLO QCD accuracy employing the matrix elements
for the fully decayed final state \proc{} at 
$\mathcal{O}(\alpha_s^3\alpha^6)$ and $\mathcal{O}(\alpha_s\alpha^8)$. This 
approach includes all double-, single-, and non-resonant top-quark and $W$-boson 
contributions as well as interference and spin-correlation effects at the matrix 
element level. Top quarks and electroweak gauge bosons are described in the 
complex-mass scheme~\cite{Denner:1999gp,Denner:2000bj,Denner:2005fg} by 
Breit-Wigner propagators. Therefore, the top-quark decay width is calculated
at NLO QCD accuracy including off-shell $W$ bosons according to formulae given in
Refs.~\cite{Jezabek:1988iv,Chetyrkin:1999ju,Denner:2012yc}, which for our study 
corresponds to:
\begin{equation}
 \Gamma_{t,\textrm{off-shell}}^\nlo = 1.33247~\GeV\;.
\end{equation}
Bottom quarks are treated consistently as massless quarks throughout the 
computation. The renormalization and factorization scales are chosen as
\begin{equation}
 \mu_0 = \frac{H_T}{3}\;, \qquad \text{with} \qquad
 H_T = \sum_{i=1}^3 p_T(\ell_i) + \sum_{i=1}^2 p_T(b_i) + p_T^{\textrm{miss}}\;.
\end{equation}
More details on the calculation can be found in Ref.~\cite{Bevilacqua:2020pzy}.
%
\item[\textbf{NWA}] 
Benefiting from the recent automation of the NWA in the \helacnlo{} 
framework~\cite{Bevilacqua:2019quz} we are able to provide results at NLO QCD 
accuracy including spin-correlated top-quark decays. In this approach, the full 
matrix elements for \proc{} are approximated 
by the double resonant $t\tb W^\pm$ contributions via a factorization of the 
cross section into a production and a decay stage by applying the following 
relation to top-quark and $W$-boson propagators
\begin{equation}
 \lim_{\Gamma/m\to 0}  \frac{1}{(p^2-m^2)^2 + m^2\Gamma^2} =
 \frac{\pi}{m\Gamma}\delta(p^2-m^2) + \mathcal{O}\left(\frac{\Gamma}{m}\right)\;.
 \label{eqn:nwa_limit}
\end{equation}
In this limit the cross section can be written as
\begin{equation}
\begin{split}
 d\sigma_\nwa = d\sigma_{pp\to t\tb W^\pm}~&\otimes d\mathcal{B}_{t\to bW^+}~
 \otimes~d\mathcal{B}_{\tb\to \bb W^-} \\
 &\otimes~d\mathcal{B}_{W^+\to\ell^+\nu}~\otimes~
 d\mathcal{B}_{W^-\to\ell^-\nu}~\otimes~
 d\mathcal{B}_{W^\pm\to\ell^\pm\nu}\;,
\end{split}
\end{equation}
where $\mathcal{B}$ denotes the corresponding branching ratios and $\otimes$ 
indicates that full spin correlations are kept. The computation is tremendously 
simplified with respect to the full off-shell calculation but still allows to 
systematically include NLO QCD corrections separately in the production and decay
stage. In this case the top-quark decay width is given by
\begin{equation}
 \Gamma_{t,\textrm{NWA}}^\nlo = 1.35355~\GeV\;.
\end{equation}
If QCD corrections to the top-quark decay are omitted, which we denote as ``NWA LO
decay'', we use the leading-order (LO) prediction for the top-quark
width given by
\begin{equation}
 \Gamma_{t,\textrm{NWA}}^\lo = 1.48063~\GeV\;.
\end{equation}
For the renormalization and factorization scales we choose as before
\begin{equation}
    \mu_0 = \frac{H_T}{3}\;.
\end{equation}
%
\item[\textbf{\powhegbox{}}] 
We employ parton-shower matched results for $pp\to t\tb W^\pm$ using the recent 
\powhegbox{} implementation presented in Ref.~\cite{Cordero:2021iau}, based on 
one-loop amplitudes provided via \nlox~\cite{Honeywell:2018fcl,Figueroa:2021txg}.
In this calculation the on-shell production of $t\tb W^\pm$ is matched to parton 
showers at NLO accuracy, meaning that NLO QCD corrections are only included in 
the production stage. The renormalization and factorization scales are chosen 
according to
\begin{equation}
 \mu_0  = \frac{E_T}{3}\;. \qquad \text{with} \qquad E_T =\sum_{i \in \{t,\tb,W\}} 
 \sqrt{m_i^2 + p_{T,i}^2}\;.
 \label{eqn:pwg_scale}
\end{equation}
In the matching procedure, the \powhegbox{} employs two damping parameters,
$h_\mathrm{damp}$ and $h_\mathrm{bornzero}$, to define a jet-function that 
classifies the soft and hard real contributions. Based on the results
of Ref.~\cite{Cordero:2021iau}, in this study we choose the following
central values
\begin{equation*}
 h_\mathrm{damp} = \frac{E_T}{2}\;, \qquad h_\mathrm{bornzero} = 5\;,
\end{equation*}
and estimate the dependence of our results on these parameters by varying their 
values independently in the ranges
\begin{equation}
 h_\mathrm{damp} = \left\{ \frac{E_T}{4}\;, \frac{E_T}{2}\;, E_T \right\}\;,
 \qquad \text{and} \qquad 
 h_\mathrm{bornzero} = \{2,5,10\}\;.
\end{equation}
The decay of the top quarks and the $W$ bosons are included at LO 
accuracy using the method introduced in Ref.~\cite{Frixione:2007zp} that allows 
to retain spin correlations and to introduce a smearing of top-quark and $W$ 
boson virtualities according to Breit-Wigner distributions. However, single- and 
non-resonant top-quark and $W$-boson contributions are still missing. They 
are not included even with LO accuracy. As the decay modeling
in the \powhegbox{} does not follow any particular fixed-order scheme we use the 
LO top-quark width including off-shell effects for the $W$ boson
\begin{equation}
 \Gamma_{t,\textrm{off-shell}}^\lo = 1.45759~\GeV\;.
\end{equation}
However, the top-quark decay width enters only the decay chain matrix elements
which employ Breit-Wigner propagators, see Ref.~\cite{Cordero:2021iau}. Thus, the
chosen value of $\Gamma_t$ has generally a very small impact on our results.
We employ NLO accurate branching ratios for the $W$ boson decays to be consistent
with the full off-shell calculation. Thus, the total leptonic branching ratio is 
given by
\begin{equation}
 \mathrm{Br}(W \to \ell\nu) = 
 \sum_{i=1}^3 \frac{\Gamma(W\to \ell_i\nu_i)}{\Gamma_W^\nlo} = 0.325036\;.
 \label{eqn:Wbranch}
\end{equation}
%
\item[\textbf{\mgfive{}}] 
We also use parton-shower matched computations for the on-shell 
$pp\to t\tb W^\pm$ process in the \mcnlo{} framework as provided by
\mgfive{}~\cite{Alwall:2014hca}. This approach has the same formal accuracy as 
the aforementioned \powhegbox{} computation. The renormalization and 
factorization scales are also chosen as
\begin{equation}
 \mu_0 = \frac{E_T}{3}\;,
\end{equation}
where $E_T$ is defined in Eq.~(\ref{eqn:pwg_scale})\footnote{The default 
dynamical scale choice for the renormalization and factorization scales in the 
\powhegbox{} and in \mgfive{} actually corresponds to the definition of 
Eq.~\eqref{eqn:mg5_scale}. We noticed that our choice of 
Eq.~\eqref{eqn:pwg_scale} increases the total cross section by approximately 
$11\%$, which is within the estimated scale uncertainties as we will see later 
on.}, while the initial shower scale $\mu_Q$ is left to the default option
\begin{equation}
 \mu_Q = \frac{1}{2} \sum_{i \in \{t,\tb,W,j\}} \sqrt{m_i^2 + p_{T,i}^2}\;.
 \label{eqn:mg5_scale}
\end{equation}
We study the dependence of our results on the initial shower scale $\mu_Q$ by 
varying it up and down by a factor of $2$. LO spin-correlated top-quark and 
$W$-boson decays are included via \madspin{}~\cite{Artoisenet:2012st} that 
automates the approach of Ref.~\cite{Frixione:2007zp}. Within \madspin{} the 
necessary branching ratios are evaluated at leading-order accuracy by default. 
However, we rescale our results to be consistent with Eq.~\eqref{eqn:Wbranch} as 
discussed in the next section.
\end{description}
In all cases, we choose for the renormalization and factorization scales
\begin{equation}
    \mu_R = \mu_F = \mu_0\;,
\end{equation}
where the specific choice of $\mu_0$ in each computational framework considered 
has been discussed above. The dependence on the scale choice is estimated from 
independent variations of renormalization and factorization scales in the range 
of
\begin{equation}
 \left(\frac{\mu_R}{\mu_0}, \frac{\mu_F}{\mu_0}\right) = \Big\{ 
 (0.5,0.5), (0.5,1), (1,0.5),(1,1), (1,2), (2,1), (2,2)\Big\}\;.
\end{equation}
By searching for the minimum and the maximum of the resulting cross sections
one obtains an uncertainty band.

For the parton-shower matched predictions, we generate Les Houches event files
\cite{Boos:2001cv,Alwall:2006yp} that are subsequently showered with \pythia{}
\cite{Sjostrand:2006za,Sjostrand:2014zea}. In the parton-shower evolution we do 
not take into account hadronization effects and multiple parton interactions.
The analysis of showered events is then performed within the \rivet{} 
framework~\cite{Buckley:2010ar,Bierlich:2019rhm} that also provides an interface 
to \fastjet{}~\cite{Cacciari:2005hq,Cacciari:2011ma}. 

Due to the complexity of the full off-shell computation we store our theoretical 
predictions obtained with the \helacnlo{} framework in the form of events, either
available in the format of modified Les Houches event files~\cite{Alwall:2006yp} 
or ROOT Ntuples~\cite{Antcheva:2009zz}. Based on ideas presented in 
Ref.~\cite{Bern:2013zja} events are stored with additional information about
matrix elements and PDFs. We improve the performance of our framework by storing
partially unweighted events, see e.g. Ref.~\cite{Bevilacqua:2016jfk} for more
information. The Ntuples allow for on-the-fly reweighting for different
PDF sets or scale choices as well as studying different fiducial phase space cuts.
The reweighting and event analysis is performed within \heplot{}~\cite{heplot}.

The shower input cards, the \rivet{} analysis code and all histogram data are 
made available in the ancillary material.

\section{Differences of the various computational approaches}
\label{sec:differences}
Before presenting our phenomenological results we want to discuss the intrinsic 
differences between the various computational approaches introduced above.

Let us start with the purely technical issue of choosing renormalization and 
factorization scales $\mu_0$ for the various approaches. In order to obtain a 
good perturbative convergence for dimensionful observables a dynamic-scale choice
has to be employed. However, due to the nature of the different approaches we can
not use a common scale choice. For instance, the off-shell and NWA predictions 
use scales evaluated on the momenta of the fully decayed final state, while 
parton-shower based predictions use the on-shell momenta of the intermediate 
unstable particles. We checked explicitly that our conclusions in the following 
sections do not depend significantly on the scale chosen. Specifically we 
performed a comparison using the fixed scale $\mu_0 = m_t + m_W/2$ that can be 
employed in all predictions. Shape differences of differential distributions, 
induced by the dynamic scale choices, are common in magnitude and for all 
predictions they are independent of the specific definition of the scale. 

Next we want to discuss differences that affect predominantly the overall 
normalization of the predictions. First, we focus on differences between 
\mgfive{} and the \powhegbox{}. The inclusive cross section for the three-lepton 
signature, as predicted by parton-shower matched calculations, is given in 
terms of production cross section times branching ratios
\begin{equation}
 \sigma_{3\ell}^\nlops = \sigma(pp\to t\tb W^\pm) 
 \times\mathrm{Br}(t\to W^+b)~\mathrm{Br}(\tb\to W^-\bb) 
 \times8\left[\mathrm{Br}(W^\pm \to \ell_i\nu_i)\right]^3\;,
\end{equation}
where the factor $8$ accounts for all combinations of $\ell=e,\mu$.
In parton-shower based predictions the top-quark decays always into a $W$ boson
and a $b$ quark, thus the branching ratio is $\mathrm{Br}(t\to Wb)=1$ and drops 
out from any inclusive or fiducial cross section. In other words, the obtained
cross section is completely independent of $\Gamma_t$, even in fiducial phase 
space volumes. However, the cross section is very sensitive to the branching 
ratio for the leptonic decay width of the $W$ boson as it depends on it cubicly. 
In \madspin{} all branching ratios are computed at LO accuracy from the given 
input parameters. On the other hand in the \powhegbox{} the total leptonic $W$ 
boson branching ratio is an independent input parameter that we set to the NLO 
value. The difference between LO and NLO branching ratios is only of the order of
$2.5\%$. Nonetheless, due to the strong dependence of the cross section on this 
parameter, we observe a $\mathcal{O}(10\%)$ higher cross section if LO branching
ratios are employed. In order to obtain a cleaner comparison we rescale the 
\mgfive{} predictions globally by a factor of
\begin{equation}
 \left( \frac{\mathrm{Br}^\nlo(W^\pm \to \ell_i\nu_i)}
 {\mathrm{Br}^\lo(W^\pm \to \ell_i\nu_i)}\right)^3 \approx 0.92716\ldots\;.
\end{equation}

Considering next the NWA approach, we observe that in this case the cross section
depends globally on the top-quark decay width via $1/\Gamma_t^2$. Therefore, 
contrary to parton-shower based calculation the inclusive cross section depends 
crucially on the chosen value of $\Gamma_t$. Moreover, to ensure consistency in 
the comparison between full off-shell results and predictions in the NWA the 
unexpanded NWA results are used. We have studied in Ref.~\cite{Bevilacqua:2020pzy}
that the impact of these higher-order terms is at the level of $3\%-4\%$.

\begin{figure}[h!]
\begin{center}
 \includegraphics[width=\textwidth]{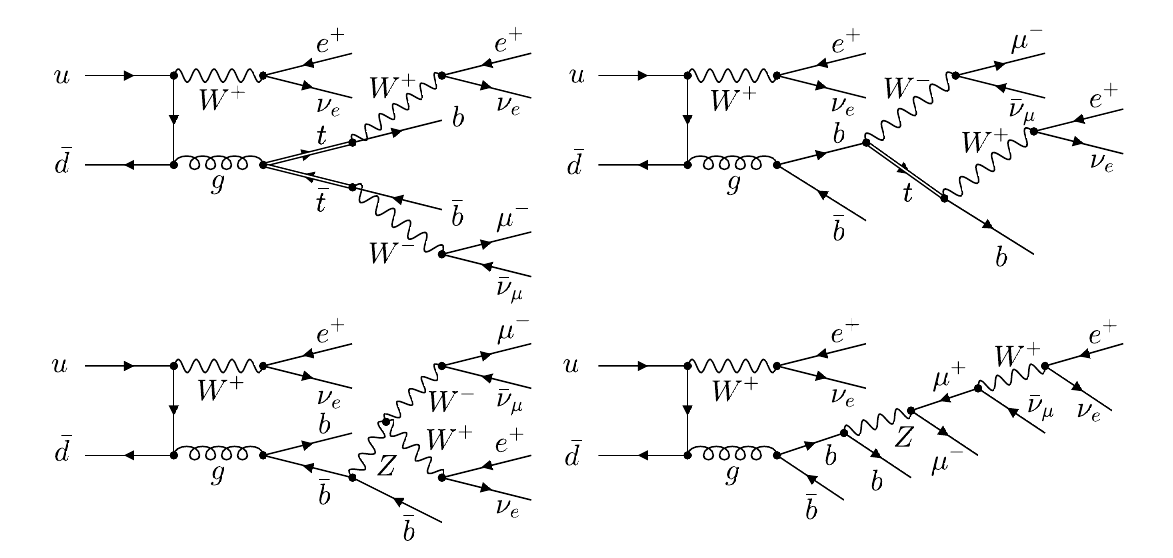}
\end{center}
\caption{Representative Feynman diagrams of the double-, single-, and non-resonant
top-quark contributions that are part of the matrix elements employed in the
full off-shell computation. Feynman diagrams were produced with the help of the
\feyngame{} program \cite{Harlander:2020cyh}.}
\label{fig:resonants}
\end{figure}
With respect to the NWA, the full off-shell calculation includes
additional contributions. Besides the double-resonant $pp\to t\tb W^\pm$ 
production the off-shell matrix elements also receive contributions from 
single-resonant $pp\to tWWb$ and non-resonant $pp\to WWWb\bb$ production modes as
well as their interferences. We show some representative Feynman diagrams for each
contribution in Fig.~\ref{fig:resonants}. We want to stress, however, that the
single and non-resonant production cross sections can not be computed separately
in a gauge invariant way, as these processes start to mix once higher-order 
corrections are included. Therefore, the only unambiguous way to account for 
these contributions is to employ the full matrix elements for \proc{}.
Notice, that the $\Gamma_t$ dependence in the full off-shell computation is due 
to Breit-Wigner propagators of the double and single-resonant contributions. 

Finally, the last conceptual difference between the fixed-order computations and
the parton-shower predictions concerns the definition of $b$ jets. 
The fixed-order computations are performed in the $5$-flavor scheme treating 
$b$ quarks as massless. In this case, we define the $b$-jet flavor using the net 
\textit{bottomness} in a jet, according to the recombination rules
\begin{equation}
 bg \to b\;, \qquad \bb g\to \bb\;, \qquad b\bb \to g\;,
\end{equation}
which renders the jet-flavor definition infrared safe at NLO. Beyond NLO the
flavored $k_T$-jet algorithm of Ref.~\cite{Banfi:2006hf} is the most adopted 
solution to define the flavor of jets for massless partons. Recently a 
flavored anti-$k_T$ jet algorithm has also been introduced~\cite{antikTflav}.
On the other hand, the $b$ quark is treated as a massive quark in the 
parton-shower evolution. As the collinear divergence in the $g\to b\bb$ splitting
is regulated by the bottom-quark mass we can define a $b$ jet to be a jet that has 
at least one $b$ quark among its constituents. Hadronization effects can have an 
impact on the definition of $b$-jet observables. However, in preliminary 
calculations where we included hadronization effects we only found marginal 
differences for the considered $b$-jet observables. We have also checked 
explicitly that a parton-shower evolution with massless $b$ quarks but employing 
the jet algorithm of Ref.~\cite{Banfi:2006hf}\footnote{We thank Gavin Salam for 
providing us with his private implementation of the flavor-$k_T$ jet algorithm in 
\fastjet{}.} yields results that are very similar to the default approach of 
massive $b$ quarks clustered via the anti-$k_T$ jet algorithm.

\section{Phenomenological results}
\label{sec:pheno}
For our comparison we focus on the three-lepton signature, where we exclude 
$\tau$ leptons as their decays produce very distinct signatures and 
can be easily distinguished from electron or muon final states. Jets are formed 
using the anti-$k_T$ jet algorithm~\cite{Cacciari:2008gp} with a separation 
parameter $R = 0.4$. We require at least 2 $b$ jets that
satisfy the conditions
\begin{equation}
 p_T(j_b) > 25~\GeV\;, \qquad |\eta(j_b)| < 2.5\;.
\end{equation}
Furthermore, we require the presence of exactly three charged leptons with 
\begin{equation}
 p_T(\ell) > 25~\GeV\;, \qquad |\eta(\ell)| < 2.5\;, \qquad 
 \Delta R(\ell\ell) > 0.4\;, \qquad \Delta R(\ell j_b) > 0.4\;.
\end{equation}
Note that in our case the three charged leptons always consist of a two-same-sign
lepton pair $\ell\ell^{ss}$ and a single lepton with opposite charge $\ell^{os}$.

\subsection{QCD production of $t\tb W^\pm$}
We start our discussion with the dominant QCD production mechanism for the 
$pp \to t\tb W^\pm$ process and its leptonic decay at 
$\mathcal{O}(\alpha_s^3\alpha^6)$.

\subsubsection{Integrated fiducial cross sections}
First we present a comparison of fiducial cross sections and 
their uncertainties. For the off-shell calculation we obtain
\begin{equation}
 \sigma^\nlo_{\textrm{off-shell}} = 1.58^{+0.05~(3\%)}_{-0.10~(6\%)}
 ~\mathrm{fb}\;,
\end{equation}
which shows a very reduced sensitivity to the choice of the renormalization and 
factorization scales with changes only between $+3\%$ and $-6\%$ around the 
central prediction. We do not investigate PDF uncertainties here, since these 
have already been addressed in Ref.~\cite{Bevilacqua:2020pzy} and were estimated 
to be of the order of $2\%-3\%$ depending on the PDF set used. For the NWA with 
and without NLO QCD corrections to the top-quark decays we obtain
\begin{equation}
 \sigma^\nlo_{\textrm{NWA}} = 1.57^{+0.05~(3\%)}_{-0.10~(6\%)}
 ~\mathrm{fb}\;, \qquad
 \sigma^\nlo_{\textrm{NWA LOdec}} = 1.66^{+0.17~(10\%)}_{-0.17~(10\%)}
 ~\mathrm{fb}\;.
\end{equation}
We observe that the full NWA is in perfect agreement with the off-shell 
prediction. However, if QCD corrections to top-quark decays are neglected we 
obtain a $5\%$ larger cross section and the residual scale sensitivity increases 
up to $\pm 10\%$. Finally, for the parton-shower matched computations we find 
\begin{equation}
 \sigma^\nlops_{\textrm{PWG}} = 1.40^{+0.16~(11\%)}_{-0.15~(11\%)}
 ~\mathrm{fb}\;, \qquad
 \sigma^\nlops_{\textrm{MG5}} = 1.40^{+0.16~(11\%)}_{-0.15~(11\%)}
 ~\mathrm{fb}\;. 
\end{equation}
The remarkable agreement between the two predictions is not a coincidence since
we have corrected for the different values of branching ratios employed
in the calculations (see Sec.~\ref{sec:differences}) and have aligned
as much as possible \pythia{} parameters between 
\mgfive{} and the \powhegbox{}. The parton-shower matched computations predict a 
$11\%$ smaller cross section if compared to the off-shell result. The reduction
of the cross section is due to multiple emissions in the top-quark decays that
are not clustered back into the corresponding $b$ jet and thus decrease the 
amount of events passing our selection cuts. We checked explicitly that if we 
turn off parton-shower emissions in the decays we obtain a larger cross section, 
closer to the result of the NWA with LO decays. Finally, the $\nlops$ results 
show, similar to the NWA LO-decay result, a scale sensitivity of the order of 
$\pm 11\%$, as expected since both computations include NLO QCD corrections only 
in the production part of the $pp\to t\tb W^\pm$ processes.

\subsubsection{Differential distributions}
At the differential level we start by comparing hadronic $b$-jet observables. 
All observables are shown as plots containing three panels. The upper panel 
always depicts the central differential distribution (i.e. the distribution 
obtained using the central values of both scale and matching parameters) for the 
various predictions employed in our study. The middle panel illustrates the scale
uncertainty bands stemming from independent variations of factorization and
renormalization scales.
All curves are normalized to the central prediction of the off-shell calculation
and for ease of readability we do not show uncertainty bands for the predictions
based on the NWA. Finally, the bottom panel shows the matching uncertainties for 
the parton-shower based predictions which are estimated by a variation of the
initial shower scale $\mu_Q$ in the case of \mgfive{} and of the various damping 
parameters in the case of the \powhegbox{}.

%
\begin{figure}[h!]
\begin{center}
 \includegraphics[width=0.49\textwidth]{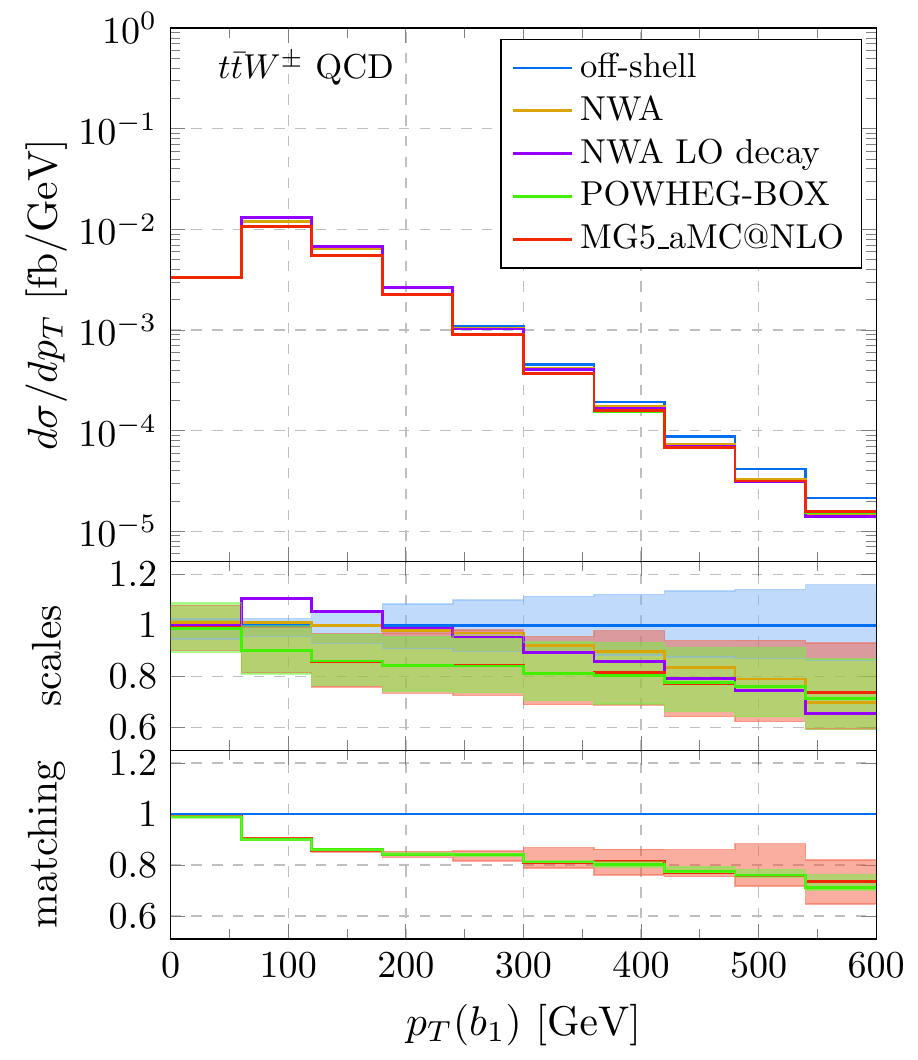}
 \includegraphics[width=0.49\textwidth]{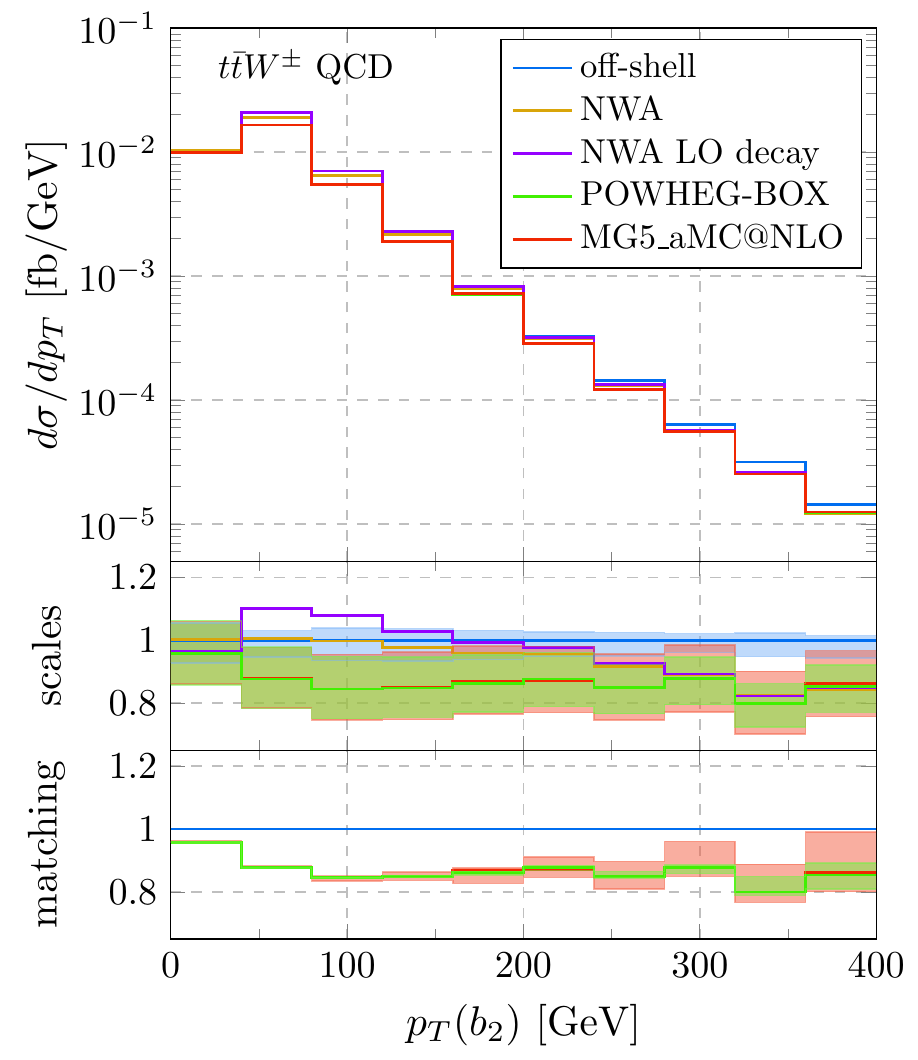}
\end{center}
\caption{Differential cross section distribution in the $3\ell$ fiducial region 
as a function of the transverse momentum of the hardest (l.h.s.) and the 
second hardest (r.h.s) $b$ jet for the $pp\to t\tb W^\pm$ QCD process. 
The uncertainty bands correspond to independent variations of the renormalization
and factorization scales (middle panel) and of the matching parameters (bottom 
panel).}
\label{fig:fig_QCD_1}
\end{figure}
On the left of Fig.~\ref{fig:fig_QCD_1} the transverse momentum distribution of 
the hardest $b$ jet is shown. We observe shape differences between the various 
predictions over the whole plotted range. Results in the full NWA have the same 
shape as those of the off-shell calculation for $p_T(b_1) \lesssim 300~\GeV$. 
However, all of them differ by $27\%-35\%$ from the full calculation at 
$p_T \sim 600~\GeV$. This is not surprising since in this region the 
single-resonant contribution from $pp\to tWWb$ becomes sizable\footnote{We 
checked this explicitly via a LO computation for $pp\to tWWb$ that removes all 
double-resonant diagrams. Even though this approach is not gauge invariant it 
provides a qualitative explanation.}. These kinds of corrections can only be 
provided by an off-shell computation and the comparison can not be improved by 
including among others NNLO QCD corrections in the NWA for the $t\tb W^\pm$ 
process or by incorporating multi-jet merging.
The dominant uncertainties for all predictions are attributed to missing 
higher-order corrections. In the case of the full off-shell calculation the scale
sensitivity is the smallest, starting from $\pm 5\%$ and increasing up to 
$\pm 15\%$ towards the end of the plotted spectrum. On the contrary, the scale 
uncertainties of parton-shower based results already start at $10\%$ and grow up 
to $21\%-26\%$ at high $p_T$. Matching uncertainties, on the other hand, are 
negligible below $300~\GeV$ but increase up to $\pm 7\%$ in the case of the 
\powhegbox{} and $\pm 12\%$ for \mgfive{} at higher $p_T$.

For the transverse-momentum distribution of the second hardest $b$ jet, displayed
on the right hand side of Fig.~\ref{fig:fig_QCD_1}, the situation changes 
slightly. While the general shape of the full NWA prediction diverges from the 
off-shell result above $p_T \geq 120~\GeV$, we observe that the shapes of the 
\powhegbox{} and \mgfive{} spectra above $80~\GeV$ are nearly identical with the 
off-shell result over the whole plotted range up to an overall shift in the 
normalization by $-15\%$. Remarkably, the scale uncertainties of the off-shell 
calculation are rather constant and below $6\%$. Comparing this to the 
corresponding $15\%$ uncertainties for $p_T(b_1)$ indicates that $p_T(b_2)$ is 
less sensitive to new LO-like contributions in the real radiation part. Also for 
the \powhegbox{} and \mgfive{} results the scale uncertainties are constant and 
of the order of $\pm 10\%$. As for the case of the hardest $b$ jet, matching 
uncertainties become comparable in size only in the tail of the distribution.

%
\begin{figure}[h!]
\begin{center}
 \includegraphics[width=0.49\textwidth]{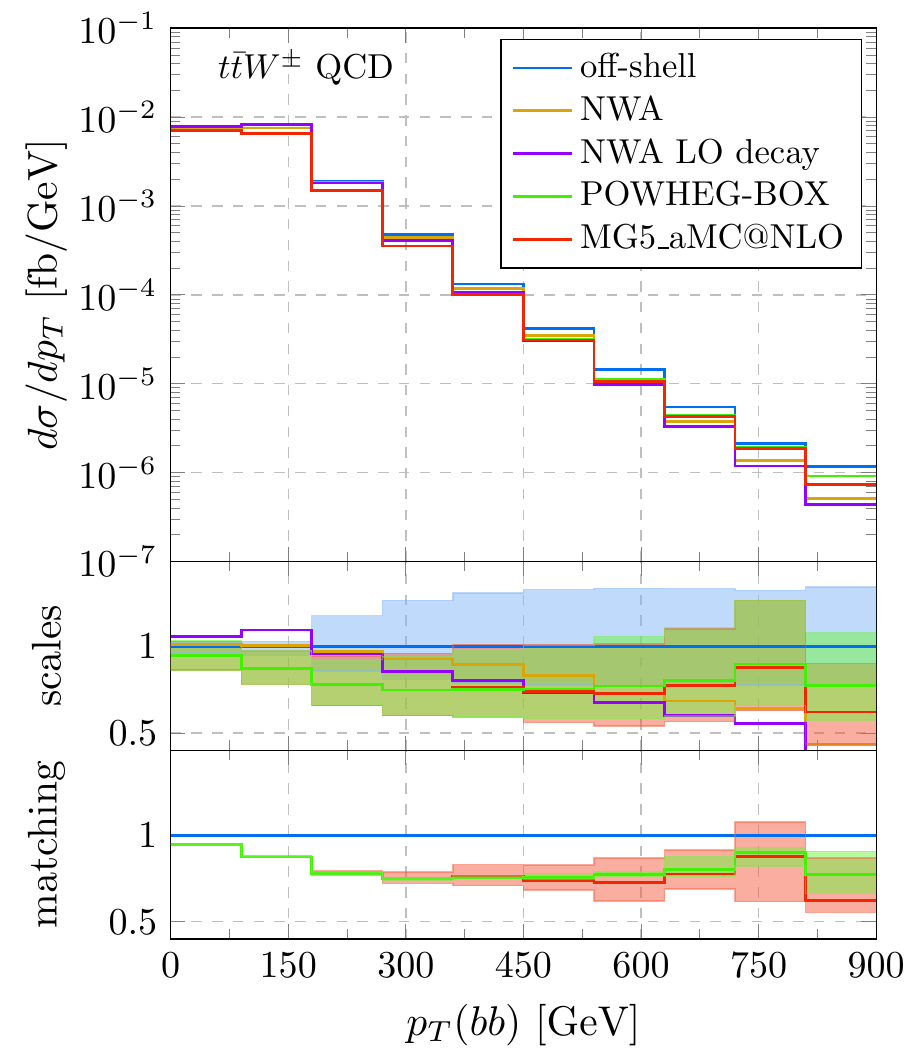}
 \includegraphics[width=0.49\textwidth]{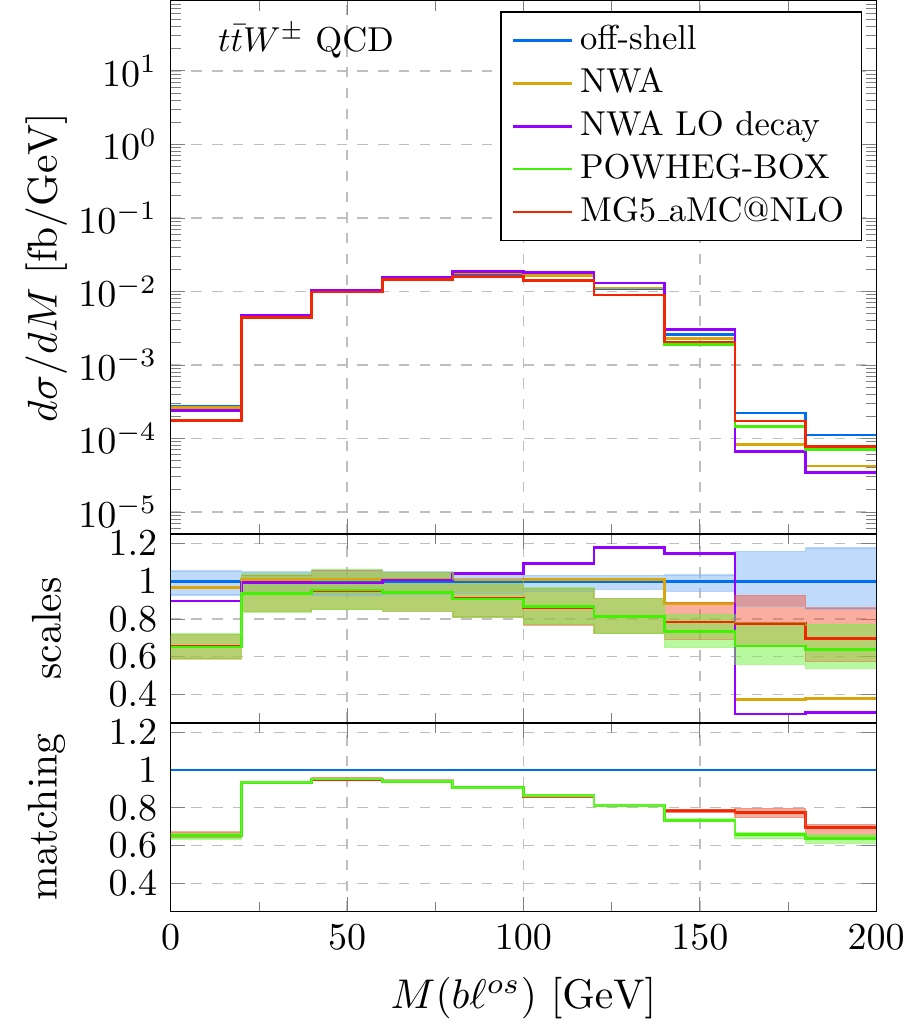}
\end{center}
\caption{Differential cross section distribution in the $3\ell$ fiducial region 
as a function of the transverse momentum of the system of the two hardest $b$ 
jets (l.h.s.) and the minimal invariant mass of a $b$ jet and the opposite-sign 
lepton $\ell^{os}$ (r.h.s.) for the $pp\to t\tb W^\pm$ QCD process. The 
uncertainty bands correspond to independent variations of the renormalization and 
factorization scales (middle panel) and of the matching parameters (bottom 
panel).}
\label{fig:fig_QCD_2}
\end{figure}
For the transverse momentum of the system of the two hardest $b$ jets shown on 
the left hand side of Fig.~\ref{fig:fig_QCD_2}, we observe first of all the 
enlarged uncertainty band, of the order of $\pm30\%$, for the off-shell 
calculation, which indicates that the observable is very sensitive to extra jet 
radiation. Again starting from $p_T \sim 300~\GeV$ the NWA increasingly deviates 
from the off-shell calculation all the way up to $55\%$ at the end of the 
spectrum. This can be understood from the fact that in the tail of the 
distribution the single-resonant contribution for $pp \to tWWb$ scattering, which
is only included in the full off-shell computation, becomes quite large. On the 
other hand, the parton-shower based predictions lie somewhat in between the NWA 
and the off-shell calculation for $p_T \gtrsim 600~\GeV$. Within the estimated 
uncertainties the parton-shower results are fully compatible with the full 
off-shell calculation over the whole considered range.

On the right of Fig.~\ref{fig:fig_QCD_2} we show the invariant mass of the 
opposite-sign lepton $\ell^{os}$ and the $b$ jet that minimizes the invariant 
mass itself. This observable is especially sensitive to off-shell effects and 
additional radiation since it has a natural kinematical edge at $M(b\ell^{os}) 
\leq \sqrt{m_t^2-m_W^2} \approx 152~\GeV$ in the case of the $t\to W^+b$ on-shell
decay. Thus, this observable is indeed particularly interesting for top-quark 
mass measurements~\cite{Biswas:2010sa,Heinrich:2013qaa,Heinrich:2017bqp,
Bevilacqua:2017ipv}. The full NWA recovers the shape of the off-shell result for 
$M(b\ell^{os}) \leq 140~\GeV$ but deviates up to $60\%$ in the tail of the 
distribution. If top-quark decays are treated at leading-order accuracy in the 
NWA also the kinematical boundary is not described well, even below the edge. 
Finally, the parton-shower based predictions that account for multiple emissions 
in the top-quark decays wash out the kinematic edge even further. A considerable 
fraction of events are pushed above the edge and populate the tail of the 
distribution. We find only a difference of the order of $30\%$ with respect to 
the off-shell calculation. However, here as well, the tail of the distribution 
receives sizable contributions from the single-resonant $tWWb$ scattering. 
Scale uncertainties for the off-shell calculation are of the order of $\pm 5\%$ 
below the kinematic edge and $\pm 15\%$ above. On the other hand, the \powhegbox{}
and \mgfive{} obtain roughly $\pm 10\%$ below and $\pm 20\%$ uncertainties above 
the edge, while matching uncertainties are negligible throughout the whole range.

%
\begin{figure}[h!]
\begin{center}
 \includegraphics[width=0.49\textwidth]{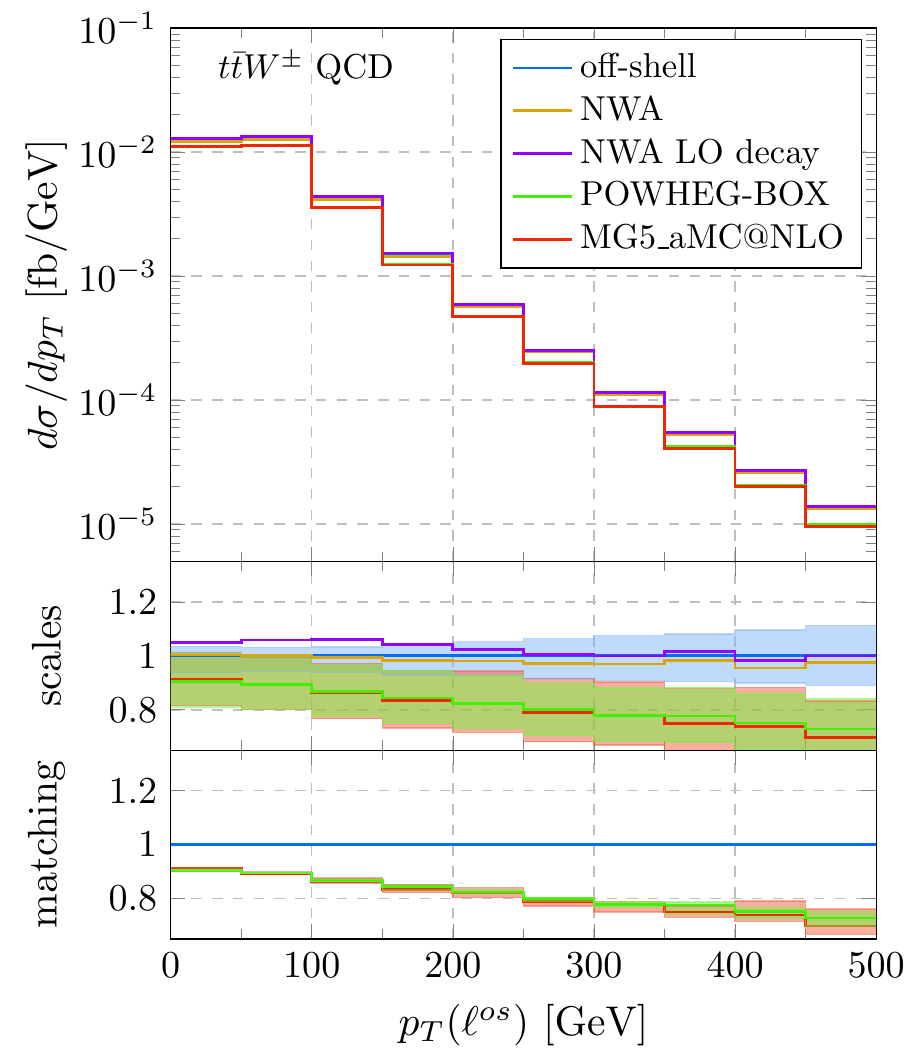}
 \includegraphics[width=0.49\textwidth]{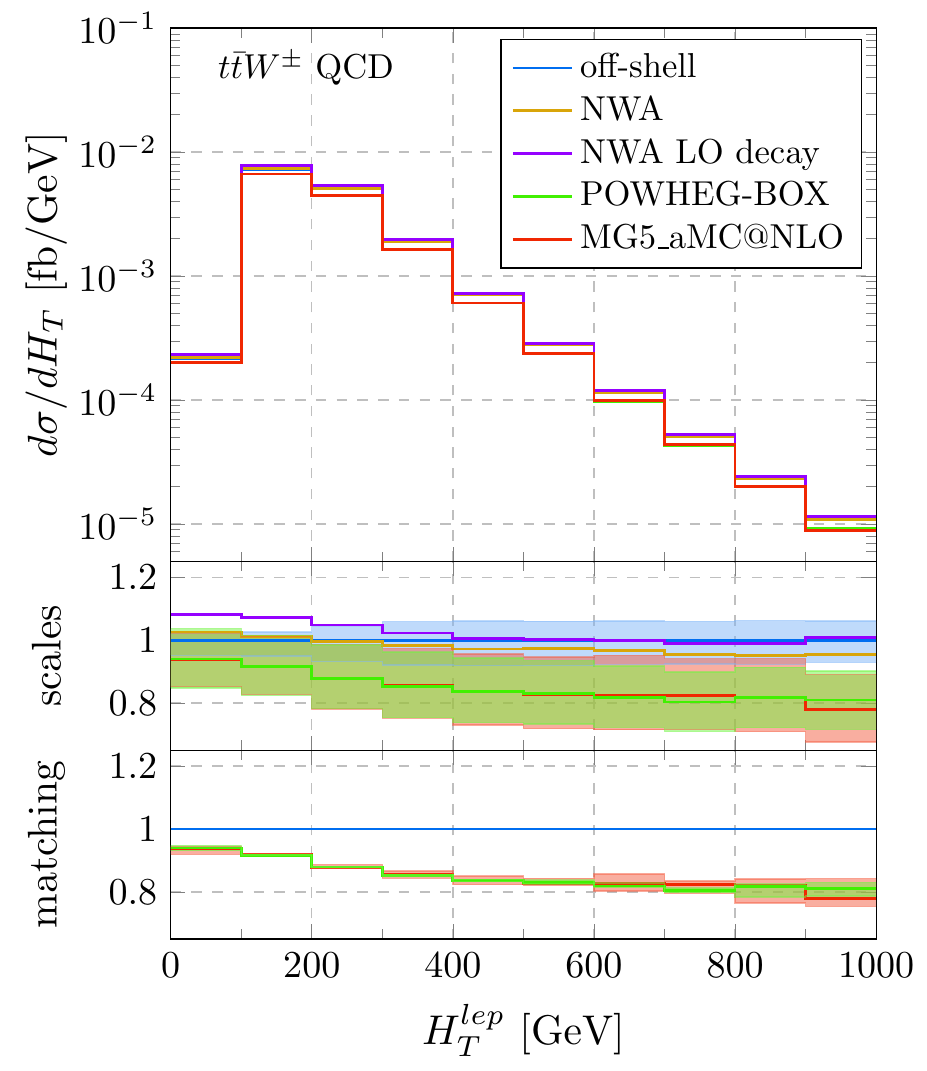}
\end{center}
\caption{Differential cross section distribution in the $3\ell$ fiducial region 
as a function of the transverse momentum of the opposite-sign lepton $\ell^{os}$
(l.h.s.) and the $H_T^{lep}$ observable (r.h.s) for the $pp\to t\tb W^\pm$ QCD
process. The uncertainty bands correspond to independent variations of the 
renormalization and factorization scales (middle panel) and of the matching 
parameters (bottom panel).}
\label{fig:fig_QCD_3}
\end{figure}
Next we discuss the transverse momentum of the opposite-sign lepton $\ell^{os}$ 
and the $H_T^{lep}$ observable, which is defined as
\begin{equation}
 H_T^{lep} = \sum_{i=1}^3 p_T(\ell_i)\;.
\end{equation}
Both observables are shown in Fig.~\ref{fig:fig_QCD_3}. In the case of the 
transverse momentum distribution we note, that the full NWA gives an equivalent 
description compared to the off-shell prediction of the observable over the whole
plotted range. If NLO QCD corrections to the top-quark decays are ignored then 
the prediction at the beginning of the spectrum with $p_T \leq 200~\GeV$ 
overshoots the off-shell results mildly by $+6\%$. The tail of the distribution 
is again in excellent agreement with the off-shell calculation. On the contrary, 
the parton-shower matched predictions obtain a very different shape compared to 
the fixed-order predictions. Starting from a difference of about $-10\%$ the 
parton-shower predictions diverge further up to $-30\%$ at the end of the plotted
range. In addition, the scale uncertainties of the off-shell result are below 
$11\%$, while they are at most $15\%$ for the \powhegbox{} and $20\%$ for 
\mgfive{}. Thus, starting from $p_T \geq 350~\GeV$ the uncertainty bands do not 
overlap anymore. In comparison, matching uncertainties are negligible in the 
considered range.

For the $H_T^{lep}$ observable on the right of Fig.~\ref{fig:fig_QCD_3} we 
observe a similar trend. Even though predictions based on the full NWA have minor 
shape differences, they are fully compatible with the off-shell calculation 
within the uncertainties. For the off-shell results the scale uncertainties are 
estimated to be below $8\%$ in the whole plotted range. The \powhegbox{} and 
\mgfive{} predictions are hardly distinguishable and show a different shape, with
deviations of up to $20\%$ in the tail of the distribution, when compared to the 
off-shell results. Only because of the larger scale uncertainty of the 
parton-shower matched predictions, which are of the order of $15\%$, the curves 
are compatible with the off-shell prediction up to $700~\GeV$. Above that point, 
the uncertainty bands do not overlap anymore.

\subsection{Electroweak production of $t\tb W^\pm$}
Let us now turn to the discussion of the electroweak production of the 
$pp\to t\tb W^\pm$ process including decays at $\mathcal{O}(\alpha_s\alpha^8)$. 
We start again by investigating the integrated fiducial cross sections and turn 
to differential distributions afterwards.

\subsubsection{Integrated fiducial cross sections}
In the case of the full off-shell computation the integrated fiducial cross 
section evaluates to
\begin{equation}
 \sigma^\nlo_{\textrm{off-shell}} = 0.206^{+0.045~(22\%)}_{-0.034~(17\%)}
 ~\mathrm{fb}\;.
\end{equation}
We observe large scale uncertainties of the order of $\pm 20\%$ at NLO, which 
reflects the fact that the $\alpha_s$ dependence only starts at NLO. For the EW 
part the NLO QCD corrections are anomalously large and of the order of 
$\mathcal{K}=\nlo{}/\lo{} \approx 18$, due to the $t$-channel $tW \to tW$ 
scattering in the $qg$ channels that opens up at NLO. Overall, the EW 
contribution amounts to $13\%$ of the dominant NLO QCD contributions to 
$t\tb W^\pm$ production, i.e. 
\begin{equation}
 \left(\frac{\sigma^\nlo_{\textrm{QCD}} + 
 \sigma^\nlo_{\textrm{EW}}}{\sigma^\nlo_{\textrm{QCD}}}
 \right)_{\textrm{off-shell}} = 1.13\;.
\end{equation}
If instead the NWA is employed we find the following cross sections 
\begin{equation}
 \sigma^\nlo_{\textrm{NWA}} = 0.190^{+0.041~(22\%)}_{-0.031~(16\%)}
 ~\mathrm{fb}\;, \qquad
 \sigma^\nlo_{\textrm{NWA LOdec}} = 0.162^{+0.035~(22\%)}_{-0.026~(16\%)}
 ~\mathrm{fb}\;.
\end{equation}
First of all, we notice that NLO QCD corrections to top-quark decays increase the 
cross section by $17\%$, but do not affect the overall scale uncertainties, which
are of the same order as in the case of the full off-shell calculation. 
Furthermore, besides the large corrections from the top-quark decays we also 
observe large off-shell effects of the order of $8\%$ at the level of integrated 
cross sections. Such effects originate from $WW \to WW$ scattering contributions 
as we have checked explicitly at LO where the off-shell effects are even larger 
and of the order of $12\%$.

Finally, the obtained cross sections via parton-shower matched calculations are
given by
\begin{equation}
 \sigma^\nlops_{\textrm{PWG}} = 0.133^{+0.028~(21\%)}_{-0.021~(16\%)}
 ~\mathrm{fb}\;, \qquad
 \sigma^\nlops_{\textrm{MG5}} = 0.136^{+0.028~(21\%)}_{-0.022~(16\%)}
 ~\mathrm{fb}\;. 
\end{equation}
The event generators have the smallest cross section among all the approaches
considered in our study. The parton-shower matched results predict a $34\%-35\%$ 
smaller cross section if compared to the full off-shell result. Also in the case 
of the EW contribution the \powhegbox{} and \mgfive{} results align very well and
are $16\%-18\%$ smaller than the predictions obtained with the NWA with LO 
top-quark decays. This is the same level of reduction of the cross section as 
observed in the QCD contribution and, as in that case, can be attributed to 
additional radiation in top decays. However, in the EW case parton-shower based 
results show the same level of uncertainties as all the other approaches. 

\subsubsection{Differential distributions}
Let us now turn again to the differential comparison of the various computational
approaches in our study.

%
\begin{figure}[h!]
\begin{center}
 \includegraphics[width=0.49\textwidth]{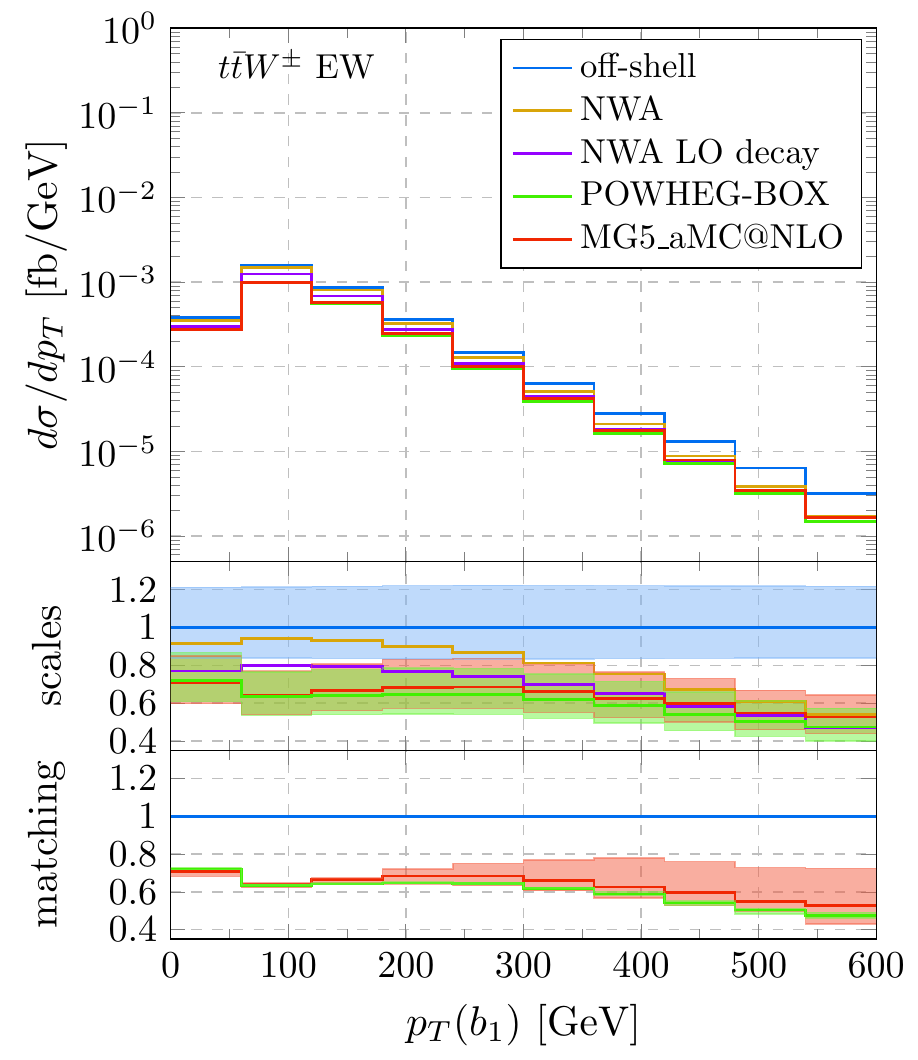}
 \includegraphics[width=0.49\textwidth]{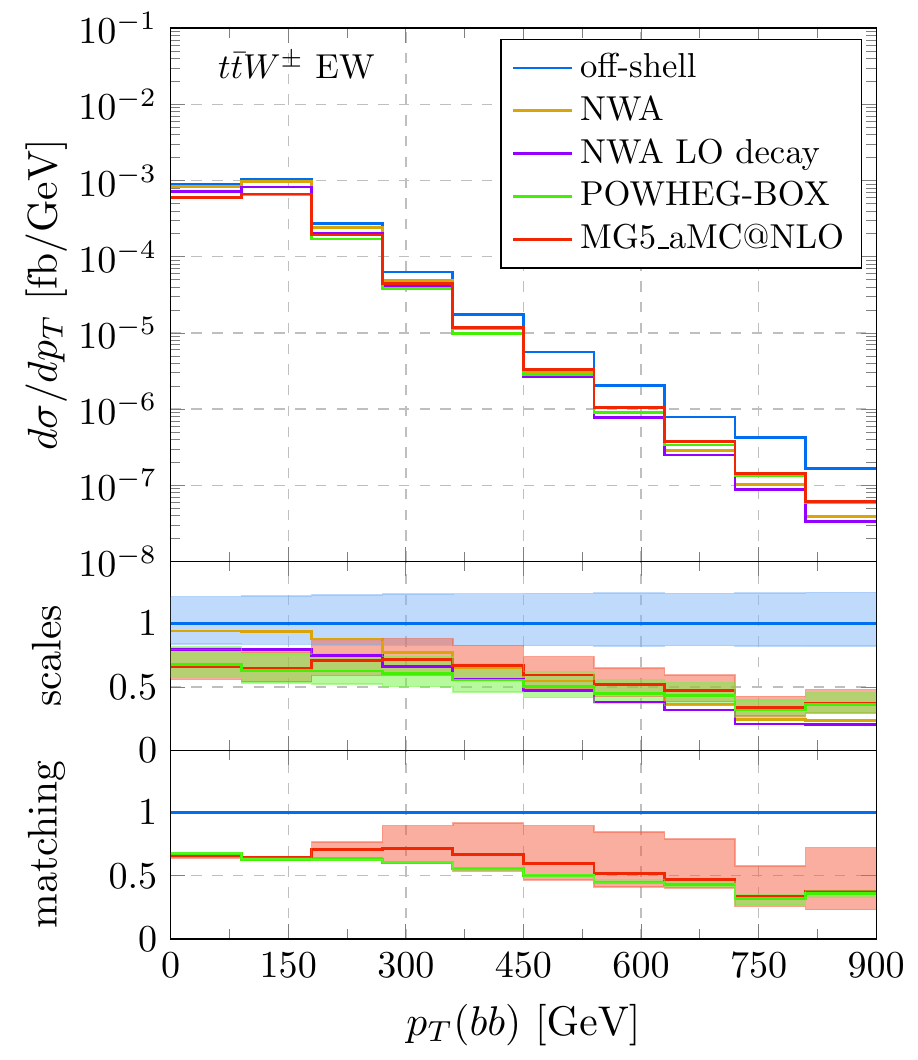}
\end{center}
\caption{Differential cross section distribution in the $3\ell$ fiducial region 
as a function of the transverse momentum of the hardest $b$ jet (l.h.s.) and of 
the system of the two hardest $b$ jets (r.h.s) for the $pp\to t\tb W^\pm$ EW 
process. The uncertainty bands correspond to independent variations of the 
renormalization and factorization scales (middle panel) and of the matching 
parameters (bottom panel).}
\label{fig:fig_EW_1}
\end{figure}
First, we focus on $b$-jet observables like the transverse momentum of the
hardest $b$ jet and the pair of the two hardest $b$ jets, as shown in 
Fig.~\ref{fig:fig_EW_1}. It is evident that, even though the scale uncertainties 
of the full off-shell prediction are of the order of $\pm 20\%$, none of the other
predictions can reproduce the shape of the spectrum. While the full NWA at least 
agrees with the off-shell prediction for low transverse momenta 
$p_T \lesssim 200~\GeV$ within scale uncertainties, the other predictions differ 
by $20\%-40\%$ already in the bulk of the distribution. Towards the end of the 
plotted spectrum all predictions deviate by $50\%$ from the full off-shell 
calculation. Missing higher-order corrections are the dominant source of 
uncertainties, while matching uncertainties stay below $5\%$ for the \powhegbox{}
they can increase up to $20\%-40\%$ for \mgfive{}.
For the transverse momentum of the system of the two hardest $b$ jets shown on 
the right hand side of Fig.~\ref{fig:fig_EW_1} we observe even stronger 
deviations. As in the case of the QCD production mode the scale uncertainties for
this observable are slightly larger and of the order of $20\%-25\%$. Nonetheless,
all predictions diverge from the off-shell result by $65\%-75\%$ at $p_T \approx 
900~\GeV$. The estimated matching uncertainties are very different between the 
\powhegbox{} and \mgfive{}. While the parton-shower starting scale $\mu_Q$ has a 
big impact for \mgfive{} and can affect the shape the distribution by up to 
$90\%$, the dependence of the various damping parameters in the \powhegbox{} 
induces differences up to $\pm 10\%$.

%
\begin{figure}[h!]
\begin{center}
 \includegraphics[width=0.49\textwidth]{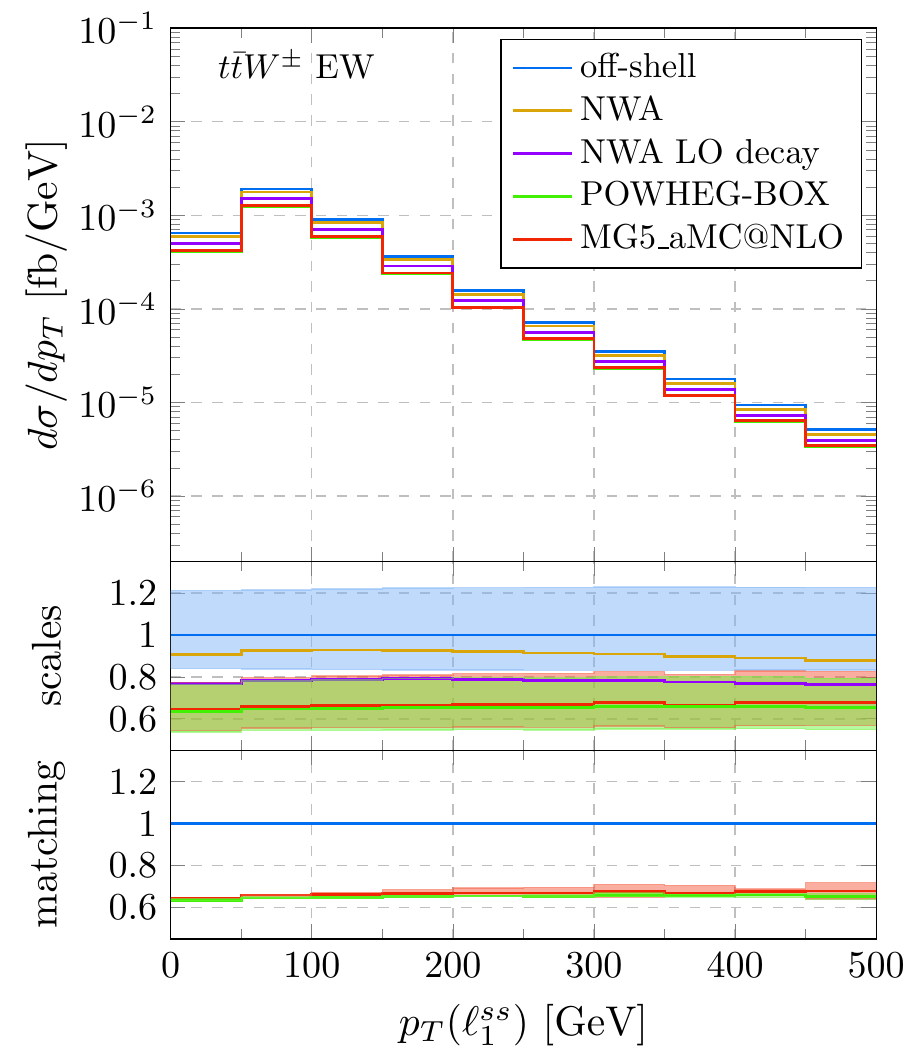}
 \includegraphics[width=0.49\textwidth]{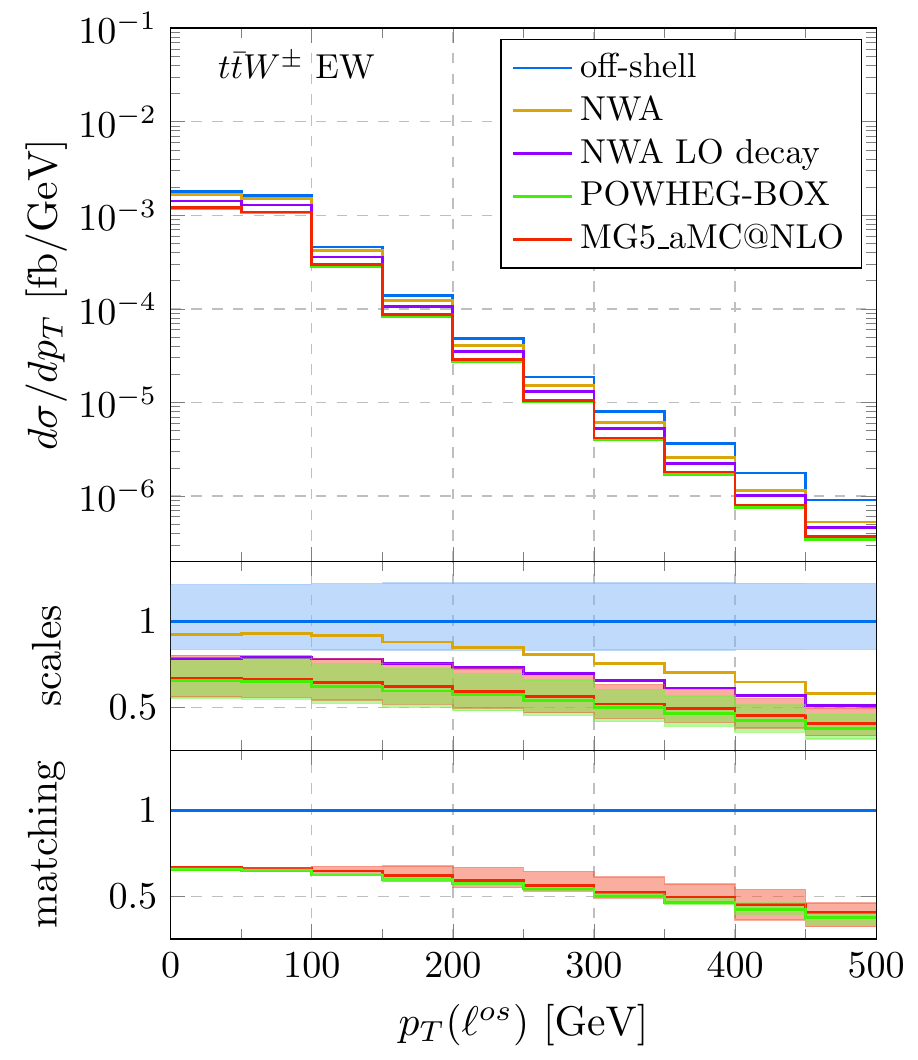}
\end{center}
\caption{Differential cross section distribution in the $3\ell$ fiducial region 
as a function of the transverse momentum of the hardest same-sign lepton 
$\ell^{ss}_1$ (l.h.s.) and the opposite-sign lepton $\ell^{os}$ (r.h.s) for the
$pp\to t\tb W^\pm$ EW process. The uncertainty bands correspond to independent 
variations of the renormalization and factorization scales (middle panel) and of 
the matching parameters (bottom panel).}
\label{fig:fig_EW_2}
\end{figure}
Next, we investigate lepton observables such as the transverse momentum of the 
hardest same-sign lepton $\ell^{ss}_1$ and the opposite-sign lepton $\ell^{os}$
as depicted in Fig.~\ref{fig:fig_EW_2}. These two observables behave very 
differently. In the case of $p_T(\ell^{ss}_1)$ the lepton can either 
originate from the top-quark decay or from the $W$ boson radiated in the initial
state. For the off-shell calculation even more topologies are possible as single-
and non-resonant contributions are present. However, we find that all predictions
generate a very similar shape of the spectrum and differences are mainly due to
overall normalizations. To be precise, the full NWA is globally $10\%$, the NWA
with LO decays $20\%$, and the parton-shower predictions $35\%$ lower than the full 
off-shell calculation. Scale uncertainties are for all predictions at the level
of $\pm 20\%$. However due to the overall shift in the normalization the 
parton-shower results barely overlap with the full off-shell results. In addition, 
matching uncertainties are negligible for both parton-shower predictions.
On the other hand, when we look at the transverse momentum of the opposite-sign 
lepton $\ell^{os}$ we expect larger off-shell effects. This is due to fact that, 
in the case of the double-resonant $t\tb W^\pm$ contribution, the lepton can only 
originate from a top-quark decay, which is not the case for the full off-shell 
computation. Indeed, we find that all approximations undershoot the tail of the 
distribution by $40\%-60\%$. Comparing the same observable to the QCD production 
mode as shown in Fig.~\ref{fig:fig_QCD_3} we notice that the single resonant 
contribution must be much larger in the EW contribution. As in the previous cases
the residual dependence on the renormalization and factorization scales are the 
dominant uncertainties, which for all predictions are of the order of $22\%$. 
Nonetheless, the matching uncertainties become sizable in the tail of the 
spectrum and can reach up to $13\%$ for the \powhegbox{} and $20\%$ for \mgfive{}.

%
\begin{figure}[h!]
\begin{center}
 \includegraphics[width=0.5\textwidth]{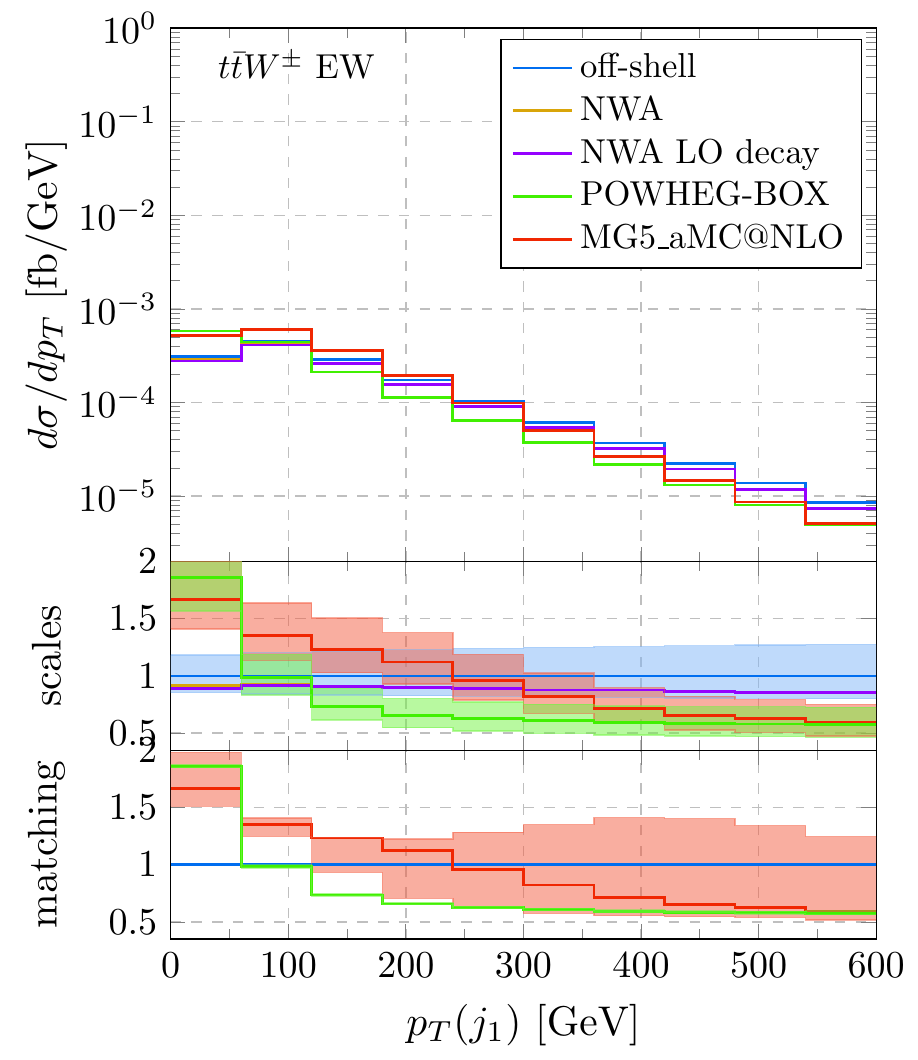}
 \includegraphics[width=0.48\textwidth]{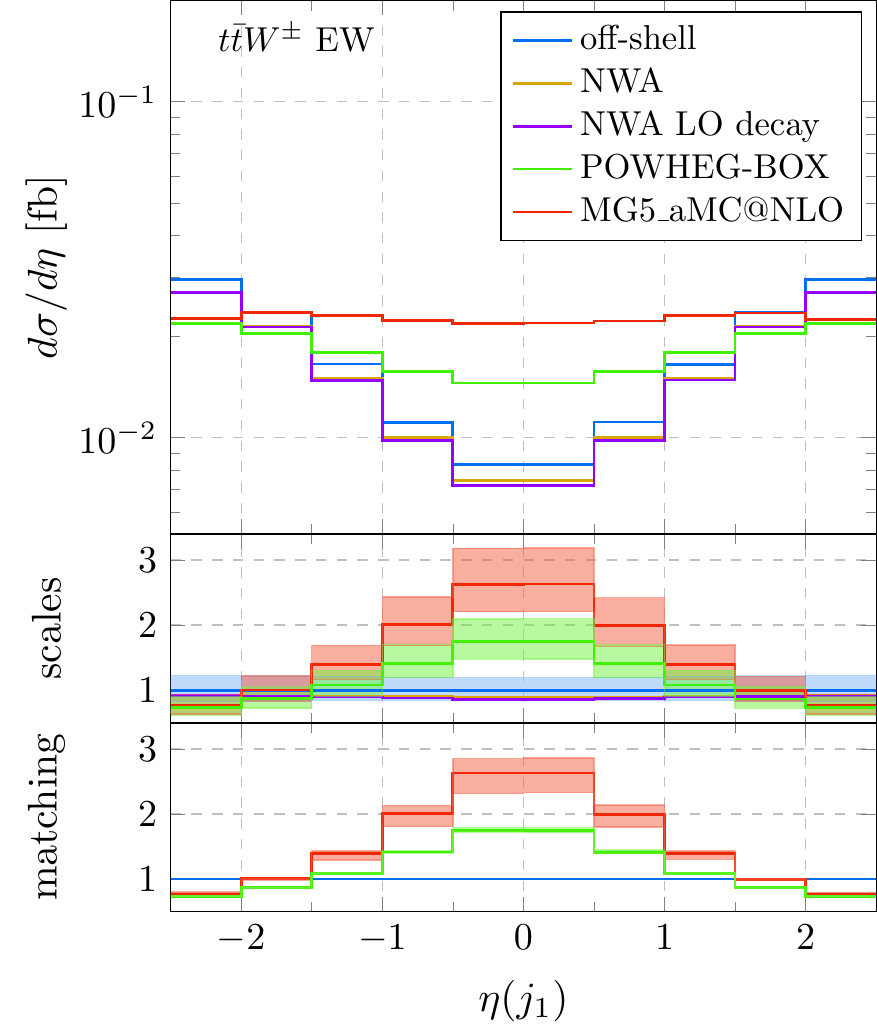}
\end{center}
\caption{Differential cross section distribution in the $3\ell$ fiducial region 
as a function of the transverse momentum (l.h.s.) and the pseudorapidity (r.h.s) 
of the leading light jet for the $pp\to t\tb W^\pm$ EW process. The uncertainty 
bands correspond to independent variations of the renormalization and 
factorization scales (middle panel) and of the matching parameters (bottom 
panel).}
\label{fig:fig_EW_3}
\end{figure}
At last we discuss the transverse momentum and the rapidity of the leading light
jet as shown in Fig.~\ref{fig:fig_EW_3}. First of all we notice that for both 
observables the NWA reproduces the shape of the full off-shell prediction 
perfectly. The curves only differ by an overall shift of $-10\%$ in the 
normalization. Additionally, we can infer that NLO QCD corrections to the decay 
are not important for this observable and that both the full NWA and NWA with LO 
decays are compatible with the full off-shell prediction within the scale 
dependence. On the other hand, the parton-shower matched calculations receive 
large corrections. Especially in the beginning of the transverse momentum 
spectrum deviations up to $85\%$ are visible. The uncertainties of the various 
predictions turn out to be very different. However, the full off-shell 
calculation has a residual scale dependence at the level of $20\%-25\%$, which is
at the same level as for the parton shower based computations. In 
addition in the \powhegbox{} the spectrum is very stable with respect to the 
various matching parameters with shape modifications below $4\%$. Thus, in the 
case of the \powhegbox{} the dominant uncertainties are attributed to missing 
higher-order corrections. Contrary, the \mgfive{} curve depends crucially on the 
choice of the initial shower scale $\mu_Q$ and can alter the tails of the 
distribution by more than $100\%$, making the matching uncertainties the dominant
source of uncertainties.
For the rapidity of the hardest light jet, shown on the right of 
Fig.~\ref{fig:fig_EW_3} we also find large parton-shower corrections. While the 
leading jet in fixed-order perturbation theory is constrained to the very forward
region, the inclusion of additional radiation in the parton-shower evolution 
fills the gap in the central rapidity region. Thus, parton-shower matched 
calculations predict significantly more events in the central regions with nearly
$200\%$ corrections. Still, the \powhegbox{} and \mgfive{} predictions are very 
different.
Similar effects for these two observables have been already observed in 
Ref.~\cite{Cordero:2021iau} in the case of an on-shell $t\tb W^\pm$ signature.

\section{Combined QCD and electroweak production of $t\tb W^\pm$}
\label{sec:improve}
In the following we provide combined results including the QCD and EW production 
modes discussed in section~\ref{sec:pheno} and propose a way to improve 
theoretical predictions in the absence of fully exclusive NLO parton-shower event
generators based on full off-shell \proc{} matrix elements.

In the previous sections we have seen that the theoretical predictions with 
parton-shower effects include many kinematical features that are not accessible 
to a fixed-order NLO computation. However, while comparing the various approaches
to the full off-shell computation, we observed that the single-resonant 
contributions start to dominate in some phase-space regions and give rise to 
sizable corrections in the tail of dimensionful observables. Naturally the 
question arises how one can combine these two different theoretical predictions 
that describe the same observables but differ in the amount and kind of physical 
effects taken into account. In the case at hand we would like to supplement a NLO
parton-shower matched computation with full off-shell effects. Our method of 
choice is the following additive combination 
\begin{equation}
 \frac{d\sigma^\mathrm{th}}{dX} = \frac{d\sigma^\nlops}{dX} + 
 \frac{d\Delta\sigma_\textrm{off-shell}}{dX}\;, \quad \text{with} \quad
 \frac{d\Delta\sigma_\textrm{off-shell}}{dX} =  
 \frac{d\sigma^\nlo_\textrm{off-shell}}{dX} - \frac{d\sigma^\nlo_\nwa}{dX}\;,
\label{eqn:improve}
\end{equation}
where fixed-order full off-shell effects are simply added to a parton-shower 
based computation. Let us comment on the choice of subtraction used in defining 
the $d\Delta\sigma_\textrm{off-shell}/dX$ contribution. The subtraction of the 
full NWA removes approximately the double counting between the double-resonant 
contributions in the full off-shell and the $\nlops$ computation. Strictly 
speaking one should subtract the $\nlops$ result expanded to the same order in 
$\alpha_s$ as the full off-shell computation. This would include parton-shower 
specific contributions due to parton-shower radiation in the top-quark decays.
We approximate this additional parton-shower radiation by subtracting the full 
NWA with NLO top-quark decays instead. We note, that our differential corrections 
$d\Delta\sigma_\textrm{off-shell}/dX$ are independent of the parton-shower event 
generator employed. In addition, $d\Delta\sigma_\textrm{off-shell}/dX$ provides 
differential corrections for the single and non-resonant contributions as well as
interference effects in an approximate way. To estimate the theoretical 
uncertainty of our improved predictions we compute the scale variations and 
matching uncertainties independently and combine them by adding them in 
quadrature
\begin{equation}
 \delta^\mathrm{th} = \sqrt{ \left(\delta^\nlops_\textrm{scale}\right)^2
 + \left(\delta^\nlops_\textrm{matching}\right)^2
 + \left(\delta^{\Delta\sigma}_\textrm{scale}\right)^2
 }\;,
\label{eqn:error}
\end{equation}
where $\delta^{\Delta\sigma}_\textrm{scale}$ is the estimated uncertainty of 
$d\Delta\sigma_\textrm{off-shell}/dX$ and computed from correlated scale 
variations. Finally, all contributions to $\delta^\mathrm{th}$ are symmetrized.

Finally, given the good agreement we found between the \powhegbox{} and 
\mgfive{} in the course of our study, and considering the independence of the 
$d\Delta\sigma_\textrm{off-shell}/dX$ of the parton shower used, we will show in 
the following results for the \powhegbox{} only. Incidentally, if theoretical
predictions including multi-jet merging are available then the aforementioned
differential corrections can also be included.

\subsubsection{Integrated fiducial cross sections}
In Tab.~\ref{tab:xsecs} we present results for the integrated fiducial cross 
section.
\begin{table}[h!]
\begin{center}
\begin{tabular}{ccccc}
 Type & QCD [fb] & EW [fb] & QCD+EW [fb] & (QCD+EW)/QCD \\
 \hline
 full off-shell       & $1.58^{+0.05~(3\%)}_{-0.10~(6\%)}$   & $0.206^{+0.045~(22\%)}_{-0.034~(16\%)}$ & $1.79^{+0.10~(6\%)}_{-0.13~(7\%)}$   & $1.13$ \\[0.2cm]
 NLOPS                & $1.40^{+0.16~(11\%)}_{-0.15~(11\%)}$ & $0.133^{+0.028~(21\%)}_{-0.021~(16\%)}$ & $1.53^{+0.19~(12\%)}_{-0.17~(11\%)}$ & $1.10$ \\[0.2cm]
 NLOPS$+\Delta\sigma$ & $1.41^{+0.16~(11\%)}_{-0.16~(11\%)}$ & $0.149^{+0.028~(19\%)}_{-0.028~(19\%)}$ & $1.56^{+0.21~(13\%)}_{-0.21~(13\%)}$ & $1.11$ \\[0.2cm]
\end{tabular}
\caption{Integrated fiducial cross sections for the $pp\to t\tb W^\pm$ process 
for various approaches. We use the \powhegbox{} to obtain the NLOPS results.}
\label{tab:xsecs}
\end{center}
\end{table}

We observe that for the combined QCD+EW result including the $\Delta\sigma$ 
correction increases the NLOPS cross section by $2\%$. On the other hand, looking
at the individual QCD and EW contributions one can notice that the impact of 
$\Delta\sigma$ on the EW part is much larger and of the order of $12\%$ while for
the QCD part it is about $1\%$.

Let us note that the full off-shell effects increase the relative contribution 
due to the electroweak production process by $3\%$ with respect to the NLOPS 
prediction.

\subsubsection{Differential distributions}
Turning to differential distributions we can now investigate how well our improved
NLOPS+$\Delta\sigma$ predictions capture off-shell effects in an approximate way.
We show our results as plots with three panels: the upper panel shows the central
predictions, the middle one the ratio to the full off-shell prediction, while the
bottom one shows for each prediction the impact of the electroweak contribution 
over the pure QCD result. Also shown in the middle panel are uncertainty bands 
for the full off-shell and the NLOPS calculation, as well as total uncertainties 
for the NLOPS+$\Delta\sigma$ results that are obtained according to 
Eq.~\eqref{eqn:error}. 

%
\begin{figure}[h!]
\begin{center}
 \includegraphics[width=0.49\textwidth]{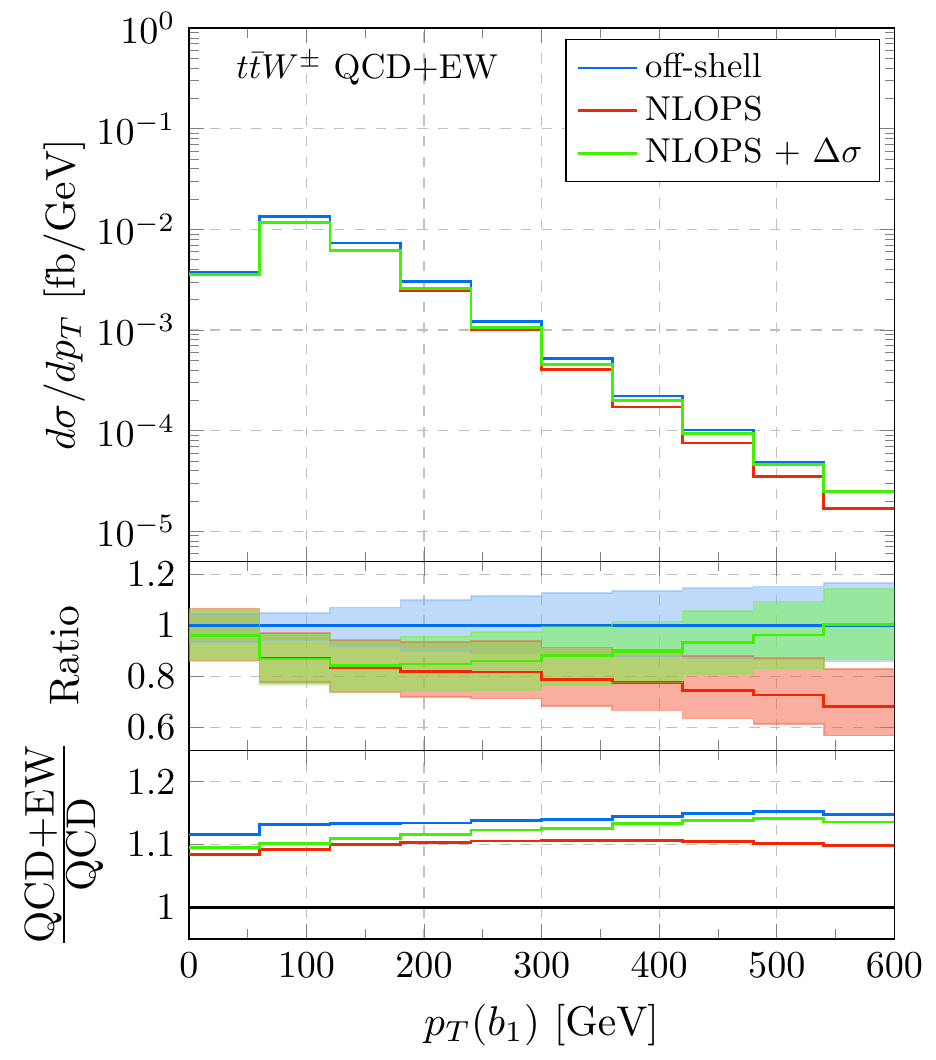}
 \includegraphics[width=0.49\textwidth]{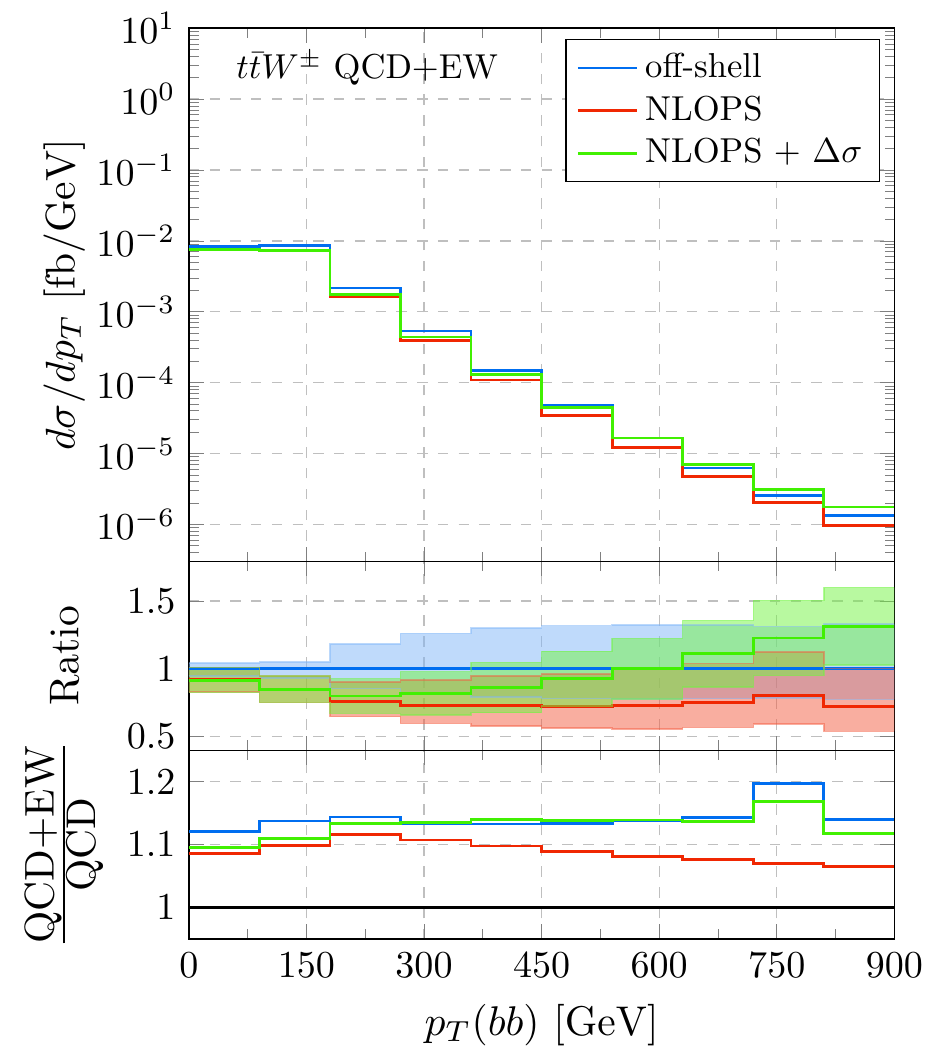}
\end{center}
\caption{Differential cross section distribution in the $3\ell$ fiducial region 
as a function of the transverse momentum of the hardest $b$ jet (l.h.s.) and of 
the system of the two hardest $b$ jets (r.h.s) for the 
$pp\to t\tb W^\pm$ QCD+EW process. The uncertainty bands correspond to scale 
variations (off-shell) and total uncertainties (NLOPS) (middle panel). The 
differential impact of the EW contribution is shown as well (bottom panel).}
\label{fig:fig_QCDEW_1}
\end{figure}
Starting with the transverse momentum of the hardest $b$ jet shown on the left of
Fig.~\ref{fig:fig_QCDEW_1}, we observe that in the bulk of the distribution up to
$250~\GeV$ the corrections due to the approximate off-shell effects are small and
the result is dominated by parton-shower effects. However, the tails of the 
distribution receives sizable corrections due to 
$d\Delta\sigma_\textrm{off-shell}/dX$ with respect to the NLOPS result. 
Specifically, they amount to $+48\%$ at the end of the plotted spectrum. As 
parton-shower corrections in the tails are small the NLOPS+$\Delta\sigma$ 
predictions agree well within the estimated uncertainties with the off-shell 
computation in this region. Furthermore, we can infer from the bottom panel of 
the plot that the EW contribution behaves differently in the off-shell 
computation than in the NLOPS prediction. In the latter case, the EW contribution
is a very flat $+10\%$ corrections on top of the dominant QCD contribution. If 
the full off-shell calculation is employed a clear trend is visible. At the 
beginning of the spectrum the EW channels contribute a $+11\%$ correction that 
steadily grows to $+15\%$ in the tails of the distribution. Our theoretical 
NLOPS+$\Delta\sigma$ prediction clearly captures the large off-shell corrections 
in the tails. The situation is qualitatively different if we consider the 
transverse momentum of the system of the two hardest $b$ jets depicted on the 
right of Fig.~\ref{fig:fig_QCDEW_1}. Here, the improved NLOPS+$\Delta\sigma$ 
prediction overshoots the off-shell tails by $30\%$. The reason for this is that 
the tails of the distribution receive large corrections from single-resonant 
top-quark contributions but also from parton-shower radiation. The theoretical 
uncertainties are estimated to be of the order of $20\%$. As in the previous case
we observe that the EW contribution behaves qualitatively differently in the full 
off-shell calculation. In the NLOPS computation the impact of the EW 
production mode peaks around $p_T \sim 225~\GeV$ and decreases slowly for larger 
transverse momenta. On the contrary, for the full off-shell calculation the EW 
contribution is a flat $+15\%$ correction above $p_T \gtrsim 150~\GeV$.

%
\begin{figure}[h!]
\begin{center}
 \includegraphics[width=0.49\textwidth]{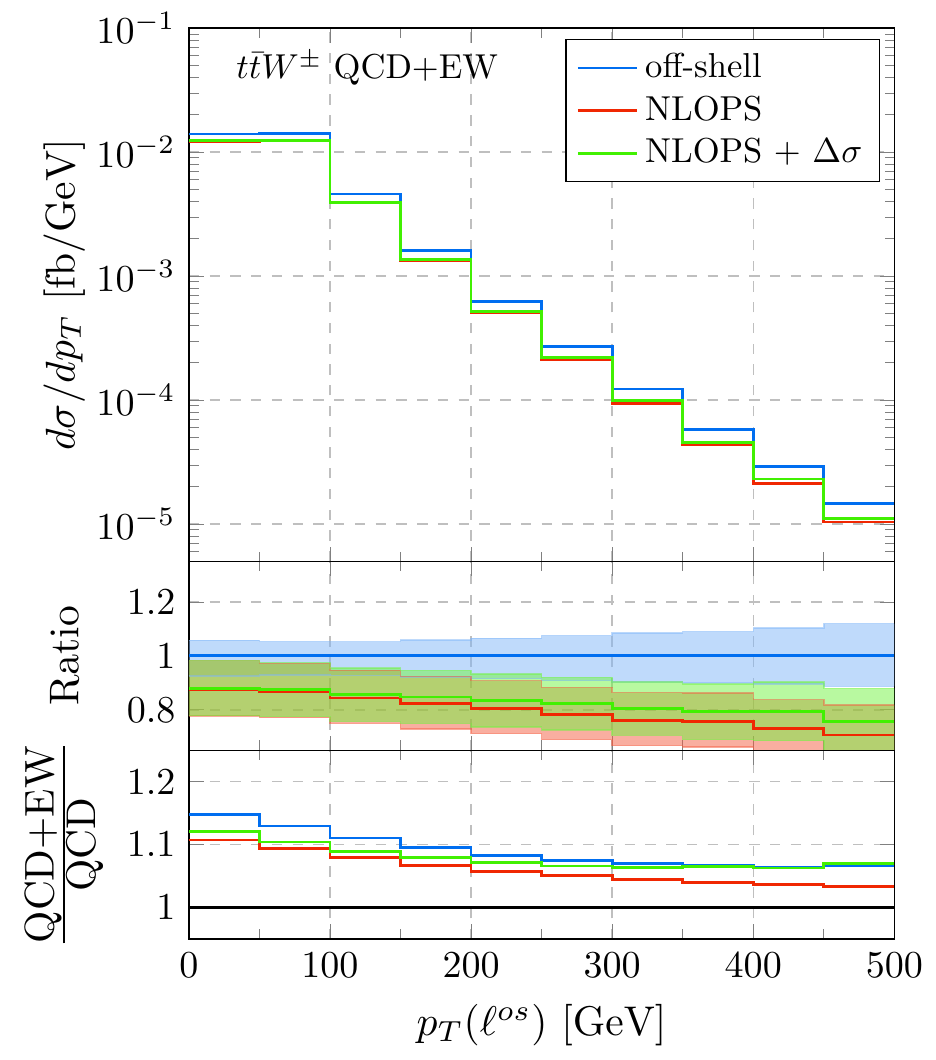}
 \includegraphics[width=0.49\textwidth]{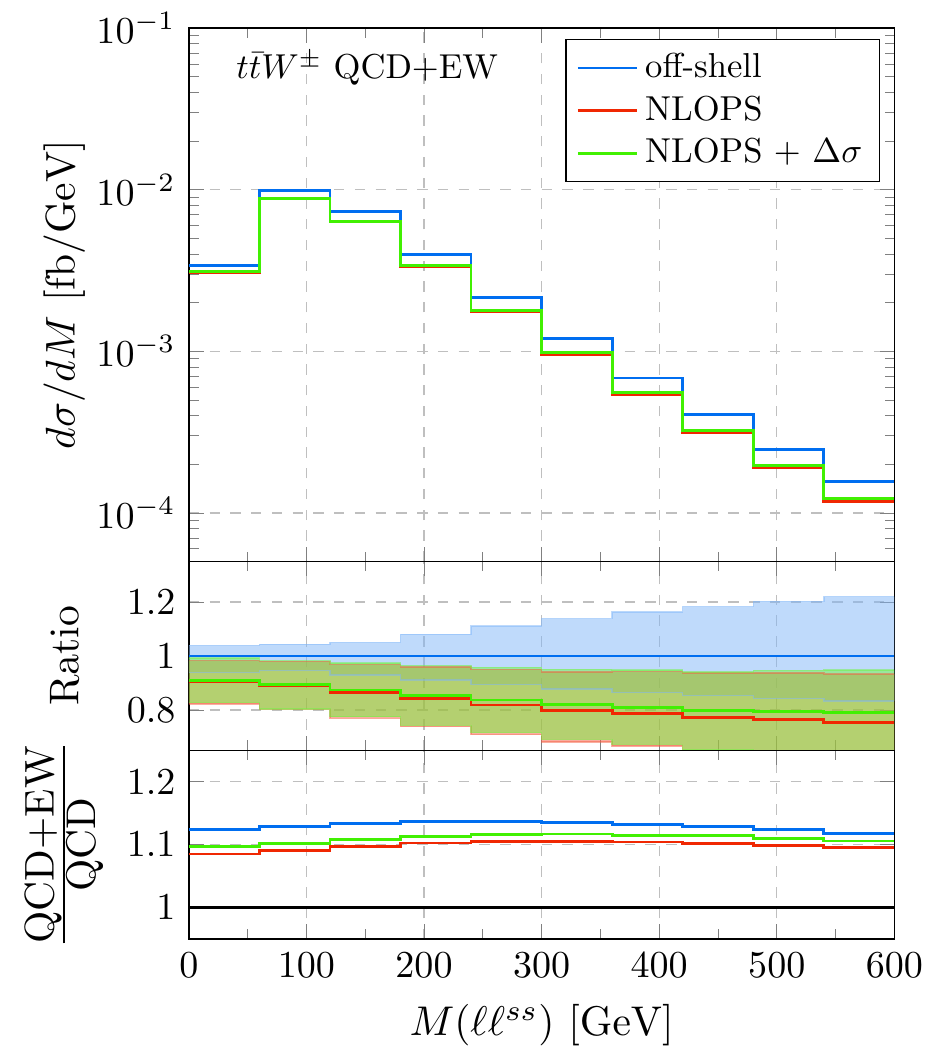}
\end{center}
\caption{Differential cross section distribution in the $3\ell$ fiducial region 
as a function of the transverse momentum of the opposite-sign lepton $\ell^{os}$
(l.h.s.) and the invariant mass of the same-sign lepton pair (r.h.s) for the
$pp\to t\tb W^\pm$ QCD+EW process. The uncertainty bands correspond to scale 
variations (off-shell) and total uncertainties (NLOPS) (middle panel). The 
differential impact of the EW contribution is shown as well (bottom panel).}
\label{fig:fig_QCDEW_2}
\end{figure}
Turning now to lepton observables, we show the transverse momentum of the 
opposite-sign lepton $\ell^{os}$ on the left and the invariant mass of the 
same-sign lepton-pair on the right of Fig.~\ref{fig:fig_QCDEW_2}. Contrary to the 
$b$-jet observables we only find minor corrections of at most $+8\%$ over the 
NLOPS results. This is expected since for the dominant QCD contribution the 
results based on the NWA and the full off-shell calculation are in very good 
agreement. For the spectrum of the transverse momentum distribution the EW 
contributions contribute $10\%-15\%$, depending on the computational approach 
employed, at the beginning of the spectrum, while they decrease to $3\%-7\%$ at 
the end of the spectrum. For the invariant mass spectrum of the same-sign 
lepton-pair the EW channels contribute a rather constant $10\%-12\%$ correction 
over the whole spectrum. In this case we do not observe sizable improvements when 
NLOPS+$\Delta\sigma$ is used instead of NLOPS.

%
\begin{figure}[h!]
\begin{center}
 \includegraphics[width=0.49\textwidth]{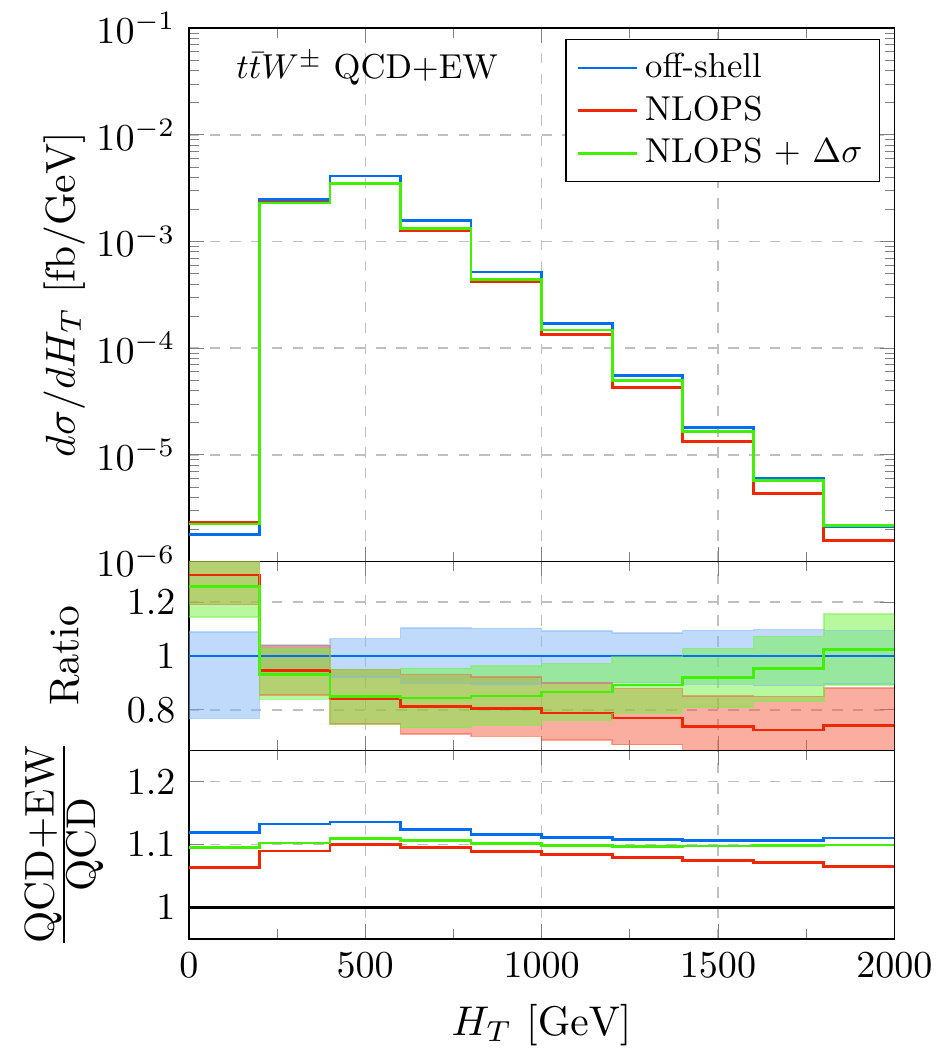}
 \includegraphics[width=0.49\textwidth]{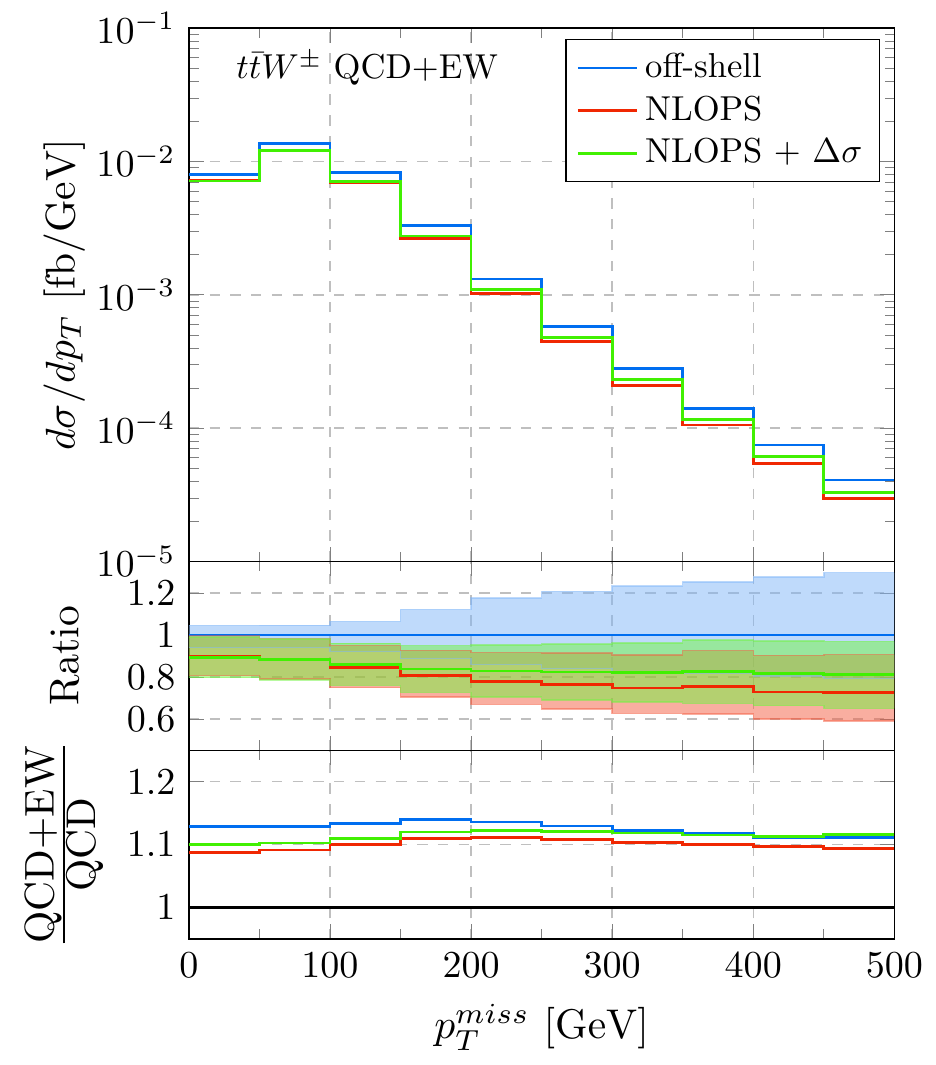}
\end{center}
\caption{Differential cross section distribution in the $3\ell$ fiducial region
as a function of the total transverse momentum $H_T$ (l.h.s.) and the missing
transverse momentum (r.h.s) for the $pp\to t\tb W^\pm$ QCD+EW process. The
uncertainty bands correspond to scale variations (off-shell) and total
uncertainties (NLOPS) (middle panel). The differential impact of the EW
contribution is shown as well (bottom panel).}
\label{fig:fig_QCDEW_3}
\end{figure}
At last, we study the differential distributions of the $H_T$ observable and 
$p_T^{miss}$ as shown in Fig.~\ref{fig:fig_QCDEW_3}. Let us remind the reader,
that the $H_T$ observable is defined as 
\begin{equation}
 H_T = \sum_{i=1}^3 p_T(\ell_i) + \sum_{i=1}^2 p_T(b_i) + p_T^{\textrm{miss}}\;,
\end{equation}
where only the two hardest $b$ jets are included. For the $H_T$ observable shown 
on the left of Fig.~\ref{fig:fig_QCDEW_3} we observe that the NLOPS+$\Delta\sigma$ 
predictions are in better agreement with the off-shell calculation. While the 
bulk of the distribution is dominated by parton-shower effects, the tail receives
large single-resonant contributions of the order of $40\%$. As in all previous 
cases the full off-shell result obtains slightly larger EW contributions than in 
the case of NLOPS predictions. On the other hand, the $p_T^{miss}$ observable only 
shows small corrections from single and non-resonant contributions below $12\%$. 
The NLOPS+$\Delta\sigma$ prediction is fully compatible with the original NLOPS 
result within the estimated uncertainties. For the pure NLOPS prediction the EW 
production mode provides a nearly constant correction of $10\%$, while in the 
case of the full off-shell result the corrections are slightly larger with 
$+14\%$ at the beginning of the distribution.

\section{Conclusions}
\label{sec:conclusions}
In this work we consider various approaches to the theoretical computation of QCD
and EW production of multi-lepton final states for the $pp\to t\tb W^\pm$ 
processes and study the impact of modeling differences in the light of recent 
tensions between Standard Model theoretical predictions and experimental 
measurements.

NLOPS event generators and full off-shell fixed-order calculations lend 
themselves better to describe different fiducial regions of phase space. For 
instance, for dimensionful observables such as invariant masses and transverse 
momenta, we find large differences in the tails of distributions between the full
off-shell calculation and any other approximate calculation based on on-shell 
top quarks. These off-shell contributions are dominantly attributed to 
single-resonant contributions that can only be included in an unambiguous way 
using the full off-shell calculation for the fully decayed final state. On the 
other hand, parton-shower effects impact the shape of various distributions over 
a broader range. In this paper we propose a method to approximately combine the 
strengths of both approaches.

We also quantified for the first time the size of off-shell effects for the EW
production process at $\mathcal{O}(\alpha_s\alpha^8)$ and found unusual large
differences of the order of $8\%$ already at the level of the integrated fiducial
cross section. At the differential level larger effects have been found as 
compared to the case of the QCD production mode at 
$\mathcal{O}(\alpha_s^3\alpha^6)$.

In the absence of NLOPS theoretical predictions that comprise resonant-aware
matched full off-shell calculations~\cite{Jezo:2015aia,Jezo:2016ujg} for $t\tb W^\pm$, we proposed in
section V to improve on the current available predictions by combining NLOPS
and full off-shell results in an additive way according to Eq.~(32).  Our
proposal is an approximation that aims at removing the double counting of
double-resonant contributions between NLOPS and fixed-order computations but
includes single and non-resonant contributions at fixed-order.

In this way, parton-shower effects can be retained and off-shell corrections can 
be included in an approximate but event generator independent way for a large 
class of observables. We investigated in detail the impact of such improved 
theoretical predictions and provide conservative uncertainty estimates for the 
fully combined QCD+EW results. 
As we have observed in our study the impact of the parton shower evolution can
have a sizable impact on differential distributions, therefore in order to
account for those shower effects also for the single and non-resonant
contributions the proper matching of the full off-shell $t\tb W^\pm$
calculation to parton showers is necessary.

Moving forward, we think that we could already learn a great deal by a detailed
comparison of $t\tb W^\pm$ unfolded data with the level of theoretical 
predictions presented in this paper. Beyond that, a fixed-order calculation of 
NNLO QCD corrections to the on-shell $t\tb W^\pm$ process could help removing
some of the theoretical uncertainty in both scale-dependence and hard-radiation 
modeling. Equally important and equally challenging would be interfacing the 
full off-shell fixed-order calculation with parton-shower event generators.

\section*{Acknowledgements}
The work of F. F. C., M.K. and L. R. is supported in part by the U.S. Department 
of Energy under grant DE-SC0010102.

The research of H.Y.B., J.N. and M.W. was supported by the Deutsche 
Forschungsgemeinschaft (DFG) under the following grants: 400140256 - GRK 2497: 
\textit{The physics of the heaviest particles at the Large Hardon Collider} and 
396021762 - TRR 257: \textit{P3H - Particle Physics Phenomenology after the Higgs 
Discovery}. Support by a grant of the Bundesministerium f\"ur Bildung und 
Forschung (BMBF) is additionally acknowledged. 

The work of G.B. was supported by grant K 125105 of the National Research, 
Development and Innovation Office in Hungary.

H.B.H. has received funding from the European Research Council (ERC) under the 
European Union’s Horizon 2020 Research and Innovation Programme (grant agreement 
no. 683211). Furthermore, the work of H.B.H has been partially 
supported by STFC consolidated HEP theory grant ST/T000694/1. 

Simulations were performed with computing resources granted by RWTH Aachen 
University under project \texttt{rwth0414}.
\bibliography{biblio.bib}

\begin{thebibliography}{86}%
\makeatletter
\providecommand \@ifxundefined [1]{%
 \@ifx{#1\undefined}
}%
\providecommand \@ifnum [1]{%
 \ifnum #1\expandafter \@firstoftwo
 \else \expandafter \@secondoftwo
 \fi
}%
\providecommand \@ifx [1]{%
 \ifx #1\expandafter \@firstoftwo
 \else \expandafter \@secondoftwo
 \fi
}%
\providecommand \natexlab [1]{#1}%
\providecommand \enquote  [1]{``#1''}%
\providecommand \bibnamefont  [1]{#1}%
\providecommand \bibfnamefont [1]{#1}%
\providecommand \citenamefont [1]{#1}%
\providecommand \href@noop [0]{\@secondoftwo}%
\providecommand \href [0]{\begingroup \@sanitize@url \@href}%
\providecommand \@href[1]{\@@startlink{#1}\@@href}%
\providecommand \@@href[1]{\endgroup#1\@@endlink}%
\providecommand \@sanitize@url [0]{\catcode `\\12\catcode `\$12\catcode
  `\&12\catcode `\#12\catcode `\^12\catcode `\_12\catcode `\%12\relax}%
\providecommand \@@startlink[1]{}%
\providecommand \@@endlink[0]{}%
\providecommand \url  [0]{\begingroup\@sanitize@url \@url }%
\providecommand \@url [1]{\endgroup\@href {#1}{\urlprefix }}%
\providecommand \urlprefix  [0]{URL }%
\providecommand \Eprint [0]{\href }%
\providecommand \doibase [0]{http://dx.doi.org/}%
\providecommand \selectlanguage [0]{\@gobble}%
\providecommand \bibinfo  [0]{\@secondoftwo}%
\providecommand \bibfield  [0]{\@secondoftwo}%
\providecommand \translation [1]{[#1]}%
\providecommand \BibitemOpen [0]{}%
\providecommand \bibitemStop [0]{}%
\providecommand \bibitemNoStop [0]{.\EOS\space}%
\providecommand \EOS [0]{\spacefactor3000\relax}%
\providecommand \BibitemShut  [1]{\csname bibitem#1\endcsname}%
\let\auto@bib@innerbib\@empty
\bibitem [{\citenamefont {Aad}\ \emph {et~al.}(2016)\citenamefont {Aad} \emph
  {et~al.}}]{ATLAS:2016dlg}%
  \BibitemOpen
  \bibfield  {author} {\bibinfo {author} {\bibfnamefont {Georges}\ \bibnamefont
  {Aad}} \emph {et~al.} (\bibinfo {collaboration} {ATLAS}),\ }\bibfield
  {title} {\enquote {\bibinfo {title} {{Search for supersymmetry at
  $\sqrt{s}=13$ TeV in final states with jets and two same-sign leptons or
  three leptons with the ATLAS detector}},}\ }\href {\doibase
  10.1140/epjc/s10052-016-4095-8} {\bibfield  {journal} {\bibinfo  {journal}
  {Eur. Phys. J. C}\ }\textbf {\bibinfo {volume} {76}},\ \bibinfo {pages} {259}
  (\bibinfo {year} {2016})},\ \Eprint {http://arxiv.org/abs/1602.09058}
  {arXiv:1602.09058 [hep-ex]} \BibitemShut {NoStop}%
\bibitem [{\citenamefont {Khachatryan}\ \emph
  {et~al.}(2016{\natexlab{a}})\citenamefont {Khachatryan} \emph
  {et~al.}}]{CMS:2016mku}%
  \BibitemOpen
  \bibfield  {author} {\bibinfo {author} {\bibfnamefont {Vardan}\ \bibnamefont
  {Khachatryan}} \emph {et~al.} (\bibinfo {collaboration} {CMS}),\ }\bibfield
  {title} {\enquote {\bibinfo {title} {{Search for new physics in same-sign
  dilepton events in proton\textendash{}proton collisions at $\sqrt{s} =
  13\,\text {TeV} $}},}\ }\href {\doibase 10.1140/epjc/s10052-016-4261-z}
  {\bibfield  {journal} {\bibinfo  {journal} {Eur. Phys. J. C}\ }\textbf
  {\bibinfo {volume} {76}},\ \bibinfo {pages} {439} (\bibinfo {year}
  {2016}{\natexlab{a}})},\ \Eprint {http://arxiv.org/abs/1605.03171}
  {arXiv:1605.03171 [hep-ex]} \BibitemShut {NoStop}%
\bibitem [{\citenamefont {Aaboud}\ \emph
  {et~al.}(2017{\natexlab{a}})\citenamefont {Aaboud} \emph
  {et~al.}}]{ATLAS:2017tmw}%
  \BibitemOpen
  \bibfield  {author} {\bibinfo {author} {\bibfnamefont {Morad}\ \bibnamefont
  {Aaboud}} \emph {et~al.} (\bibinfo {collaboration} {ATLAS}),\ }\bibfield
  {title} {\enquote {\bibinfo {title} {{Search for supersymmetry in final
  states with two same-sign or three leptons and jets using 36 fb$^{-1}$ of
  $\sqrt{s}=13$ TeV $pp$ collision data with the ATLAS detector}},}\ }\href
  {\doibase 10.1007/JHEP09(2017)084} {\bibfield  {journal} {\bibinfo  {journal}
  {JHEP}\ }\textbf {\bibinfo {volume} {09}},\ \bibinfo {pages} {084} (\bibinfo
  {year} {2017}{\natexlab{a}})},\ \bibinfo {note} {[Erratum: JHEP 08, 121
  (2019)]},\ \Eprint {http://arxiv.org/abs/1706.03731} {arXiv:1706.03731
  [hep-ex]} \BibitemShut {NoStop}%
\bibitem [{\citenamefont {Sirunyan}\ \emph {et~al.}(2017)\citenamefont
  {Sirunyan} \emph {et~al.}}]{CMS:2017tec}%
  \BibitemOpen
  \bibfield  {author} {\bibinfo {author} {\bibfnamefont {Albert~M}\
  \bibnamefont {Sirunyan}} \emph {et~al.} (\bibinfo {collaboration} {CMS}),\
  }\bibfield  {title} {\enquote {\bibinfo {title} {{Search for physics beyond
  the standard model in events with two leptons of same sign, missing
  transverse momentum, and jets in proton\textendash{}proton collisions at
  $\sqrt{s} = 13\,\text {TeV} $}},}\ }\href {\doibase
  10.1140/epjc/s10052-017-5079-z} {\bibfield  {journal} {\bibinfo  {journal}
  {Eur. Phys. J. C}\ }\textbf {\bibinfo {volume} {77}},\ \bibinfo {pages} {578}
  (\bibinfo {year} {2017})},\ \Eprint {http://arxiv.org/abs/1704.07323}
  {arXiv:1704.07323 [hep-ex]} \BibitemShut {NoStop}%
\bibitem [{\citenamefont {Aaboud}\ \emph {et~al.}(2018)\citenamefont {Aaboud}
  \emph {et~al.}}]{ATLAS:2018mme}%
  \BibitemOpen
  \bibfield  {author} {\bibinfo {author} {\bibfnamefont {M.}~\bibnamefont
  {Aaboud}} \emph {et~al.} (\bibinfo {collaboration} {ATLAS}),\ }\bibfield
  {title} {\enquote {\bibinfo {title} {{Observation of Higgs boson production
  in association with a top quark pair at the LHC with the ATLAS detector}},}\
  }\href {\doibase 10.1016/j.physletb.2018.07.035} {\bibfield  {journal}
  {\bibinfo  {journal} {Phys. Lett. B}\ }\textbf {\bibinfo {volume} {784}},\
  \bibinfo {pages} {173--191} (\bibinfo {year} {2018})},\ \Eprint
  {http://arxiv.org/abs/1806.00425} {arXiv:1806.00425 [hep-ex]} \BibitemShut
  {NoStop}%
\bibitem [{\citenamefont {Sirunyan}\ \emph
  {et~al.}(2018{\natexlab{a}})\citenamefont {Sirunyan} \emph
  {et~al.}}]{CMS:2018uxb}%
  \BibitemOpen
  \bibfield  {author} {\bibinfo {author} {\bibfnamefont {Albert~M}\
  \bibnamefont {Sirunyan}} \emph {et~al.} (\bibinfo {collaboration} {CMS}),\
  }\bibfield  {title} {\enquote {\bibinfo {title} {{Observation of
  $\mathrm{t\overline{t}}$H production}},}\ }\href {\doibase
  10.1103/PhysRevLett.120.231801} {\bibfield  {journal} {\bibinfo  {journal}
  {Phys. Rev. Lett.}\ }\textbf {\bibinfo {volume} {120}},\ \bibinfo {pages}
  {231801} (\bibinfo {year} {2018}{\natexlab{a}})},\ \Eprint
  {http://arxiv.org/abs/1804.02610} {arXiv:1804.02610 [hep-ex]} \BibitemShut
  {NoStop}%
\bibitem [{ATL(2019)}]{ATLAS:2019nvo}%
  \BibitemOpen
  \href {http://cds.cern.ch/record/2693930} {\emph {\bibinfo {title} {{Analysis
  of $t\bar{t}H$ and $t\bar{t}W$ production in multilepton final states with
  the ATLAS detector}}}},\ \bibinfo {type} {Tech. Rep.}\ \bibinfo {number}
  {ATLAS-CONF-2019-045}\ (\bibinfo  {institution} {CERN},\ \bibinfo {address}
  {Geneva},\ \bibinfo {year} {2019})\BibitemShut {NoStop}%
\bibitem [{CMS(2020)}]{CMS:2020iwy}%
  \BibitemOpen
  \bibfield  {title} {\enquote {\bibinfo {title} {{Higgs boson production in
  association with top quarks in final states with electrons, muons, and
  hadronically decaying tau leptons at $\sqrt{s} = 13~\mathrm{TeV}$}},}\
  }\href@noop {} {\  (\bibinfo {year} {2020})}\BibitemShut {NoStop}%
\bibitem [{\citenamefont {Aaboud}\ \emph
  {et~al.}(2019{\natexlab{a}})\citenamefont {Aaboud} \emph
  {et~al.}}]{ATLAS:2018kxv}%
  \BibitemOpen
  \bibfield  {author} {\bibinfo {author} {\bibfnamefont {Morad}\ \bibnamefont
  {Aaboud}} \emph {et~al.} (\bibinfo {collaboration} {ATLAS}),\ }\bibfield
  {title} {\enquote {\bibinfo {title} {{Search for four-top-quark production in
  the single-lepton and opposite-sign dilepton final states in pp collisions at
  $\sqrt{s}$ = 13 TeV with the ATLAS detector}},}\ }\href {\doibase
  10.1103/PhysRevD.99.052009} {\bibfield  {journal} {\bibinfo  {journal} {Phys.
  Rev. D}\ }\textbf {\bibinfo {volume} {99}},\ \bibinfo {pages} {052009}
  (\bibinfo {year} {2019}{\natexlab{a}})},\ \Eprint
  {http://arxiv.org/abs/1811.02305} {arXiv:1811.02305 [hep-ex]} \BibitemShut
  {NoStop}%
\bibitem [{\citenamefont {Sirunyan}\ \emph {et~al.}(2020)\citenamefont
  {Sirunyan} \emph {et~al.}}]{CMS:2019rvj}%
  \BibitemOpen
  \bibfield  {author} {\bibinfo {author} {\bibfnamefont {Albert~M}\
  \bibnamefont {Sirunyan}} \emph {et~al.} (\bibinfo {collaboration} {CMS}),\
  }\bibfield  {title} {\enquote {\bibinfo {title} {{Search for production of
  four top quarks in final states with same-sign or multiple leptons in
  proton-proton collisions at $\sqrt{s}=$ 13 TeV}},}\ }\href {\doibase
  10.1140/epjc/s10052-019-7593-7} {\bibfield  {journal} {\bibinfo  {journal}
  {Eur. Phys. J. C}\ }\textbf {\bibinfo {volume} {80}},\ \bibinfo {pages} {75}
  (\bibinfo {year} {2020})},\ \Eprint {http://arxiv.org/abs/1908.06463}
  {arXiv:1908.06463 [hep-ex]} \BibitemShut {NoStop}%
\bibitem [{\citenamefont {Aad}\ \emph {et~al.}(2020)\citenamefont {Aad} \emph
  {et~al.}}]{ATLAS:2020hpj}%
  \BibitemOpen
  \bibfield  {author} {\bibinfo {author} {\bibfnamefont {Georges}\ \bibnamefont
  {Aad}} \emph {et~al.} (\bibinfo {collaboration} {ATLAS}),\ }\bibfield
  {title} {\enquote {\bibinfo {title} {{Evidence for $t\bar{t}t\bar{t}$
  production in the multilepton final state in proton\textendash{}proton
  collisions at $\sqrt{s}=13$ $\text {TeV}$ with the ATLAS detector}},}\ }\href
  {\doibase 10.1140/epjc/s10052-020-08509-3} {\bibfield  {journal} {\bibinfo
  {journal} {Eur. Phys. J. C}\ }\textbf {\bibinfo {volume} {80}},\ \bibinfo
  {pages} {1085} (\bibinfo {year} {2020})},\ \Eprint
  {http://arxiv.org/abs/2007.14858} {arXiv:2007.14858 [hep-ex]} \BibitemShut
  {NoStop}%
\bibitem [{\citenamefont {Aad}\ \emph {et~al.}(2021)\citenamefont {Aad} \emph
  {et~al.}}]{ATLAS:2021kqb}%
  \BibitemOpen
  \bibfield  {author} {\bibinfo {author} {\bibfnamefont {Georges}\ \bibnamefont
  {Aad}} \emph {et~al.} (\bibinfo {collaboration} {ATLAS}),\ }\bibfield
  {title} {\enquote {\bibinfo {title} {{Measurement of the $t\bar{t}t\bar{t}$
  production cross section in $pp$ collisions at $\sqrt{s}$=13 TeV with the
  ATLAS detector}},}\ }\href@noop {} {\  (\bibinfo {year} {2021})},\ \Eprint
  {http://arxiv.org/abs/2106.11683} {arXiv:2106.11683 [hep-ex]} \BibitemShut
  {NoStop}%
\bibitem [{\citenamefont {Aad}\ \emph {et~al.}(2015)\citenamefont {Aad} \emph
  {et~al.}}]{ATLAS:2015qtq}%
  \BibitemOpen
  \bibfield  {author} {\bibinfo {author} {\bibfnamefont {Georges}\ \bibnamefont
  {Aad}} \emph {et~al.} (\bibinfo {collaboration} {ATLAS}),\ }\bibfield
  {title} {\enquote {\bibinfo {title} {{Measurement of the $ t\overline{t}W $
  and $ t\overline{t}Z $ production cross sections in pp collisions at $
  \sqrt{s}=8 $ TeV with the ATLAS detector}},}\ }\href {\doibase
  10.1007/JHEP11(2015)172} {\bibfield  {journal} {\bibinfo  {journal} {JHEP}\
  }\textbf {\bibinfo {volume} {11}},\ \bibinfo {pages} {172} (\bibinfo {year}
  {2015})},\ \Eprint {http://arxiv.org/abs/1509.05276} {arXiv:1509.05276
  [hep-ex]} \BibitemShut {NoStop}%
\bibitem [{\citenamefont {Khachatryan}\ \emph
  {et~al.}(2016{\natexlab{b}})\citenamefont {Khachatryan} \emph
  {et~al.}}]{CMS:2015uvn}%
  \BibitemOpen
  \bibfield  {author} {\bibinfo {author} {\bibfnamefont {Vardan}\ \bibnamefont
  {Khachatryan}} \emph {et~al.} (\bibinfo {collaboration} {CMS}),\ }\bibfield
  {title} {\enquote {\bibinfo {title} {{Observation of top quark pairs produced
  in association with a vector boson in pp collisions at $ \sqrt{s}=8 $
  TeV}},}\ }\href {\doibase 10.1007/JHEP01(2016)096} {\bibfield  {journal}
  {\bibinfo  {journal} {JHEP}\ }\textbf {\bibinfo {volume} {01}},\ \bibinfo
  {pages} {096} (\bibinfo {year} {2016}{\natexlab{b}})},\ \Eprint
  {http://arxiv.org/abs/1510.01131} {arXiv:1510.01131 [hep-ex]} \BibitemShut
  {NoStop}%
\bibitem [{\citenamefont {Aaboud}\ \emph
  {et~al.}(2017{\natexlab{b}})\citenamefont {Aaboud} \emph
  {et~al.}}]{ATLAS:2016wgc}%
  \BibitemOpen
  \bibfield  {author} {\bibinfo {author} {\bibfnamefont {Morad}\ \bibnamefont
  {Aaboud}} \emph {et~al.} (\bibinfo {collaboration} {ATLAS}),\ }\bibfield
  {title} {\enquote {\bibinfo {title} {{Measurement of the $t\bar{t}Z$ and
  $t\bar{t}W$ production cross sections in multilepton final states using 3.2
  fb$^{-1}$ of $pp$ collisions at $\sqrt{s}$ = 13 TeV with the ATLAS
  detector}},}\ }\href {\doibase 10.1140/epjc/s10052-016-4574-y} {\bibfield
  {journal} {\bibinfo  {journal} {Eur. Phys. J. C}\ }\textbf {\bibinfo {volume}
  {77}},\ \bibinfo {pages} {40} (\bibinfo {year} {2017}{\natexlab{b}})},\
  \Eprint {http://arxiv.org/abs/1609.01599} {arXiv:1609.01599 [hep-ex]}
  \BibitemShut {NoStop}%
\bibitem [{\citenamefont {Sirunyan}\ \emph
  {et~al.}(2018{\natexlab{b}})\citenamefont {Sirunyan} \emph
  {et~al.}}]{CMS:2017ugv}%
  \BibitemOpen
  \bibfield  {author} {\bibinfo {author} {\bibfnamefont {Albert~M}\
  \bibnamefont {Sirunyan}} \emph {et~al.} (\bibinfo {collaboration} {CMS}),\
  }\bibfield  {title} {\enquote {\bibinfo {title} {{Measurement of the cross
  section for top quark pair production in association with a W or Z boson in
  proton-proton collisions at $\sqrt{s} =$ 13 TeV}},}\ }\href {\doibase
  10.1007/JHEP08(2018)011} {\bibfield  {journal} {\bibinfo  {journal} {JHEP}\
  }\textbf {\bibinfo {volume} {08}},\ \bibinfo {pages} {011} (\bibinfo {year}
  {2018}{\natexlab{b}})},\ \Eprint {http://arxiv.org/abs/1711.02547}
  {arXiv:1711.02547 [hep-ex]} \BibitemShut {NoStop}%
\bibitem [{\citenamefont {Aaboud}\ \emph
  {et~al.}(2019{\natexlab{b}})\citenamefont {Aaboud} \emph
  {et~al.}}]{ATLAS:2019fwo}%
  \BibitemOpen
  \bibfield  {author} {\bibinfo {author} {\bibfnamefont {Morad}\ \bibnamefont
  {Aaboud}} \emph {et~al.} (\bibinfo {collaboration} {ATLAS}),\ }\bibfield
  {title} {\enquote {\bibinfo {title} {{Measurement of the $t\bar{t}Z$ and
  $t\bar{t}W$ cross sections in proton-proton collisions at $\sqrt{s}=13$ TeV
  with the ATLAS detector}},}\ }\href {\doibase 10.1103/PhysRevD.99.072009}
  {\bibfield  {journal} {\bibinfo  {journal} {Phys. Rev. D}\ }\textbf {\bibinfo
  {volume} {99}},\ \bibinfo {pages} {072009} (\bibinfo {year}
  {2019}{\natexlab{b}})},\ \Eprint {http://arxiv.org/abs/1901.03584}
  {arXiv:1901.03584 [hep-ex]} \BibitemShut {NoStop}%
\bibitem [{\citenamefont {Badger}\ \emph {et~al.}(2011)\citenamefont {Badger},
  \citenamefont {Campbell},\ and\ \citenamefont {Ellis}}]{Badger:2010mg}%
  \BibitemOpen
  \bibfield  {author} {\bibinfo {author} {\bibfnamefont {Simon}\ \bibnamefont
  {Badger}}, \bibinfo {author} {\bibfnamefont {John~M.}\ \bibnamefont
  {Campbell}}, \ and\ \bibinfo {author} {\bibfnamefont {R.~K.}\ \bibnamefont
  {Ellis}},\ }\bibfield  {title} {\enquote {\bibinfo {title} {{QCD Corrections
  to the Hadronic Production of a Heavy Quark Pair and a W-Boson Including
  Decay Correlations}},}\ }\href {\doibase 10.1007/JHEP03(2011)027} {\bibfield
  {journal} {\bibinfo  {journal} {JHEP}\ }\textbf {\bibinfo {volume} {03}},\
  \bibinfo {pages} {027} (\bibinfo {year} {2011})},\ \Eprint
  {http://arxiv.org/abs/1011.6647} {arXiv:1011.6647 [hep-ph]} \BibitemShut
  {NoStop}%
\bibitem [{\citenamefont {Campbell}\ and\ \citenamefont
  {Ellis}(2012)}]{Campbell:2012dh}%
  \BibitemOpen
  \bibfield  {author} {\bibinfo {author} {\bibfnamefont {John~M.}\ \bibnamefont
  {Campbell}}\ and\ \bibinfo {author} {\bibfnamefont {R.~Keith}\ \bibnamefont
  {Ellis}},\ }\bibfield  {title} {\enquote {\bibinfo {title} {{$t \bar{t}
  W^{+-}$ production and decay at NLO}},}\ }\href {\doibase
  10.1007/JHEP07(2012)052} {\bibfield  {journal} {\bibinfo  {journal} {JHEP}\
  }\textbf {\bibinfo {volume} {07}},\ \bibinfo {pages} {052} (\bibinfo {year}
  {2012})},\ \Eprint {http://arxiv.org/abs/1204.5678} {arXiv:1204.5678
  [hep-ph]} \BibitemShut {NoStop}%
\bibitem [{\citenamefont {Dror}\ \emph {et~al.}(2016)\citenamefont {Dror},
  \citenamefont {Farina}, \citenamefont {Salvioni},\ and\ \citenamefont
  {Serra}}]{Dror:2015nkp}%
  \BibitemOpen
  \bibfield  {author} {\bibinfo {author} {\bibfnamefont {Jeff~Asaf}\
  \bibnamefont {Dror}}, \bibinfo {author} {\bibfnamefont {Marco}\ \bibnamefont
  {Farina}}, \bibinfo {author} {\bibfnamefont {Ennio}\ \bibnamefont
  {Salvioni}}, \ and\ \bibinfo {author} {\bibfnamefont {Javi}\ \bibnamefont
  {Serra}},\ }\bibfield  {title} {\enquote {\bibinfo {title} {{Strong tW
  Scattering at the LHC}},}\ }\href {\doibase 10.1007/JHEP01(2016)071}
  {\bibfield  {journal} {\bibinfo  {journal} {JHEP}\ }\textbf {\bibinfo
  {volume} {01}},\ \bibinfo {pages} {071} (\bibinfo {year} {2016})},\ \Eprint
  {http://arxiv.org/abs/1511.03674} {arXiv:1511.03674 [hep-ph]} \BibitemShut
  {NoStop}%
\bibitem [{\citenamefont {Frixione}\ \emph {et~al.}(2015)\citenamefont
  {Frixione}, \citenamefont {Hirschi}, \citenamefont {Pagani}, \citenamefont
  {Shao},\ and\ \citenamefont {Zaro}}]{Frixione:2015zaa}%
  \BibitemOpen
  \bibfield  {author} {\bibinfo {author} {\bibfnamefont {S.}~\bibnamefont
  {Frixione}}, \bibinfo {author} {\bibfnamefont {V.}~\bibnamefont {Hirschi}},
  \bibinfo {author} {\bibfnamefont {D.}~\bibnamefont {Pagani}}, \bibinfo
  {author} {\bibfnamefont {H.~S.}\ \bibnamefont {Shao}}, \ and\ \bibinfo
  {author} {\bibfnamefont {M.}~\bibnamefont {Zaro}},\ }\bibfield  {title}
  {\enquote {\bibinfo {title} {{Electroweak and QCD corrections to top-pair
  hadroproduction in association with heavy bosons}},}\ }\href {\doibase
  10.1007/JHEP06(2015)184} {\bibfield  {journal} {\bibinfo  {journal} {JHEP}\
  }\textbf {\bibinfo {volume} {06}},\ \bibinfo {pages} {184} (\bibinfo {year}
  {2015})},\ \Eprint {http://arxiv.org/abs/1504.03446} {arXiv:1504.03446
  [hep-ph]} \BibitemShut {NoStop}%
\bibitem [{\citenamefont {Frederix}\ \emph {et~al.}(2018)\citenamefont
  {Frederix}, \citenamefont {Pagani},\ and\ \citenamefont
  {Zaro}}]{Frederix:2017wme}%
  \BibitemOpen
  \bibfield  {author} {\bibinfo {author} {\bibfnamefont {Rikkert}\ \bibnamefont
  {Frederix}}, \bibinfo {author} {\bibfnamefont {Davide}\ \bibnamefont
  {Pagani}}, \ and\ \bibinfo {author} {\bibfnamefont {Marco}\ \bibnamefont
  {Zaro}},\ }\bibfield  {title} {\enquote {\bibinfo {title} {{Large NLO
  corrections in $t\bar{t}W^{\pm}$ and $t\bar{t}t\bar{t}$ hadroproduction from
  supposedly subleading EW contributions}},}\ }\href {\doibase
  10.1007/JHEP02(2018)031} {\bibfield  {journal} {\bibinfo  {journal} {JHEP}\
  }\textbf {\bibinfo {volume} {02}},\ \bibinfo {pages} {031} (\bibinfo {year}
  {2018})},\ \Eprint {http://arxiv.org/abs/1711.02116} {arXiv:1711.02116
  [hep-ph]} \BibitemShut {NoStop}%
\bibitem [{\citenamefont {Li}\ \emph {et~al.}(2014)\citenamefont {Li},
  \citenamefont {Li},\ and\ \citenamefont {Li}}]{Li:2014ula}%
  \BibitemOpen
  \bibfield  {author} {\bibinfo {author} {\bibfnamefont {Hai~Tao}\ \bibnamefont
  {Li}}, \bibinfo {author} {\bibfnamefont {Chong~Sheng}\ \bibnamefont {Li}}, \
  and\ \bibinfo {author} {\bibfnamefont {Shi~Ang}\ \bibnamefont {Li}},\
  }\bibfield  {title} {\enquote {\bibinfo {title} {{Renormalization group
  improved predictions for $t\bar{t}W^\pm$ production at hadron colliders}},}\
  }\href {\doibase 10.1103/PhysRevD.90.094009} {\bibfield  {journal} {\bibinfo
  {journal} {Phys. Rev. D}\ }\textbf {\bibinfo {volume} {90}},\ \bibinfo
  {pages} {094009} (\bibinfo {year} {2014})},\ \Eprint
  {http://arxiv.org/abs/1409.1460} {arXiv:1409.1460 [hep-ph]} \BibitemShut
  {NoStop}%
\bibitem [{\citenamefont {Broggio}\ \emph {et~al.}(2016)\citenamefont
  {Broggio}, \citenamefont {Ferroglia}, \citenamefont {Ossola},\ and\
  \citenamefont {Pecjak}}]{Broggio:2016zgg}%
  \BibitemOpen
  \bibfield  {author} {\bibinfo {author} {\bibfnamefont {Alessandro}\
  \bibnamefont {Broggio}}, \bibinfo {author} {\bibfnamefont {Andrea}\
  \bibnamefont {Ferroglia}}, \bibinfo {author} {\bibfnamefont {Giovanni}\
  \bibnamefont {Ossola}}, \ and\ \bibinfo {author} {\bibfnamefont {Ben~D.}\
  \bibnamefont {Pecjak}},\ }\bibfield  {title} {\enquote {\bibinfo {title}
  {{Associated production of a top pair and a W boson at
  next-to-next-to-leading logarithmic accuracy}},}\ }\href {\doibase
  10.1007/JHEP09(2016)089} {\bibfield  {journal} {\bibinfo  {journal} {JHEP}\
  }\textbf {\bibinfo {volume} {09}},\ \bibinfo {pages} {089} (\bibinfo {year}
  {2016})},\ \Eprint {http://arxiv.org/abs/1607.05303} {arXiv:1607.05303
  [hep-ph]} \BibitemShut {NoStop}%
\bibitem [{\citenamefont {Kulesza}\ \emph {et~al.}(2019)\citenamefont
  {Kulesza}, \citenamefont {Motyka}, \citenamefont {Schwartl\"ander},
  \citenamefont {Stebel},\ and\ \citenamefont {Theeuwes}}]{Kulesza:2018tqz}%
  \BibitemOpen
  \bibfield  {author} {\bibinfo {author} {\bibfnamefont {Anna}\ \bibnamefont
  {Kulesza}}, \bibinfo {author} {\bibfnamefont {Leszek}\ \bibnamefont
  {Motyka}}, \bibinfo {author} {\bibfnamefont {Daniel}\ \bibnamefont
  {Schwartl\"ander}}, \bibinfo {author} {\bibfnamefont {Tomasz}\ \bibnamefont
  {Stebel}}, \ and\ \bibinfo {author} {\bibfnamefont {Vincent}\ \bibnamefont
  {Theeuwes}},\ }\bibfield  {title} {\enquote {\bibinfo {title} {{Associated
  production of a top quark pair with a heavy electroweak gauge boson at
  NLO$+$NNLL accuracy}},}\ }\href {\doibase 10.1140/epjc/s10052-019-6746-z}
  {\bibfield  {journal} {\bibinfo  {journal} {Eur. Phys. J. C}\ }\textbf
  {\bibinfo {volume} {79}},\ \bibinfo {pages} {249} (\bibinfo {year} {2019})},\
  \Eprint {http://arxiv.org/abs/1812.08622} {arXiv:1812.08622 [hep-ph]}
  \BibitemShut {NoStop}%
\bibitem [{\citenamefont {Broggio}\ \emph {et~al.}(2019)\citenamefont
  {Broggio}, \citenamefont {Ferroglia}, \citenamefont {Frederix}, \citenamefont
  {Pagani}, \citenamefont {Pecjak},\ and\ \citenamefont
  {Tsinikos}}]{Broggio:2019ewu}%
  \BibitemOpen
  \bibfield  {author} {\bibinfo {author} {\bibfnamefont {Alessandro}\
  \bibnamefont {Broggio}}, \bibinfo {author} {\bibfnamefont {Andrea}\
  \bibnamefont {Ferroglia}}, \bibinfo {author} {\bibfnamefont {Rikkert}\
  \bibnamefont {Frederix}}, \bibinfo {author} {\bibfnamefont {Davide}\
  \bibnamefont {Pagani}}, \bibinfo {author} {\bibfnamefont {Benjamin~D.}\
  \bibnamefont {Pecjak}}, \ and\ \bibinfo {author} {\bibfnamefont {Ioannis}\
  \bibnamefont {Tsinikos}},\ }\bibfield  {title} {\enquote {\bibinfo {title}
  {{Top-quark pair hadroproduction in association with a heavy boson at
  NLO+NNLL including EW corrections}},}\ }\href {\doibase
  10.1007/JHEP08(2019)039} {\bibfield  {journal} {\bibinfo  {journal} {JHEP}\
  }\textbf {\bibinfo {volume} {08}},\ \bibinfo {pages} {039} (\bibinfo {year}
  {2019})},\ \Eprint {http://arxiv.org/abs/1907.04343} {arXiv:1907.04343
  [hep-ph]} \BibitemShut {NoStop}%
\bibitem [{\citenamefont {Kulesza}\ \emph {et~al.}(2020)\citenamefont
  {Kulesza}, \citenamefont {Motyka}, \citenamefont {Schwartl\"ander},
  \citenamefont {Stebel},\ and\ \citenamefont {Theeuwes}}]{Kulesza:2020nfh}%
  \BibitemOpen
  \bibfield  {author} {\bibinfo {author} {\bibfnamefont {Anna}\ \bibnamefont
  {Kulesza}}, \bibinfo {author} {\bibfnamefont {Leszek}\ \bibnamefont
  {Motyka}}, \bibinfo {author} {\bibfnamefont {Daniel}\ \bibnamefont
  {Schwartl\"ander}}, \bibinfo {author} {\bibfnamefont {Tomasz}\ \bibnamefont
  {Stebel}}, \ and\ \bibinfo {author} {\bibfnamefont {Vincent}\ \bibnamefont
  {Theeuwes}},\ }\bibfield  {title} {\enquote {\bibinfo {title} {{Associated
  top quark pair production with a heavy boson: differential cross sections at
  NLO+NNLL accuracy}},}\ }\href {\doibase 10.1140/epjc/s10052-020-7987-6}
  {\bibfield  {journal} {\bibinfo  {journal} {Eur. Phys. J. C}\ }\textbf
  {\bibinfo {volume} {80}},\ \bibinfo {pages} {428} (\bibinfo {year} {2020})},\
  \Eprint {http://arxiv.org/abs/2001.03031} {arXiv:2001.03031 [hep-ph]}
  \BibitemShut {NoStop}%
\bibitem [{\citenamefont {Frixione}\ and\ \citenamefont
  {Webber}(2002)}]{Frixione:2002ik}%
  \BibitemOpen
  \bibfield  {author} {\bibinfo {author} {\bibfnamefont {Stefano}\ \bibnamefont
  {Frixione}}\ and\ \bibinfo {author} {\bibfnamefont {Bryan~R.}\ \bibnamefont
  {Webber}},\ }\bibfield  {title} {\enquote {\bibinfo {title} {{Matching NLO
  QCD computations and parton shower simulations}},}\ }\href {\doibase
  10.1088/1126-6708/2002/06/029} {\bibfield  {journal} {\bibinfo  {journal}
  {JHEP}\ }\textbf {\bibinfo {volume} {06}},\ \bibinfo {pages} {029} (\bibinfo
  {year} {2002})},\ \Eprint {http://arxiv.org/abs/hep-ph/0204244}
  {arXiv:hep-ph/0204244} \BibitemShut {NoStop}%
\bibitem [{\citenamefont {Frixione}\ \emph {et~al.}(2003)\citenamefont
  {Frixione}, \citenamefont {Nason},\ and\ \citenamefont
  {Webber}}]{Frixione:2003ei}%
  \BibitemOpen
  \bibfield  {author} {\bibinfo {author} {\bibfnamefont {Stefano}\ \bibnamefont
  {Frixione}}, \bibinfo {author} {\bibfnamefont {Paolo}\ \bibnamefont {Nason}},
  \ and\ \bibinfo {author} {\bibfnamefont {Bryan~R.}\ \bibnamefont {Webber}},\
  }\bibfield  {title} {\enquote {\bibinfo {title} {{Matching NLO QCD and parton
  showers in heavy flavor production}},}\ }\href {\doibase
  10.1088/1126-6708/2003/08/007} {\bibfield  {journal} {\bibinfo  {journal}
  {JHEP}\ }\textbf {\bibinfo {volume} {08}},\ \bibinfo {pages} {007} (\bibinfo
  {year} {2003})},\ \Eprint {http://arxiv.org/abs/hep-ph/0305252}
  {arXiv:hep-ph/0305252} \BibitemShut {NoStop}%
\bibitem [{\citenamefont {Maltoni}\ \emph {et~al.}(2014)\citenamefont
  {Maltoni}, \citenamefont {Mangano}, \citenamefont {Tsinikos},\ and\
  \citenamefont {Zaro}}]{Maltoni:2014zpa}%
  \BibitemOpen
  \bibfield  {author} {\bibinfo {author} {\bibfnamefont {F.}~\bibnamefont
  {Maltoni}}, \bibinfo {author} {\bibfnamefont {M.~L.}\ \bibnamefont
  {Mangano}}, \bibinfo {author} {\bibfnamefont {I.}~\bibnamefont {Tsinikos}}, \
  and\ \bibinfo {author} {\bibfnamefont {M.}~\bibnamefont {Zaro}},\ }\bibfield
  {title} {\enquote {\bibinfo {title} {{Top-quark charge asymmetry and
  polarization in $t\overline{t}W^\pm$ production at the LHC}},}\ }\href
  {\doibase 10.1016/j.physletb.2014.07.033} {\bibfield  {journal} {\bibinfo
  {journal} {Phys. Lett. B}\ }\textbf {\bibinfo {volume} {736}},\ \bibinfo
  {pages} {252--260} (\bibinfo {year} {2014})},\ \Eprint
  {http://arxiv.org/abs/1406.3262} {arXiv:1406.3262 [hep-ph]} \BibitemShut
  {NoStop}%
\bibitem [{\citenamefont {Maltoni}\ \emph {et~al.}(2016)\citenamefont
  {Maltoni}, \citenamefont {Pagani},\ and\ \citenamefont
  {Tsinikos}}]{Maltoni:2015ena}%
  \BibitemOpen
  \bibfield  {author} {\bibinfo {author} {\bibfnamefont {Fabio}\ \bibnamefont
  {Maltoni}}, \bibinfo {author} {\bibfnamefont {Davide}\ \bibnamefont
  {Pagani}}, \ and\ \bibinfo {author} {\bibfnamefont {Ioannis}\ \bibnamefont
  {Tsinikos}},\ }\bibfield  {title} {\enquote {\bibinfo {title} {{Associated
  production of a top-quark pair with vector bosons at NLO in QCD: impact on $
  \mathrm{t}\overline{\mathrm{t}}\mathrm{H} $ searches at the LHC}},}\ }\href
  {\doibase 10.1007/JHEP02(2016)113} {\bibfield  {journal} {\bibinfo  {journal}
  {JHEP}\ }\textbf {\bibinfo {volume} {02}},\ \bibinfo {pages} {113} (\bibinfo
  {year} {2016})},\ \Eprint {http://arxiv.org/abs/1507.05640} {arXiv:1507.05640
  [hep-ph]} \BibitemShut {NoStop}%
\bibitem [{\citenamefont {Frederix}\ and\ \citenamefont
  {Tsinikos}(2020)}]{Frederix:2020jzp}%
  \BibitemOpen
  \bibfield  {author} {\bibinfo {author} {\bibfnamefont {Rikkert}\ \bibnamefont
  {Frederix}}\ and\ \bibinfo {author} {\bibfnamefont {Ioannis}\ \bibnamefont
  {Tsinikos}},\ }\bibfield  {title} {\enquote {\bibinfo {title} {{Subleading EW
  corrections and spin-correlation effects in $t\bar{t}W$ multi-lepton
  signatures}},}\ }\href {\doibase 10.1140/epjc/s10052-020-8388-6} {\bibfield
  {journal} {\bibinfo  {journal} {Eur. Phys. J. C}\ }\textbf {\bibinfo {volume}
  {80}},\ \bibinfo {pages} {803} (\bibinfo {year} {2020})},\ \Eprint
  {http://arxiv.org/abs/2004.09552} {arXiv:2004.09552 [hep-ph]} \BibitemShut
  {NoStop}%
\bibitem [{\citenamefont {Nason}(2004)}]{Nason:2004rx}%
  \BibitemOpen
  \bibfield  {author} {\bibinfo {author} {\bibfnamefont {Paolo}\ \bibnamefont
  {Nason}},\ }\bibfield  {title} {\enquote {\bibinfo {title} {{A New method for
  combining NLO QCD with shower Monte Carlo algorithms}},}\ }\href {\doibase
  10.1088/1126-6708/2004/11/040} {\bibfield  {journal} {\bibinfo  {journal}
  {JHEP}\ }\textbf {\bibinfo {volume} {11}},\ \bibinfo {pages} {040} (\bibinfo
  {year} {2004})},\ \Eprint {http://arxiv.org/abs/hep-ph/0409146}
  {arXiv:hep-ph/0409146} \BibitemShut {NoStop}%
\bibitem [{\citenamefont {Frixione}\ \emph
  {et~al.}(2007{\natexlab{a}})\citenamefont {Frixione}, \citenamefont {Nason},\
  and\ \citenamefont {Oleari}}]{Frixione:2007vw}%
  \BibitemOpen
  \bibfield  {author} {\bibinfo {author} {\bibfnamefont {Stefano}\ \bibnamefont
  {Frixione}}, \bibinfo {author} {\bibfnamefont {Paolo}\ \bibnamefont {Nason}},
  \ and\ \bibinfo {author} {\bibfnamefont {Carlo}\ \bibnamefont {Oleari}},\
  }\bibfield  {title} {\enquote {\bibinfo {title} {{Matching NLO QCD
  computations with Parton Shower simulations: the POWHEG method}},}\ }\href
  {\doibase 10.1088/1126-6708/2007/11/070} {\bibfield  {journal} {\bibinfo
  {journal} {JHEP}\ }\textbf {\bibinfo {volume} {11}},\ \bibinfo {pages} {070}
  (\bibinfo {year} {2007}{\natexlab{a}})},\ \Eprint
  {http://arxiv.org/abs/0709.2092} {arXiv:0709.2092 [hep-ph]} \BibitemShut
  {NoStop}%
\bibitem [{\citenamefont {Garzelli}\ \emph {et~al.}(2012)\citenamefont
  {Garzelli}, \citenamefont {Kardos}, \citenamefont {Papadopoulos},\ and\
  \citenamefont {Trocsanyi}}]{Garzelli:2012bn}%
  \BibitemOpen
  \bibfield  {author} {\bibinfo {author} {\bibfnamefont {M.~V.}\ \bibnamefont
  {Garzelli}}, \bibinfo {author} {\bibfnamefont {A.}~\bibnamefont {Kardos}},
  \bibinfo {author} {\bibfnamefont {C.~G.}\ \bibnamefont {Papadopoulos}}, \
  and\ \bibinfo {author} {\bibfnamefont {Z.}~\bibnamefont {Trocsanyi}},\
  }\bibfield  {title} {\enquote {\bibinfo {title} {{t $\bar{t}$ $W^{+-}$ and t
  $\bar{t}$ Z Hadroproduction at NLO accuracy in QCD with Parton Shower and
  Hadronization effects}},}\ }\href {\doibase 10.1007/JHEP11(2012)056}
  {\bibfield  {journal} {\bibinfo  {journal} {JHEP}\ }\textbf {\bibinfo
  {volume} {11}},\ \bibinfo {pages} {056} (\bibinfo {year} {2012})},\ \Eprint
  {http://arxiv.org/abs/1208.2665} {arXiv:1208.2665 [hep-ph]} \BibitemShut
  {NoStop}%
\bibitem [{\citenamefont {Cordero}\ \emph {et~al.}(2021)\citenamefont
  {Cordero}, \citenamefont {Kraus},\ and\ \citenamefont
  {Reina}}]{Cordero:2021iau}%
  \BibitemOpen
  \bibfield  {author} {\bibinfo {author} {\bibfnamefont {F.~Febres}\
  \bibnamefont {Cordero}}, \bibinfo {author} {\bibfnamefont {M.}~\bibnamefont
  {Kraus}}, \ and\ \bibinfo {author} {\bibfnamefont {L.}~\bibnamefont
  {Reina}},\ }\bibfield  {title} {\enquote {\bibinfo {title} {{Top-quark pair
  production in association with a $W^\pm$ gauge boson in the POWHEG-BOX}},}\
  }\href {\doibase 10.1103/PhysRevD.103.094014} {\bibfield  {journal} {\bibinfo
   {journal} {Phys. Rev. D}\ }\textbf {\bibinfo {volume} {103}},\ \bibinfo
  {pages} {094014} (\bibinfo {year} {2021})},\ \Eprint
  {http://arxiv.org/abs/2101.11808} {arXiv:2101.11808 [hep-ph]} \BibitemShut
  {NoStop}%
\bibitem [{\citenamefont {von Buddenbrock}\ \emph {et~al.}(2020)\citenamefont
  {von Buddenbrock}, \citenamefont {Ruiz},\ and\ \citenamefont
  {Mellado}}]{vonBuddenbrock:2020ter}%
  \BibitemOpen
  \bibfield  {author} {\bibinfo {author} {\bibfnamefont {Stefan}\ \bibnamefont
  {von Buddenbrock}}, \bibinfo {author} {\bibfnamefont {Richard}\ \bibnamefont
  {Ruiz}}, \ and\ \bibinfo {author} {\bibfnamefont {Bruce}\ \bibnamefont
  {Mellado}},\ }\bibfield  {title} {\enquote {\bibinfo {title} {{Anatomy of
  inclusive $t\bar t W$ production at hadron colliders}},}\ }\href {\doibase
  10.1016/j.physletb.2020.135964} {\bibfield  {journal} {\bibinfo  {journal}
  {Phys. Lett. B}\ }\textbf {\bibinfo {volume} {811}},\ \bibinfo {pages}
  {135964} (\bibinfo {year} {2020})},\ \Eprint
  {http://arxiv.org/abs/2009.00032} {arXiv:2009.00032 [hep-ph]} \BibitemShut
  {NoStop}%
\bibitem [{ATL(2020)}]{ATLAS:2020esn}%
  \BibitemOpen
  \bibfield  {title} {\enquote {\bibinfo {title} {{Modelling of rare top quark
  processes at $\sqrt{s}$ = 13 TeV in ATLAS}},}\ }\href@noop {} {\  (\bibinfo
  {year} {2020})}\BibitemShut {NoStop}%
\bibitem [{\citenamefont {Frederix}\ and\ \citenamefont
  {Tsinikos}(2021)}]{Frederix:2021agh}%
  \BibitemOpen
  \bibfield  {author} {\bibinfo {author} {\bibfnamefont {Rikkert}\ \bibnamefont
  {Frederix}}\ and\ \bibinfo {author} {\bibfnamefont {Ioannis}\ \bibnamefont
  {Tsinikos}},\ }\bibfield  {title} {\enquote {\bibinfo {title} {{On improving
  NLO merging for $t \bar t W$ production}},}\ }\href@noop {} {\  (\bibinfo
  {year} {2021})},\ \Eprint {http://arxiv.org/abs/2108.07826} {arXiv:2108.07826
  [hep-ph]} \BibitemShut {NoStop}%
\bibitem [{\citenamefont {Bevilacqua}\ \emph
  {et~al.}(2020{\natexlab{a}})\citenamefont {Bevilacqua}, \citenamefont {Bi},
  \citenamefont {Hartanto}, \citenamefont {Kraus},\ and\ \citenamefont
  {Worek}}]{Bevilacqua:2020pzy}%
  \BibitemOpen
  \bibfield  {author} {\bibinfo {author} {\bibfnamefont {Giuseppe}\
  \bibnamefont {Bevilacqua}}, \bibinfo {author} {\bibfnamefont {Huan-Yu}\
  \bibnamefont {Bi}}, \bibinfo {author} {\bibfnamefont {Heribertus~Bayu}\
  \bibnamefont {Hartanto}}, \bibinfo {author} {\bibfnamefont {Manfred}\
  \bibnamefont {Kraus}}, \ and\ \bibinfo {author} {\bibfnamefont {Malgorzata}\
  \bibnamefont {Worek}},\ }\bibfield  {title} {\enquote {\bibinfo {title} {{The
  simplest of them all: $t\bar{t} W^\pm$ at NLO accuracy in QCD}},}\ }\href
  {\doibase 10.1007/JHEP08(2020)043} {\bibfield  {journal} {\bibinfo  {journal}
  {JHEP}\ }\textbf {\bibinfo {volume} {08}},\ \bibinfo {pages} {043} (\bibinfo
  {year} {2020}{\natexlab{a}})},\ \Eprint {http://arxiv.org/abs/2005.09427}
  {arXiv:2005.09427 [hep-ph]} \BibitemShut {NoStop}%
\bibitem [{\citenamefont {Denner}\ and\ \citenamefont
  {Pelliccioli}(2020)}]{Denner:2020hgg}%
  \BibitemOpen
  \bibfield  {author} {\bibinfo {author} {\bibfnamefont {Ansgar}\ \bibnamefont
  {Denner}}\ and\ \bibinfo {author} {\bibfnamefont {Giovanni}\ \bibnamefont
  {Pelliccioli}},\ }\bibfield  {title} {\enquote {\bibinfo {title} {{NLO QCD
  corrections to off-shell $\text{t}\bar{\text{t}}\text{W}^+$ production at the
  LHC}},}\ }\href {\doibase 10.1007/JHEP11(2020)069} {\bibfield  {journal}
  {\bibinfo  {journal} {JHEP}\ }\textbf {\bibinfo {volume} {11}},\ \bibinfo
  {pages} {069} (\bibinfo {year} {2020})},\ \Eprint
  {http://arxiv.org/abs/2007.12089} {arXiv:2007.12089 [hep-ph]} \BibitemShut
  {NoStop}%
\bibitem [{\citenamefont {Bevilacqua}\ \emph {et~al.}(2021)\citenamefont
  {Bevilacqua}, \citenamefont {Bi}, \citenamefont {Hartanto}, \citenamefont
  {Kraus}, \citenamefont {Nasufi},\ and\ \citenamefont
  {Worek}}]{Bevilacqua:2020srb}%
  \BibitemOpen
  \bibfield  {author} {\bibinfo {author} {\bibfnamefont {Giuseppe}\
  \bibnamefont {Bevilacqua}}, \bibinfo {author} {\bibfnamefont {Huan-Yu}\
  \bibnamefont {Bi}}, \bibinfo {author} {\bibfnamefont {Heribertus~Bayu}\
  \bibnamefont {Hartanto}}, \bibinfo {author} {\bibfnamefont {Manfred}\
  \bibnamefont {Kraus}}, \bibinfo {author} {\bibfnamefont {Jasmina}\
  \bibnamefont {Nasufi}}, \ and\ \bibinfo {author} {\bibfnamefont {Malgorzata}\
  \bibnamefont {Worek}},\ }\bibfield  {title} {\enquote {\bibinfo {title} {{NLO
  QCD corrections to off-shell ${t\bar{t}W^\pm}$ production at the LHC:
  Correlations and Asymmetries}},}\ }\href {\doibase
  10.1140/epjc/s10052-021-09478-x} {\bibfield  {journal} {\bibinfo  {journal}
  {Eur. Phys. J. C}\ }\textbf {\bibinfo {volume} {81}},\ \bibinfo {pages} {675}
  (\bibinfo {year} {2021})},\ \Eprint {http://arxiv.org/abs/2012.01363}
  {arXiv:2012.01363 [hep-ph]} \BibitemShut {NoStop}%
\bibitem [{\citenamefont {Denner}\ and\ \citenamefont
  {Pelliccioli}(2021)}]{Denner:2021hqi}%
  \BibitemOpen
  \bibfield  {author} {\bibinfo {author} {\bibfnamefont {Ansgar}\ \bibnamefont
  {Denner}}\ and\ \bibinfo {author} {\bibfnamefont {Giovanni}\ \bibnamefont
  {Pelliccioli}},\ }\bibfield  {title} {\enquote {\bibinfo {title} {{Combined
  NLO EW and QCD corrections to off-shell $\text {t} \overline{\text {t}}\text
  {W} $ production at the LHC}},}\ }\href {\doibase
  10.1140/epjc/s10052-021-09143-3} {\bibfield  {journal} {\bibinfo  {journal}
  {Eur. Phys. J. C}\ }\textbf {\bibinfo {volume} {81}},\ \bibinfo {pages} {354}
  (\bibinfo {year} {2021})},\ \Eprint {http://arxiv.org/abs/2102.03246}
  {arXiv:2102.03246 [hep-ph]} \BibitemShut {NoStop}%
\bibitem [{\citenamefont {Ball}\ \emph {et~al.}(2017)\citenamefont {Ball} \emph
  {et~al.}}]{NNPDF:2017mvq}%
  \BibitemOpen
  \bibfield  {author} {\bibinfo {author} {\bibfnamefont {Richard~D.}\
  \bibnamefont {Ball}} \emph {et~al.} (\bibinfo {collaboration} {NNPDF}),\
  }\bibfield  {title} {\enquote {\bibinfo {title} {{Parton distributions from
  high-precision collider data}},}\ }\href {\doibase
  10.1140/epjc/s10052-017-5199-5} {\bibfield  {journal} {\bibinfo  {journal}
  {Eur. Phys. J. C}\ }\textbf {\bibinfo {volume} {77}},\ \bibinfo {pages} {663}
  (\bibinfo {year} {2017})},\ \Eprint {http://arxiv.org/abs/1706.00428}
  {arXiv:1706.00428 [hep-ph]} \BibitemShut {NoStop}%
\bibitem [{\citenamefont {Buckley}\ \emph {et~al.}(2015)\citenamefont
  {Buckley}, \citenamefont {Ferrando}, \citenamefont {Lloyd}, \citenamefont
  {Nordstr\"om}, \citenamefont {Page}, \citenamefont {R\"ufenacht},
  \citenamefont {Sch\"onherr},\ and\ \citenamefont {Watt}}]{Buckley:2014ana}%
  \BibitemOpen
  \bibfield  {author} {\bibinfo {author} {\bibfnamefont {Andy}\ \bibnamefont
  {Buckley}}, \bibinfo {author} {\bibfnamefont {James}\ \bibnamefont
  {Ferrando}}, \bibinfo {author} {\bibfnamefont {Stephen}\ \bibnamefont
  {Lloyd}}, \bibinfo {author} {\bibfnamefont {Karl}\ \bibnamefont
  {Nordstr\"om}}, \bibinfo {author} {\bibfnamefont {Ben}\ \bibnamefont {Page}},
  \bibinfo {author} {\bibfnamefont {Martin}\ \bibnamefont {R\"ufenacht}},
  \bibinfo {author} {\bibfnamefont {Marek}\ \bibnamefont {Sch\"onherr}}, \ and\
  \bibinfo {author} {\bibfnamefont {Graeme}\ \bibnamefont {Watt}},\ }\bibfield
  {title} {\enquote {\bibinfo {title} {{LHAPDF6: parton density access in the
  LHC precision era}},}\ }\href {\doibase 10.1140/epjc/s10052-015-3318-8}
  {\bibfield  {journal} {\bibinfo  {journal} {Eur. Phys. J. C}\ }\textbf
  {\bibinfo {volume} {75}},\ \bibinfo {pages} {132} (\bibinfo {year} {2015})},\
  \Eprint {http://arxiv.org/abs/1412.7420} {arXiv:1412.7420 [hep-ph]}
  \BibitemShut {NoStop}%
\bibitem [{\citenamefont {Denner}\ \emph {et~al.}(2000)\citenamefont {Denner},
  \citenamefont {Dittmaier}, \citenamefont {Roth},\ and\ \citenamefont
  {Wackeroth}}]{Denner:2000bj}%
  \BibitemOpen
  \bibfield  {author} {\bibinfo {author} {\bibfnamefont {Ansgar}\ \bibnamefont
  {Denner}}, \bibinfo {author} {\bibfnamefont {S.}~\bibnamefont {Dittmaier}},
  \bibinfo {author} {\bibfnamefont {M.}~\bibnamefont {Roth}}, \ and\ \bibinfo
  {author} {\bibfnamefont {D.}~\bibnamefont {Wackeroth}},\ }\bibfield  {title}
  {\enquote {\bibinfo {title} {{Electroweak radiative corrections to e+ e-
  ---\ensuremath{>} W W ---\ensuremath{>} 4 fermions in double pole
  approximation: The RACOONWW approach}},}\ }\href {\doibase
  10.1016/S0550-3213(00)00511-3} {\bibfield  {journal} {\bibinfo  {journal}
  {Nucl. Phys. B}\ }\textbf {\bibinfo {volume} {587}},\ \bibinfo {pages}
  {67--117} (\bibinfo {year} {2000})},\ \Eprint
  {http://arxiv.org/abs/hep-ph/0006307} {arXiv:hep-ph/0006307} \BibitemShut
  {NoStop}%
\bibitem [{\citenamefont {Bevilacqua}\ \emph
  {et~al.}(2013{\natexlab{a}})\citenamefont {Bevilacqua}, \citenamefont
  {Czakon}, \citenamefont {Garzelli}, \citenamefont {van Hameren},
  \citenamefont {Kardos}, \citenamefont {Papadopoulos}, \citenamefont
  {Pittau},\ and\ \citenamefont {Worek}}]{Bevilacqua:2011xh}%
  \BibitemOpen
  \bibfield  {author} {\bibinfo {author} {\bibfnamefont {G.}~\bibnamefont
  {Bevilacqua}}, \bibinfo {author} {\bibfnamefont {M.}~\bibnamefont {Czakon}},
  \bibinfo {author} {\bibfnamefont {M.~V.}\ \bibnamefont {Garzelli}}, \bibinfo
  {author} {\bibfnamefont {A.}~\bibnamefont {van Hameren}}, \bibinfo {author}
  {\bibfnamefont {A.}~\bibnamefont {Kardos}}, \bibinfo {author} {\bibfnamefont
  {C.~G.}\ \bibnamefont {Papadopoulos}}, \bibinfo {author} {\bibfnamefont
  {R.}~\bibnamefont {Pittau}}, \ and\ \bibinfo {author} {\bibfnamefont
  {M.}~\bibnamefont {Worek}},\ }\bibfield  {title} {\enquote {\bibinfo {title}
  {{HELAC-NLO}},}\ }\href {\doibase 10.1016/j.cpc.2012.10.033} {\bibfield
  {journal} {\bibinfo  {journal} {Comput. Phys. Commun.}\ }\textbf {\bibinfo
  {volume} {184}},\ \bibinfo {pages} {986--997} (\bibinfo {year}
  {2013}{\natexlab{a}})},\ \Eprint {http://arxiv.org/abs/1110.1499}
  {arXiv:1110.1499 [hep-ph]} \BibitemShut {NoStop}%
\bibitem [{\citenamefont {Cafarella}\ \emph {et~al.}(2009)\citenamefont
  {Cafarella}, \citenamefont {Papadopoulos},\ and\ \citenamefont
  {Worek}}]{Cafarella:2007pc}%
  \BibitemOpen
  \bibfield  {author} {\bibinfo {author} {\bibfnamefont {Alessandro}\
  \bibnamefont {Cafarella}}, \bibinfo {author} {\bibfnamefont {Costas~G.}\
  \bibnamefont {Papadopoulos}}, \ and\ \bibinfo {author} {\bibfnamefont
  {Malgorzata}\ \bibnamefont {Worek}},\ }\bibfield  {title} {\enquote {\bibinfo
  {title} {{Helac-Phegas: A Generator for all parton level processes}},}\
  }\href {\doibase 10.1016/j.cpc.2009.04.023} {\bibfield  {journal} {\bibinfo
  {journal} {Comput. Phys. Commun.}\ }\textbf {\bibinfo {volume} {180}},\
  \bibinfo {pages} {1941--1955} (\bibinfo {year} {2009})},\ \Eprint
  {http://arxiv.org/abs/0710.2427} {arXiv:0710.2427 [hep-ph]} \BibitemShut
  {NoStop}%
\bibitem [{\citenamefont {van Hameren}\ \emph {et~al.}(2009)\citenamefont {van
  Hameren}, \citenamefont {Papadopoulos},\ and\ \citenamefont
  {Pittau}}]{vanHameren:2009dr}%
  \BibitemOpen
  \bibfield  {author} {\bibinfo {author} {\bibfnamefont {A.}~\bibnamefont {van
  Hameren}}, \bibinfo {author} {\bibfnamefont {C.~G.}\ \bibnamefont
  {Papadopoulos}}, \ and\ \bibinfo {author} {\bibfnamefont {R.}~\bibnamefont
  {Pittau}},\ }\bibfield  {title} {\enquote {\bibinfo {title} {{Automated
  one-loop calculations: A Proof of concept}},}\ }\href {\doibase
  10.1088/1126-6708/2009/09/106} {\bibfield  {journal} {\bibinfo  {journal}
  {JHEP}\ }\textbf {\bibinfo {volume} {09}},\ \bibinfo {pages} {106} (\bibinfo
  {year} {2009})},\ \Eprint {http://arxiv.org/abs/0903.4665} {arXiv:0903.4665
  [hep-ph]} \BibitemShut {NoStop}%
\bibitem [{\citenamefont {Czakon}\ \emph {et~al.}(2009)\citenamefont {Czakon},
  \citenamefont {Papadopoulos},\ and\ \citenamefont {Worek}}]{Czakon:2009ss}%
  \BibitemOpen
  \bibfield  {author} {\bibinfo {author} {\bibfnamefont {M.}~\bibnamefont
  {Czakon}}, \bibinfo {author} {\bibfnamefont {C.~G.}\ \bibnamefont
  {Papadopoulos}}, \ and\ \bibinfo {author} {\bibfnamefont {M.}~\bibnamefont
  {Worek}},\ }\bibfield  {title} {\enquote {\bibinfo {title} {{Polarizing the
  Dipoles}},}\ }\href {\doibase 10.1088/1126-6708/2009/08/085} {\bibfield
  {journal} {\bibinfo  {journal} {JHEP}\ }\textbf {\bibinfo {volume} {08}},\
  \bibinfo {pages} {085} (\bibinfo {year} {2009})},\ \Eprint
  {http://arxiv.org/abs/0905.0883} {arXiv:0905.0883 [hep-ph]} \BibitemShut
  {NoStop}%
\bibitem [{\citenamefont {Ossola}\ \emph {et~al.}(2008)\citenamefont {Ossola},
  \citenamefont {Papadopoulos},\ and\ \citenamefont {Pittau}}]{Ossola:2007ax}%
  \BibitemOpen
  \bibfield  {author} {\bibinfo {author} {\bibfnamefont {Giovanni}\
  \bibnamefont {Ossola}}, \bibinfo {author} {\bibfnamefont {Costas~G.}\
  \bibnamefont {Papadopoulos}}, \ and\ \bibinfo {author} {\bibfnamefont
  {Roberto}\ \bibnamefont {Pittau}},\ }\bibfield  {title} {\enquote {\bibinfo
  {title} {{CutTools: A Program implementing the OPP reduction method to
  compute one-loop amplitudes}},}\ }\href {\doibase
  10.1088/1126-6708/2008/03/042} {\bibfield  {journal} {\bibinfo  {journal}
  {JHEP}\ }\textbf {\bibinfo {volume} {03}},\ \bibinfo {pages} {042} (\bibinfo
  {year} {2008})},\ \Eprint {http://arxiv.org/abs/0711.3596} {arXiv:0711.3596
  [hep-ph]} \BibitemShut {NoStop}%
\bibitem [{\citenamefont {Bevilacqua}\ \emph
  {et~al.}(2013{\natexlab{b}})\citenamefont {Bevilacqua}, \citenamefont
  {Czakon}, \citenamefont {Kubocz},\ and\ \citenamefont
  {Worek}}]{Bevilacqua:2013iha}%
  \BibitemOpen
  \bibfield  {author} {\bibinfo {author} {\bibfnamefont {G.}~\bibnamefont
  {Bevilacqua}}, \bibinfo {author} {\bibfnamefont {M.}~\bibnamefont {Czakon}},
  \bibinfo {author} {\bibfnamefont {M.}~\bibnamefont {Kubocz}}, \ and\ \bibinfo
  {author} {\bibfnamefont {M.}~\bibnamefont {Worek}},\ }\bibfield  {title}
  {\enquote {\bibinfo {title} {{Complete Nagy-Soper subtraction for
  next-to-leading order calculations in QCD}},}\ }\href {\doibase
  10.1007/JHEP10(2013)204} {\bibfield  {journal} {\bibinfo  {journal} {JHEP}\
  }\textbf {\bibinfo {volume} {10}},\ \bibinfo {pages} {204} (\bibinfo {year}
  {2013}{\natexlab{b}})},\ \Eprint {http://arxiv.org/abs/1308.5605}
  {arXiv:1308.5605 [hep-ph]} \BibitemShut {NoStop}%
\bibitem [{\citenamefont {Czakon}\ \emph {et~al.}(2015)\citenamefont {Czakon},
  \citenamefont {Hartanto}, \citenamefont {Kraus},\ and\ \citenamefont
  {Worek}}]{Czakon:2015cla}%
  \BibitemOpen
  \bibfield  {author} {\bibinfo {author} {\bibfnamefont {M.}~\bibnamefont
  {Czakon}}, \bibinfo {author} {\bibfnamefont {H.~B.}\ \bibnamefont
  {Hartanto}}, \bibinfo {author} {\bibfnamefont {M.}~\bibnamefont {Kraus}}, \
  and\ \bibinfo {author} {\bibfnamefont {M.}~\bibnamefont {Worek}},\ }\bibfield
   {title} {\enquote {\bibinfo {title} {{Matching the Nagy-Soper parton shower
  at next-to-leading order}},}\ }\href {\doibase 10.1007/JHEP06(2015)033}
  {\bibfield  {journal} {\bibinfo  {journal} {JHEP}\ }\textbf {\bibinfo
  {volume} {06}},\ \bibinfo {pages} {033} (\bibinfo {year} {2015})},\ \Eprint
  {http://arxiv.org/abs/1502.00925} {arXiv:1502.00925 [hep-ph]} \BibitemShut
  {NoStop}%
\bibitem [{\citenamefont {Denner}\ \emph {et~al.}(1999)\citenamefont {Denner},
  \citenamefont {Dittmaier}, \citenamefont {Roth},\ and\ \citenamefont
  {Wackeroth}}]{Denner:1999gp}%
  \BibitemOpen
  \bibfield  {author} {\bibinfo {author} {\bibfnamefont {Ansgar}\ \bibnamefont
  {Denner}}, \bibinfo {author} {\bibfnamefont {S.}~\bibnamefont {Dittmaier}},
  \bibinfo {author} {\bibfnamefont {M.}~\bibnamefont {Roth}}, \ and\ \bibinfo
  {author} {\bibfnamefont {D.}~\bibnamefont {Wackeroth}},\ }\bibfield  {title}
  {\enquote {\bibinfo {title} {{Predictions for all processes e+ e-
  ---\ensuremath{>} 4 fermions + gamma}},}\ }\href {\doibase
  10.1016/S0550-3213(99)00437-X} {\bibfield  {journal} {\bibinfo  {journal}
  {Nucl. Phys. B}\ }\textbf {\bibinfo {volume} {560}},\ \bibinfo {pages}
  {33--65} (\bibinfo {year} {1999})},\ \Eprint
  {http://arxiv.org/abs/hep-ph/9904472} {arXiv:hep-ph/9904472} \BibitemShut
  {NoStop}%
\bibitem [{\citenamefont {Denner}\ \emph {et~al.}(2005)\citenamefont {Denner},
  \citenamefont {Dittmaier}, \citenamefont {Roth},\ and\ \citenamefont
  {Wieders}}]{Denner:2005fg}%
  \BibitemOpen
  \bibfield  {author} {\bibinfo {author} {\bibfnamefont {Ansgar}\ \bibnamefont
  {Denner}}, \bibinfo {author} {\bibfnamefont {S.}~\bibnamefont {Dittmaier}},
  \bibinfo {author} {\bibfnamefont {M.}~\bibnamefont {Roth}}, \ and\ \bibinfo
  {author} {\bibfnamefont {L.~H.}\ \bibnamefont {Wieders}},\ }\bibfield
  {title} {\enquote {\bibinfo {title} {{Electroweak corrections to
  charged-current e+ e- ---\ensuremath{>} 4 fermion processes: Technical
  details and further results}},}\ }\href {\doibase
  10.1016/j.nuclphysb.2011.09.001} {\bibfield  {journal} {\bibinfo  {journal}
  {Nucl. Phys. B}\ }\textbf {\bibinfo {volume} {724}},\ \bibinfo {pages}
  {247--294} (\bibinfo {year} {2005})},\ \bibinfo {note} {[Erratum: Nucl.Phys.B
  854, 504--507 (2012)]},\ \Eprint {http://arxiv.org/abs/hep-ph/0505042}
  {arXiv:hep-ph/0505042} \BibitemShut {NoStop}%
\bibitem [{\citenamefont {Jezabek}\ and\ \citenamefont
  {Kuhn}(1989)}]{Jezabek:1988iv}%
  \BibitemOpen
  \bibfield  {author} {\bibinfo {author} {\bibfnamefont {M.}~\bibnamefont
  {Jezabek}}\ and\ \bibinfo {author} {\bibfnamefont {Johann~H.}\ \bibnamefont
  {Kuhn}},\ }\bibfield  {title} {\enquote {\bibinfo {title} {{QCD Corrections
  to Semileptonic Decays of Heavy Quarks}},}\ }\href {\doibase
  10.1016/0550-3213(89)90108-9} {\bibfield  {journal} {\bibinfo  {journal}
  {Nucl. Phys. B}\ }\textbf {\bibinfo {volume} {314}},\ \bibinfo {pages} {1--6}
  (\bibinfo {year} {1989})}\BibitemShut {NoStop}%
\bibitem [{\citenamefont {Chetyrkin}\ \emph {et~al.}(1999)\citenamefont
  {Chetyrkin}, \citenamefont {Harlander}, \citenamefont {Seidensticker},\ and\
  \citenamefont {Steinhauser}}]{Chetyrkin:1999ju}%
  \BibitemOpen
  \bibfield  {author} {\bibinfo {author} {\bibfnamefont {K.~G.}\ \bibnamefont
  {Chetyrkin}}, \bibinfo {author} {\bibfnamefont {R.}~\bibnamefont
  {Harlander}}, \bibinfo {author} {\bibfnamefont {T.}~\bibnamefont
  {Seidensticker}}, \ and\ \bibinfo {author} {\bibfnamefont {M.}~\bibnamefont
  {Steinhauser}},\ }\bibfield  {title} {\enquote {\bibinfo {title} {{Second
  order QCD corrections to Gamma(t ---\ensuremath{>} W b)}},}\ }\href {\doibase
  10.1103/PhysRevD.60.114015} {\bibfield  {journal} {\bibinfo  {journal} {Phys.
  Rev. D}\ }\textbf {\bibinfo {volume} {60}},\ \bibinfo {pages} {114015}
  (\bibinfo {year} {1999})},\ \Eprint {http://arxiv.org/abs/hep-ph/9906273}
  {arXiv:hep-ph/9906273} \BibitemShut {NoStop}%
\bibitem [{\citenamefont {Denner}\ \emph {et~al.}(2012)\citenamefont {Denner},
  \citenamefont {Dittmaier}, \citenamefont {Kallweit},\ and\ \citenamefont
  {Pozzorini}}]{Denner:2012yc}%
  \BibitemOpen
  \bibfield  {author} {\bibinfo {author} {\bibfnamefont {Ansgar}\ \bibnamefont
  {Denner}}, \bibinfo {author} {\bibfnamefont {Stefan}\ \bibnamefont
  {Dittmaier}}, \bibinfo {author} {\bibfnamefont {Stefan}\ \bibnamefont
  {Kallweit}}, \ and\ \bibinfo {author} {\bibfnamefont {Stefano}\ \bibnamefont
  {Pozzorini}},\ }\bibfield  {title} {\enquote {\bibinfo {title} {{NLO QCD
  corrections to off-shell top-antitop production with leptonic decays at
  hadron colliders}},}\ }\href {\doibase 10.1007/JHEP10(2012)110} {\bibfield
  {journal} {\bibinfo  {journal} {JHEP}\ }\textbf {\bibinfo {volume} {10}},\
  \bibinfo {pages} {110} (\bibinfo {year} {2012})},\ \Eprint
  {http://arxiv.org/abs/1207.5018} {arXiv:1207.5018 [hep-ph]} \BibitemShut
  {NoStop}%
\bibitem [{\citenamefont {Bevilacqua}\ \emph
  {et~al.}(2020{\natexlab{b}})\citenamefont {Bevilacqua}, \citenamefont
  {Hartanto}, \citenamefont {Kraus}, \citenamefont {Weber},\ and\ \citenamefont
  {Worek}}]{Bevilacqua:2019quz}%
  \BibitemOpen
  \bibfield  {author} {\bibinfo {author} {\bibfnamefont {G.}~\bibnamefont
  {Bevilacqua}}, \bibinfo {author} {\bibfnamefont {H.~B.}\ \bibnamefont
  {Hartanto}}, \bibinfo {author} {\bibfnamefont {M.}~\bibnamefont {Kraus}},
  \bibinfo {author} {\bibfnamefont {T.}~\bibnamefont {Weber}}, \ and\ \bibinfo
  {author} {\bibfnamefont {M.}~\bibnamefont {Worek}},\ }\bibfield  {title}
  {\enquote {\bibinfo {title} {{Off-shell vs on-shell modelling of top quarks
  in photon associated production}},}\ }\href {\doibase
  10.1007/JHEP03(2020)154} {\bibfield  {journal} {\bibinfo  {journal} {JHEP}\
  }\textbf {\bibinfo {volume} {03}},\ \bibinfo {pages} {154} (\bibinfo {year}
  {2020}{\natexlab{b}})},\ \Eprint {http://arxiv.org/abs/1912.09999}
  {arXiv:1912.09999 [hep-ph]} \BibitemShut {NoStop}%
\bibitem [{\citenamefont {Honeywell}\ \emph {et~al.}(2020)\citenamefont
  {Honeywell}, \citenamefont {Quackenbush}, \citenamefont {Reina},\ and\
  \citenamefont {Reuschle}}]{Honeywell:2018fcl}%
  \BibitemOpen
  \bibfield  {author} {\bibinfo {author} {\bibfnamefont {Steve}\ \bibnamefont
  {Honeywell}}, \bibinfo {author} {\bibfnamefont {Seth}\ \bibnamefont
  {Quackenbush}}, \bibinfo {author} {\bibfnamefont {Laura}\ \bibnamefont
  {Reina}}, \ and\ \bibinfo {author} {\bibfnamefont {Christian}\ \bibnamefont
  {Reuschle}},\ }\bibfield  {title} {\enquote {\bibinfo {title} {{NLOX, a
  one-loop provider for Standard Model processes}},}\ }\href {\doibase
  10.1016/j.cpc.2020.107284} {\bibfield  {journal} {\bibinfo  {journal}
  {Comput. Phys. Commun.}\ }\textbf {\bibinfo {volume} {257}},\ \bibinfo
  {pages} {107284} (\bibinfo {year} {2020})},\ \Eprint
  {http://arxiv.org/abs/1812.11925} {arXiv:1812.11925 [hep-ph]} \BibitemShut
  {NoStop}%
\bibitem [{\citenamefont {Figueroa}\ \emph {et~al.}(2022)\citenamefont
  {Figueroa}, \citenamefont {Quackenbush}, \citenamefont {Reina},\ and\
  \citenamefont {Reuschle}}]{Figueroa:2021txg}%
  \BibitemOpen
  \bibfield  {author} {\bibinfo {author} {\bibfnamefont {Diogenes}\
  \bibnamefont {Figueroa}}, \bibinfo {author} {\bibfnamefont {Seth}\
  \bibnamefont {Quackenbush}}, \bibinfo {author} {\bibfnamefont {Laura}\
  \bibnamefont {Reina}}, \ and\ \bibinfo {author} {\bibfnamefont {Christian}\
  \bibnamefont {Reuschle}},\ }\bibfield  {title} {\enquote {\bibinfo {title}
  {{Updates to the one-loop provider NLOX}},}\ }\href {\doibase
  10.1016/j.cpc.2021.108150} {\bibfield  {journal} {\bibinfo  {journal}
  {Comput. Phys. Commun.}\ }\textbf {\bibinfo {volume} {270}},\ \bibinfo
  {pages} {108150} (\bibinfo {year} {2022})},\ \Eprint
  {http://arxiv.org/abs/2101.01305} {arXiv:2101.01305 [hep-ph]} \BibitemShut
  {NoStop}%
\bibitem [{\citenamefont {Frixione}\ \emph
  {et~al.}(2007{\natexlab{b}})\citenamefont {Frixione}, \citenamefont {Laenen},
  \citenamefont {Motylinski},\ and\ \citenamefont {Webber}}]{Frixione:2007zp}%
  \BibitemOpen
  \bibfield  {author} {\bibinfo {author} {\bibfnamefont {Stefano}\ \bibnamefont
  {Frixione}}, \bibinfo {author} {\bibfnamefont {Eric}\ \bibnamefont {Laenen}},
  \bibinfo {author} {\bibfnamefont {Patrick}\ \bibnamefont {Motylinski}}, \
  and\ \bibinfo {author} {\bibfnamefont {Bryan~R.}\ \bibnamefont {Webber}},\
  }\bibfield  {title} {\enquote {\bibinfo {title} {{Angular correlations of
  lepton pairs from vector boson and top quark decays in Monte Carlo
  simulations}},}\ }\href {\doibase 10.1088/1126-6708/2007/04/081} {\bibfield
  {journal} {\bibinfo  {journal} {JHEP}\ }\textbf {\bibinfo {volume} {04}},\
  \bibinfo {pages} {081} (\bibinfo {year} {2007}{\natexlab{b}})},\ \Eprint
  {http://arxiv.org/abs/hep-ph/0702198} {arXiv:hep-ph/0702198} \BibitemShut
  {NoStop}%
\bibitem [{\citenamefont {Alwall}\ \emph {et~al.}(2014)\citenamefont {Alwall},
  \citenamefont {Frederix}, \citenamefont {Frixione}, \citenamefont {Hirschi},
  \citenamefont {Maltoni}, \citenamefont {Mattelaer}, \citenamefont {Shao},
  \citenamefont {Stelzer}, \citenamefont {Torrielli},\ and\ \citenamefont
  {Zaro}}]{Alwall:2014hca}%
  \BibitemOpen
  \bibfield  {author} {\bibinfo {author} {\bibfnamefont {J.}~\bibnamefont
  {Alwall}}, \bibinfo {author} {\bibfnamefont {R.}~\bibnamefont {Frederix}},
  \bibinfo {author} {\bibfnamefont {S.}~\bibnamefont {Frixione}}, \bibinfo
  {author} {\bibfnamefont {V.}~\bibnamefont {Hirschi}}, \bibinfo {author}
  {\bibfnamefont {F.}~\bibnamefont {Maltoni}}, \bibinfo {author} {\bibfnamefont
  {O.}~\bibnamefont {Mattelaer}}, \bibinfo {author} {\bibfnamefont {H.~S.}\
  \bibnamefont {Shao}}, \bibinfo {author} {\bibfnamefont {T.}~\bibnamefont
  {Stelzer}}, \bibinfo {author} {\bibfnamefont {P.}~\bibnamefont {Torrielli}},
  \ and\ \bibinfo {author} {\bibfnamefont {M.}~\bibnamefont {Zaro}},\
  }\bibfield  {title} {\enquote {\bibinfo {title} {{The automated computation
  of tree-level and next-to-leading order differential cross sections, and
  their matching to parton shower simulations}},}\ }\href {\doibase
  10.1007/JHEP07(2014)079} {\bibfield  {journal} {\bibinfo  {journal} {JHEP}\
  }\textbf {\bibinfo {volume} {07}},\ \bibinfo {pages} {079} (\bibinfo {year}
  {2014})},\ \Eprint {http://arxiv.org/abs/1405.0301} {arXiv:1405.0301
  [hep-ph]} \BibitemShut {NoStop}%
\bibitem [{\citenamefont {Artoisenet}\ \emph {et~al.}(2013)\citenamefont
  {Artoisenet}, \citenamefont {Frederix}, \citenamefont {Mattelaer},\ and\
  \citenamefont {Rietkerk}}]{Artoisenet:2012st}%
  \BibitemOpen
  \bibfield  {author} {\bibinfo {author} {\bibfnamefont {Pierre}\ \bibnamefont
  {Artoisenet}}, \bibinfo {author} {\bibfnamefont {Rikkert}\ \bibnamefont
  {Frederix}}, \bibinfo {author} {\bibfnamefont {Olivier}\ \bibnamefont
  {Mattelaer}}, \ and\ \bibinfo {author} {\bibfnamefont {Robbert}\ \bibnamefont
  {Rietkerk}},\ }\bibfield  {title} {\enquote {\bibinfo {title} {{Automatic
  spin-entangled decays of heavy resonances in Monte Carlo simulations}},}\
  }\href {\doibase 10.1007/JHEP03(2013)015} {\bibfield  {journal} {\bibinfo
  {journal} {JHEP}\ }\textbf {\bibinfo {volume} {03}},\ \bibinfo {pages} {015}
  (\bibinfo {year} {2013})},\ \Eprint {http://arxiv.org/abs/1212.3460}
  {arXiv:1212.3460 [hep-ph]} \BibitemShut {NoStop}%
\bibitem [{\citenamefont {Boos}\ \emph {et~al.}(2001)\citenamefont {Boos} \emph
  {et~al.}}]{Boos:2001cv}%
  \BibitemOpen
  \bibfield  {author} {\bibinfo {author} {\bibfnamefont {E.}~\bibnamefont
  {Boos}} \emph {et~al.},\ }\bibfield  {title} {\enquote {\bibinfo {title}
  {{Generic User Process Interface for Event Generators}},}\ }in\ \href@noop {}
  {\emph {\bibinfo {booktitle} {{2nd Les Houches Workshop on Physics at TeV
  Colliders}}}}\ (\bibinfo {year} {2001})\ \Eprint
  {http://arxiv.org/abs/hep-ph/0109068} {arXiv:hep-ph/0109068} \BibitemShut
  {NoStop}%
\bibitem [{\citenamefont {Alwall}\ \emph {et~al.}(2007)\citenamefont {Alwall}
  \emph {et~al.}}]{Alwall:2006yp}%
  \BibitemOpen
  \bibfield  {author} {\bibinfo {author} {\bibfnamefont {Johan}\ \bibnamefont
  {Alwall}} \emph {et~al.},\ }\bibfield  {title} {\enquote {\bibinfo {title}
  {{A Standard format for Les Houches event files}},}\ }\href {\doibase
  10.1016/j.cpc.2006.11.010} {\bibfield  {journal} {\bibinfo  {journal}
  {Comput. Phys. Commun.}\ }\textbf {\bibinfo {volume} {176}},\ \bibinfo
  {pages} {300--304} (\bibinfo {year} {2007})},\ \Eprint
  {http://arxiv.org/abs/hep-ph/0609017} {arXiv:hep-ph/0609017} \BibitemShut
  {NoStop}%
\bibitem [{\citenamefont {Sjostrand}\ \emph {et~al.}(2006)\citenamefont
  {Sjostrand}, \citenamefont {Mrenna},\ and\ \citenamefont
  {Skands}}]{Sjostrand:2006za}%
  \BibitemOpen
  \bibfield  {author} {\bibinfo {author} {\bibfnamefont {Torbjorn}\
  \bibnamefont {Sjostrand}}, \bibinfo {author} {\bibfnamefont {Stephen}\
  \bibnamefont {Mrenna}}, \ and\ \bibinfo {author} {\bibfnamefont {Peter~Z.}\
  \bibnamefont {Skands}},\ }\bibfield  {title} {\enquote {\bibinfo {title}
  {{PYTHIA 6.4 Physics and Manual}},}\ }\href {\doibase
  10.1088/1126-6708/2006/05/026} {\bibfield  {journal} {\bibinfo  {journal}
  {JHEP}\ }\textbf {\bibinfo {volume} {05}},\ \bibinfo {pages} {026} (\bibinfo
  {year} {2006})},\ \Eprint {http://arxiv.org/abs/hep-ph/0603175}
  {arXiv:hep-ph/0603175} \BibitemShut {NoStop}%
\bibitem [{\citenamefont {Sj\"ostrand}\ \emph {et~al.}(2015)\citenamefont
  {Sj\"ostrand}, \citenamefont {Ask}, \citenamefont {Christiansen},
  \citenamefont {Corke}, \citenamefont {Desai}, \citenamefont {Ilten},
  \citenamefont {Mrenna}, \citenamefont {Prestel}, \citenamefont {Rasmussen},\
  and\ \citenamefont {Skands}}]{Sjostrand:2014zea}%
  \BibitemOpen
  \bibfield  {author} {\bibinfo {author} {\bibfnamefont {Torbj\"orn}\
  \bibnamefont {Sj\"ostrand}}, \bibinfo {author} {\bibfnamefont {Stefan}\
  \bibnamefont {Ask}}, \bibinfo {author} {\bibfnamefont {Jesper~R.}\
  \bibnamefont {Christiansen}}, \bibinfo {author} {\bibfnamefont {Richard}\
  \bibnamefont {Corke}}, \bibinfo {author} {\bibfnamefont {Nishita}\
  \bibnamefont {Desai}}, \bibinfo {author} {\bibfnamefont {Philip}\
  \bibnamefont {Ilten}}, \bibinfo {author} {\bibfnamefont {Stephen}\
  \bibnamefont {Mrenna}}, \bibinfo {author} {\bibfnamefont {Stefan}\
  \bibnamefont {Prestel}}, \bibinfo {author} {\bibfnamefont {Christine~O.}\
  \bibnamefont {Rasmussen}}, \ and\ \bibinfo {author} {\bibfnamefont
  {Peter~Z.}\ \bibnamefont {Skands}},\ }\bibfield  {title} {\enquote {\bibinfo
  {title} {{An introduction to PYTHIA 8.2}},}\ }\href {\doibase
  10.1016/j.cpc.2015.01.024} {\bibfield  {journal} {\bibinfo  {journal}
  {Comput. Phys. Commun.}\ }\textbf {\bibinfo {volume} {191}},\ \bibinfo
  {pages} {159--177} (\bibinfo {year} {2015})},\ \Eprint
  {http://arxiv.org/abs/1410.3012} {arXiv:1410.3012 [hep-ph]} \BibitemShut
  {NoStop}%
\bibitem [{\citenamefont {Buckley}\ \emph {et~al.}(2013)\citenamefont
  {Buckley}, \citenamefont {Butterworth}, \citenamefont {Grellscheid},
  \citenamefont {Hoeth}, \citenamefont {Lonnblad}, \citenamefont {Monk},
  \citenamefont {Schulz},\ and\ \citenamefont {Siegert}}]{Buckley:2010ar}%
  \BibitemOpen
  \bibfield  {author} {\bibinfo {author} {\bibfnamefont {Andy}\ \bibnamefont
  {Buckley}}, \bibinfo {author} {\bibfnamefont {Jonathan}\ \bibnamefont
  {Butterworth}}, \bibinfo {author} {\bibfnamefont {David}\ \bibnamefont
  {Grellscheid}}, \bibinfo {author} {\bibfnamefont {Hendrik}\ \bibnamefont
  {Hoeth}}, \bibinfo {author} {\bibfnamefont {Leif}\ \bibnamefont {Lonnblad}},
  \bibinfo {author} {\bibfnamefont {James}\ \bibnamefont {Monk}}, \bibinfo
  {author} {\bibfnamefont {Holger}\ \bibnamefont {Schulz}}, \ and\ \bibinfo
  {author} {\bibfnamefont {Frank}\ \bibnamefont {Siegert}},\ }\bibfield
  {title} {\enquote {\bibinfo {title} {{Rivet user manual}},}\ }\href {\doibase
  10.1016/j.cpc.2013.05.021} {\bibfield  {journal} {\bibinfo  {journal}
  {Comput. Phys. Commun.}\ }\textbf {\bibinfo {volume} {184}},\ \bibinfo
  {pages} {2803--2819} (\bibinfo {year} {2013})},\ \Eprint
  {http://arxiv.org/abs/1003.0694} {arXiv:1003.0694 [hep-ph]} \BibitemShut
  {NoStop}%
\bibitem [{\citenamefont {Bierlich}\ \emph {et~al.}(2020)\citenamefont
  {Bierlich} \emph {et~al.}}]{Bierlich:2019rhm}%
  \BibitemOpen
  \bibfield  {author} {\bibinfo {author} {\bibfnamefont {Christian}\
  \bibnamefont {Bierlich}} \emph {et~al.},\ }\bibfield  {title} {\enquote
  {\bibinfo {title} {{Robust Independent Validation of Experiment and Theory:
  Rivet version 3}},}\ }\href {\doibase 10.21468/SciPostPhys.8.2.026}
  {\bibfield  {journal} {\bibinfo  {journal} {SciPost Phys.}\ }\textbf
  {\bibinfo {volume} {8}},\ \bibinfo {pages} {026} (\bibinfo {year} {2020})},\
  \Eprint {http://arxiv.org/abs/1912.05451} {arXiv:1912.05451 [hep-ph]}
  \BibitemShut {NoStop}%
\bibitem [{\citenamefont {Cacciari}\ and\ \citenamefont
  {Salam}(2006)}]{Cacciari:2005hq}%
  \BibitemOpen
  \bibfield  {author} {\bibinfo {author} {\bibfnamefont {Matteo}\ \bibnamefont
  {Cacciari}}\ and\ \bibinfo {author} {\bibfnamefont {Gavin~P.}\ \bibnamefont
  {Salam}},\ }\bibfield  {title} {\enquote {\bibinfo {title} {{Dispelling the
  $N^{3}$ myth for the $k_t$ jet-finder}},}\ }\href {\doibase
  10.1016/j.physletb.2006.08.037} {\bibfield  {journal} {\bibinfo  {journal}
  {Phys. Lett. B}\ }\textbf {\bibinfo {volume} {641}},\ \bibinfo {pages}
  {57--61} (\bibinfo {year} {2006})},\ \Eprint
  {http://arxiv.org/abs/hep-ph/0512210} {arXiv:hep-ph/0512210} \BibitemShut
  {NoStop}%
\bibitem [{\citenamefont {Cacciari}\ \emph {et~al.}(2012)\citenamefont
  {Cacciari}, \citenamefont {Salam},\ and\ \citenamefont
  {Soyez}}]{Cacciari:2011ma}%
  \BibitemOpen
  \bibfield  {author} {\bibinfo {author} {\bibfnamefont {Matteo}\ \bibnamefont
  {Cacciari}}, \bibinfo {author} {\bibfnamefont {Gavin~P.}\ \bibnamefont
  {Salam}}, \ and\ \bibinfo {author} {\bibfnamefont {Gregory}\ \bibnamefont
  {Soyez}},\ }\bibfield  {title} {\enquote {\bibinfo {title} {{FastJet User
  Manual}},}\ }\href {\doibase 10.1140/epjc/s10052-012-1896-2} {\bibfield
  {journal} {\bibinfo  {journal} {Eur. Phys. J. C}\ }\textbf {\bibinfo {volume}
  {72}},\ \bibinfo {pages} {1896} (\bibinfo {year} {2012})},\ \Eprint
  {http://arxiv.org/abs/1111.6097} {arXiv:1111.6097 [hep-ph]} \BibitemShut
  {NoStop}%
\bibitem [{\citenamefont {Antcheva}\ \emph {et~al.}(2009)\citenamefont
  {Antcheva} \emph {et~al.}}]{Antcheva:2009zz}%
  \BibitemOpen
  \bibfield  {author} {\bibinfo {author} {\bibfnamefont {I.}~\bibnamefont
  {Antcheva}} \emph {et~al.},\ }\bibfield  {title} {\enquote {\bibinfo {title}
  {{ROOT: A C++ framework for petabyte data storage, statistical analysis and
  visualization}},}\ }\href {\doibase 10.1016/j.cpc.2009.08.005} {\bibfield
  {journal} {\bibinfo  {journal} {Comput. Phys. Commun.}\ }\textbf {\bibinfo
  {volume} {180}},\ \bibinfo {pages} {2499--2512} (\bibinfo {year} {2009})},\
  \Eprint {http://arxiv.org/abs/1508.07749} {arXiv:1508.07749
  [physics.data-an]} \BibitemShut {NoStop}%
\bibitem [{\citenamefont {Bern}\ \emph {et~al.}(2014)\citenamefont {Bern},
  \citenamefont {Dixon}, \citenamefont {Febres~Cordero}, \citenamefont
  {H\"oche}, \citenamefont {Ita}, \citenamefont {Kosower},\ and\ \citenamefont
  {Maitre}}]{Bern:2013zja}%
  \BibitemOpen
  \bibfield  {author} {\bibinfo {author} {\bibfnamefont {Z.}~\bibnamefont
  {Bern}}, \bibinfo {author} {\bibfnamefont {L.~J.}\ \bibnamefont {Dixon}},
  \bibinfo {author} {\bibfnamefont {F.}~\bibnamefont {Febres~Cordero}},
  \bibinfo {author} {\bibfnamefont {S.}~\bibnamefont {H\"oche}}, \bibinfo
  {author} {\bibfnamefont {H.}~\bibnamefont {Ita}}, \bibinfo {author}
  {\bibfnamefont {D.~A.}\ \bibnamefont {Kosower}}, \ and\ \bibinfo {author}
  {\bibfnamefont {D.}~\bibnamefont {Maitre}},\ }\bibfield  {title} {\enquote
  {\bibinfo {title} {{Ntuples for NLO Events at Hadron Colliders}},}\ }\href
  {\doibase 10.1016/j.cpc.2014.01.011} {\bibfield  {journal} {\bibinfo
  {journal} {Comput. Phys. Commun.}\ }\textbf {\bibinfo {volume} {185}},\
  \bibinfo {pages} {1443--1460} (\bibinfo {year} {2014})},\ \Eprint
  {http://arxiv.org/abs/1310.7439} {arXiv:1310.7439 [hep-ph]} \BibitemShut
  {NoStop}%
\bibitem [{\citenamefont {Bevilacqua}\ \emph {et~al.}(2016)\citenamefont
  {Bevilacqua}, \citenamefont {Hartanto}, \citenamefont {Kraus},\ and\
  \citenamefont {Worek}}]{Bevilacqua:2016jfk}%
  \BibitemOpen
  \bibfield  {author} {\bibinfo {author} {\bibfnamefont {G.}~\bibnamefont
  {Bevilacqua}}, \bibinfo {author} {\bibfnamefont {H.~B.}\ \bibnamefont
  {Hartanto}}, \bibinfo {author} {\bibfnamefont {M.}~\bibnamefont {Kraus}}, \
  and\ \bibinfo {author} {\bibfnamefont {M.}~\bibnamefont {Worek}},\ }\bibfield
   {title} {\enquote {\bibinfo {title} {{Off-shell Top Quarks with One Jet at
  the LHC: A comprehensive analysis at NLO QCD}},}\ }\href {\doibase
  10.1007/JHEP11(2016)098} {\bibfield  {journal} {\bibinfo  {journal} {JHEP}\
  }\textbf {\bibinfo {volume} {11}},\ \bibinfo {pages} {098} (\bibinfo {year}
  {2016})},\ \Eprint {http://arxiv.org/abs/1609.01659} {arXiv:1609.01659
  [hep-ph]} \BibitemShut {NoStop}%
\bibitem [{\citenamefont {Bevilacqua}(2019)}]{heplot}%
  \BibitemOpen
  \bibfield  {author} {\bibinfo {author} {\bibfnamefont {G.}~\bibnamefont
  {Bevilacqua}},\ }\href@noop {} {\enquote {\bibinfo {title} {{Heplot}},}\
  }\bibinfo {howpublished} {unpublished} (\bibinfo {year} {2019})\BibitemShut
  {NoStop}%
\bibitem [{\citenamefont {Harlander}\ \emph {et~al.}(2020)\citenamefont
  {Harlander}, \citenamefont {Klein},\ and\ \citenamefont
  {Lipp}}]{Harlander:2020cyh}%
  \BibitemOpen
  \bibfield  {author} {\bibinfo {author} {\bibfnamefont {R.~V.}\ \bibnamefont
  {Harlander}}, \bibinfo {author} {\bibfnamefont {S.~Y.}\ \bibnamefont
  {Klein}}, \ and\ \bibinfo {author} {\bibfnamefont {M.}~\bibnamefont {Lipp}},\
  }\bibfield  {title} {\enquote {\bibinfo {title} {{FeynGame}},}\ }\href
  {\doibase 10.1016/j.cpc.2020.107465} {\bibfield  {journal} {\bibinfo
  {journal} {Comput. Phys. Commun.}\ }\textbf {\bibinfo {volume} {256}},\
  \bibinfo {pages} {107465} (\bibinfo {year} {2020})},\ \Eprint
  {http://arxiv.org/abs/2003.00896} {arXiv:2003.00896 [physics.ed-ph]}
  \BibitemShut {NoStop}%
\bibitem [{\citenamefont {Banfi}\ \emph {et~al.}(2006)\citenamefont {Banfi},
  \citenamefont {Salam},\ and\ \citenamefont {Zanderighi}}]{Banfi:2006hf}%
  \BibitemOpen
  \bibfield  {author} {\bibinfo {author} {\bibfnamefont {Andrea}\ \bibnamefont
  {Banfi}}, \bibinfo {author} {\bibfnamefont {Gavin~P.}\ \bibnamefont {Salam}},
  \ and\ \bibinfo {author} {\bibfnamefont {Giulia}\ \bibnamefont
  {Zanderighi}},\ }\bibfield  {title} {\enquote {\bibinfo {title} {{Infrared
  safe definition of jet flavor}},}\ }\href {\doibase
  10.1140/epjc/s2006-02552-4} {\bibfield  {journal} {\bibinfo  {journal} {Eur.
  Phys. J. C}\ }\textbf {\bibinfo {volume} {47}},\ \bibinfo {pages} {113--124}
  (\bibinfo {year} {2006})},\ \Eprint {http://arxiv.org/abs/hep-ph/0601139}
  {arXiv:hep-ph/0601139} \BibitemShut {NoStop}%
\bibitem [{\citenamefont {Czakon}()}]{antikTflav}%
  \BibitemOpen
  \bibfield  {author} {\bibinfo {author} {\bibfnamefont {M.}~\bibnamefont
  {Czakon}},\ }\href
  {https://indico.cern.ch/event/1018454/contributions/4339316/} {\enquote
  {\bibinfo {title} {{Flavored jets in top physics and beyond}},}\ }\bibinfo
  {note} {{14th International Workshop on Top Quark Physics (TOP2021), 13-17
  September 2021 (Online workshop)}}\BibitemShut {NoStop}%
\bibitem [{\citenamefont {Cacciari}\ \emph {et~al.}(2008)\citenamefont
  {Cacciari}, \citenamefont {Salam},\ and\ \citenamefont
  {Soyez}}]{Cacciari:2008gp}%
  \BibitemOpen
  \bibfield  {author} {\bibinfo {author} {\bibfnamefont {Matteo}\ \bibnamefont
  {Cacciari}}, \bibinfo {author} {\bibfnamefont {Gavin~P.}\ \bibnamefont
  {Salam}}, \ and\ \bibinfo {author} {\bibfnamefont {Gregory}\ \bibnamefont
  {Soyez}},\ }\bibfield  {title} {\enquote {\bibinfo {title} {{The anti-$k_t$
  jet clustering algorithm}},}\ }\href {\doibase 10.1088/1126-6708/2008/04/063}
  {\bibfield  {journal} {\bibinfo  {journal} {JHEP}\ }\textbf {\bibinfo
  {volume} {04}},\ \bibinfo {pages} {063} (\bibinfo {year} {2008})},\ \Eprint
  {http://arxiv.org/abs/0802.1189} {arXiv:0802.1189 [hep-ph]} \BibitemShut
  {NoStop}%
\bibitem [{\citenamefont {Biswas}\ \emph {et~al.}(2010)\citenamefont {Biswas},
  \citenamefont {Melnikov},\ and\ \citenamefont {Schulze}}]{Biswas:2010sa}%
  \BibitemOpen
  \bibfield  {author} {\bibinfo {author} {\bibfnamefont {Sandip}\ \bibnamefont
  {Biswas}}, \bibinfo {author} {\bibfnamefont {Kirill}\ \bibnamefont
  {Melnikov}}, \ and\ \bibinfo {author} {\bibfnamefont {Markus}\ \bibnamefont
  {Schulze}},\ }\bibfield  {title} {\enquote {\bibinfo {title}
  {{Next-to-leading order QCD effects and the top quark mass measurements at
  the LHC}},}\ }\href {\doibase 10.1007/JHEP08(2010)048} {\bibfield  {journal}
  {\bibinfo  {journal} {JHEP}\ }\textbf {\bibinfo {volume} {08}},\ \bibinfo
  {pages} {048} (\bibinfo {year} {2010})},\ \Eprint
  {http://arxiv.org/abs/1006.0910} {arXiv:1006.0910 [hep-ph]} \BibitemShut
  {NoStop}%
\bibitem [{\citenamefont {Heinrich}\ \emph {et~al.}(2014)\citenamefont
  {Heinrich}, \citenamefont {Maier}, \citenamefont {Nisius}, \citenamefont
  {Schlenk},\ and\ \citenamefont {Winter}}]{Heinrich:2013qaa}%
  \BibitemOpen
  \bibfield  {author} {\bibinfo {author} {\bibfnamefont {Gudrun}\ \bibnamefont
  {Heinrich}}, \bibinfo {author} {\bibfnamefont {Andreas}\ \bibnamefont
  {Maier}}, \bibinfo {author} {\bibfnamefont {Richard}\ \bibnamefont {Nisius}},
  \bibinfo {author} {\bibfnamefont {Johannes}\ \bibnamefont {Schlenk}}, \ and\
  \bibinfo {author} {\bibfnamefont {Jan}\ \bibnamefont {Winter}},\ }\bibfield
  {title} {\enquote {\bibinfo {title} {{NLO QCD corrections to $W^{+}
  W^{-}b\bar{b}$ production with leptonic decays in the light of top quark mass
  and asymmetry measurements}},}\ }\href {\doibase 10.1007/JHEP06(2014)158}
  {\bibfield  {journal} {\bibinfo  {journal} {JHEP}\ }\textbf {\bibinfo
  {volume} {06}},\ \bibinfo {pages} {158} (\bibinfo {year} {2014})},\ \Eprint
  {http://arxiv.org/abs/1312.6659} {arXiv:1312.6659 [hep-ph]} \BibitemShut
  {NoStop}%
\bibitem [{\citenamefont {Heinrich}\ \emph {et~al.}(2018)\citenamefont
  {Heinrich}, \citenamefont {Maier}, \citenamefont {Nisius}, \citenamefont
  {Schlenk}, \citenamefont {Schulze}, \citenamefont {Scyboz},\ and\
  \citenamefont {Winter}}]{Heinrich:2017bqp}%
  \BibitemOpen
  \bibfield  {author} {\bibinfo {author} {\bibfnamefont {G.}~\bibnamefont
  {Heinrich}}, \bibinfo {author} {\bibfnamefont {Andreas}\ \bibnamefont
  {Maier}}, \bibinfo {author} {\bibfnamefont {Richard}\ \bibnamefont {Nisius}},
  \bibinfo {author} {\bibfnamefont {Johannes}\ \bibnamefont {Schlenk}},
  \bibinfo {author} {\bibfnamefont {Markus}\ \bibnamefont {Schulze}}, \bibinfo
  {author} {\bibfnamefont {Ludovic}\ \bibnamefont {Scyboz}}, \ and\ \bibinfo
  {author} {\bibfnamefont {Jan}\ \bibnamefont {Winter}},\ }\bibfield  {title}
  {\enquote {\bibinfo {title} {{NLO and off-shell effects in top quark mass
  determinations}},}\ }\href {\doibase 10.1007/JHEP07(2018)129} {\bibfield
  {journal} {\bibinfo  {journal} {JHEP}\ }\textbf {\bibinfo {volume} {07}},\
  \bibinfo {pages} {129} (\bibinfo {year} {2018})},\ \Eprint
  {http://arxiv.org/abs/1709.08615} {arXiv:1709.08615 [hep-ph]} \BibitemShut
  {NoStop}%
\bibitem [{\citenamefont {Bevilacqua}\ \emph {et~al.}(2018)\citenamefont
  {Bevilacqua}, \citenamefont {Hartanto}, \citenamefont {Kraus}, \citenamefont
  {Schulze},\ and\ \citenamefont {Worek}}]{Bevilacqua:2017ipv}%
  \BibitemOpen
  \bibfield  {author} {\bibinfo {author} {\bibfnamefont {G.}~\bibnamefont
  {Bevilacqua}}, \bibinfo {author} {\bibfnamefont {H.~B.}\ \bibnamefont
  {Hartanto}}, \bibinfo {author} {\bibfnamefont {M.}~\bibnamefont {Kraus}},
  \bibinfo {author} {\bibfnamefont {M.}~\bibnamefont {Schulze}}, \ and\
  \bibinfo {author} {\bibfnamefont {M.}~\bibnamefont {Worek}},\ }\bibfield
  {title} {\enquote {\bibinfo {title} {{Top quark mass studies with $
  t\overline{t}j $ at the LHC}},}\ }\href {\doibase 10.1007/JHEP03(2018)169}
  {\bibfield  {journal} {\bibinfo  {journal} {JHEP}\ }\textbf {\bibinfo
  {volume} {03}},\ \bibinfo {pages} {169} (\bibinfo {year} {2018})},\ \Eprint
  {http://arxiv.org/abs/1710.07515} {arXiv:1710.07515 [hep-ph]} \BibitemShut
  {NoStop}%
\bibitem [{\citenamefont {Je\v{z}o}\ and\ \citenamefont
  {Nason}(2015)}]{Jezo:2015aia}%
  \BibitemOpen
  \bibfield  {author} {\bibinfo {author} {\bibfnamefont {Tom\'a\v{s}}\
  \bibnamefont {Je\v{z}o}}\ and\ \bibinfo {author} {\bibfnamefont {Paolo}\
  \bibnamefont {Nason}},\ }\bibfield  {title} {\enquote {\bibinfo {title} {{On
  the Treatment of Resonances in Next-to-Leading Order Calculations Matched to
  a Parton Shower}},}\ }\href {\doibase 10.1007/JHEP12(2015)065} {\bibfield
  {journal} {\bibinfo  {journal} {JHEP}\ }\textbf {\bibinfo {volume} {12}},\
  \bibinfo {pages} {065} (\bibinfo {year} {2015})},\ \Eprint
  {http://arxiv.org/abs/1509.09071} {arXiv:1509.09071 [hep-ph]} \BibitemShut
  {NoStop}%
\bibitem [{\citenamefont {Je\v{z}o}\ \emph {et~al.}(2016)\citenamefont
  {Je\v{z}o}, \citenamefont {Lindert}, \citenamefont {Nason}, \citenamefont
  {Oleari},\ and\ \citenamefont {Pozzorini}}]{Jezo:2016ujg}%
  \BibitemOpen
  \bibfield  {author} {\bibinfo {author} {\bibfnamefont {Tom\'a\v{s}}\
  \bibnamefont {Je\v{z}o}}, \bibinfo {author} {\bibfnamefont {Jonas~M.}\
  \bibnamefont {Lindert}}, \bibinfo {author} {\bibfnamefont {Paolo}\
  \bibnamefont {Nason}}, \bibinfo {author} {\bibfnamefont {Carlo}\ \bibnamefont
  {Oleari}}, \ and\ \bibinfo {author} {\bibfnamefont {Stefano}\ \bibnamefont
  {Pozzorini}},\ }\bibfield  {title} {\enquote {\bibinfo {title} {{An NLO+PS
  generator for $t\bar{t}$ and $Wt$ production and decay including non-resonant
  and interference effects}},}\ }\href {\doibase
  10.1140/epjc/s10052-016-4538-2} {\bibfield  {journal} {\bibinfo  {journal}
  {Eur. Phys. J. C}\ }\textbf {\bibinfo {volume} {76}},\ \bibinfo {pages} {691}
  (\bibinfo {year} {2016})},\ \Eprint {http://arxiv.org/abs/1607.04538}
  {arXiv:1607.04538 [hep-ph]} \BibitemShut {NoStop}%
\end{thebibliography}%
\end{document}